\documentclass[fleqn,usenatbib]{mnras}

\usepackage{newtxtext,newtxmath}

\usepackage[T1]{fontenc}
\usepackage{ae,aecompl}

\usepackage{graphicx}	
\usepackage{amsmath}	
\usepackage{csvsimple}
\usepackage{lscape}
\usepackage{gensymb}
\usepackage{pgfplotstable,booktabs}

\title[Halo mass dependence of galaxy populations]{The GOGREEN survey: Dependence of galaxy properties on halo mass at $z>1$ and implications for environmental quenching}

\author[Andrew M. M. Reeves et al.]{\newauthor
Andrew M. M. Reeves$^{1,2}$\thanks{E-mail: andrew.reeves@uwaterloo.ca},
Michael L. Balogh$^{1,2}$,
Remco F. J. van der Burg$^{3}$,
\newauthor Alexis Finoguenov$^{4}$,
Egidijus Kukstas$^{5}$,
Ian G. McCarthy$^{5}$,
Kristi Webb$^{1,2}$, 
\newauthor Adam Muzzin$^{6}$,
Sean McGee$^{7}$,
Gregory Rudnick$^{8}$,
Andrea Biviano$^{9,10}$,
\newauthor Pierluigi Cerulo$^{11}$,
Jeffrey C. C. Chan$^{12}$,
M. C. Cooper$^{13}$,
Ricardo Demarco$^{14}$,
\newauthor Pascale Jablonka$^{15,16}$,
Gabriella De Lucia$^{9}$,
Benedetta Vulcani$^{17}$,
Gillian Wilson$^{12}$,
\newauthor Howard K. C. Yee$^{18}$,
Dennis Zaritsky$^{19}$\\
\\
$^{1}$Department of Physics and Astronomy, University of Waterloo, Waterloo, Ontario N2L 3G1, Canada \\
$^{2}$Waterloo Centre for Astrophysics, University of Waterloo, Waterloo, Ontario, N2L3G1, Canada \\
$^{3}$European Southern Observatory, Karl-Schwarzschild-Str. 2, 85748, Garching, Germany \\
$^{4}$Department of Physics, University of Helsinki, Gustaf H\"allstr\"omin katu 2a, FI-00014 Helsinki, Finland \\
$^{5}$Astrophysics Research Institute, Liverpool John Moores University, 146 Brownlow Hill, Liverpool L3 5RF, UK \\
$^{6}$Department of Physics and Astronomy, York University, 4700 Keele Street, Toronto, Ontario, ON MJ3 1P3, Canada \\
$^{7}$School of Physics and Astronomy, University of Birmingham, Edgbaston, Birmingham B15 2TT, England \\
$^{8}$Department of Physics and Astronomy, The University of Kansas, 1251 Wescoe Hall Drive, Lawrence, KS 66045, USA \\
$^{9}$INAF -- Osservatorio Astronomico di Trieste, via G. B. Tiepolo 11, I-34143 Trieste, Italy \\
$^{10}$IFPU -- Institute for Fundamental Physics of the Universe, via Beirut 2, 34014 Trieste, Italy \\
$^{11}$Departamento de Astronom\'ia, Facultad de Ciencias F\'isicas y Matem\'aticas, Universidad de Concepci\'on, Concepci\'on, Chile \\
$^{12}$Department of Physics and Astronomy, University of California, Riverside, 900 University Avenue, Riverside, CA 92521, USA \\
$^{13}$Department of Physics and Astronomy, University of California, Irvine, 4129 Frederick Reines Hall, Irvine, CA 92697, USA \\
$^{14}$Departamento de Astronom\'ia, Facultad de Ciencias F\'isicas y Matem\'aticas, Universidad de Concepci\'on, Concepci\'on, Chile \\
$^{15}$Laboratoire d'astrophysique, \'Ecole Polytechnique F\'ed\'erale de Lausanne (EPFL), 1290 Sauverny, Switzerland  \\
$^{16}$GEPI, Observatoire de Paris, Universit\'e PSL, CNRS, Place Jules Janssen, F-92190 Meudon, France \\
$^{17}$INAF -- Osservatorio astronomico di Padova, Vicolo Osservatorio 5, IT-35122 Padova, Italy \\
$^{18}$Department of Astronomy \& Astrophysics, University of Toronto, Toronto, Canada \\
$^{19}$Steward Observatory and Department of Astronomy, University of Arizona, Tucson, AZ 85721, USA \\
}

\date{Accepted XXX. Received YYY; in original form ZZZ}

\pubyear{2021}
\pgfplotsset{compat=1.16}

\begin{document}
\label{firstpage}
\pagerange{\pageref{firstpage}--\pageref{lastpage}}
\maketitle

\begin{abstract}
{We use photometric redshifts and statistical background subtraction to measure stellar mass functions in galaxy group-mass ($4.5-8\times10^{13}~\mathrm{M}_\odot$) haloes at $1<z<1.5$. Groups are selected from COSMOS and SXDF, based on X-ray imaging and sparse spectroscopy. Stellar mass ($M_{\mathrm{stellar}}$) functions are computed for quiescent and star-forming galaxies separately, based on their rest-frame $UVJ$ colours. From these we compute the quiescent fraction and quiescent fraction excess (QFE) relative to the field as a function of $M_{\mathrm{stellar}}$. QFE increases with $M_{\mathrm{stellar}}$, similar to more massive clusters at $1<z<1.5$. This contrasts with the apparent separability of $M_{\mathrm{stellar}}$ and environmental factors on galaxy quiescent fractions at $z\sim 0$. We then compare our results with higher mass clusters at $1<z<1.5$ and lower redshifts.
We find a strong QFE dependence on halo mass at fixed $M_{\mathrm{stellar}}$; well fit by a logarithmic slope of $\mathrm{d}(\mathrm{QFE})/\mathrm{d}\log (M_{\mathrm{halo}}) \sim 0.24 \pm 0.04$ for all $M_{\mathrm{stellar}}$ and redshift bins. This dependence is in remarkably good qualitative agreement with the hydrodynamic simulation BAHAMAS, but contradicts the observed dependence of QFE on $M_{\mathrm{stellar}}$. We interpret the results using two toy models: one where a time delay until rapid (instantaneous) quenching begins upon accretion to the main progenitor (“no pre-processing”) and one where it starts upon first becoming a satellite (“pre-processing”). Delay times appear to be halo mass dependent, with a significantly stronger dependence required without pre-processing. We conclude that our results support models in which environmental quenching begins in low-mass ($<10^{14}M_\odot$) haloes at $z>1$.}

\end{abstract}




\section{Introduction}

It is well established in the standard Lambda Cold Dark Matter ($\Lambda$CDM) cosmology that gravitationally bound dark matter haloes build up hierarchically through a combination of smooth accretion of surrounding matter, as well as merging with smaller structures \citep{WhiteRees1978, navarro1997universal, Qu2017}. Complex baryonic physics, such as radiative cooling and feedback from stars and accreting supermassive black holes \citep[e.g.][]{WhiteFrenk1991, Finlator2008, Bouche+2010accretion, Schaye2010, Dave2012, Bower2017} 
drives the formation of galaxies within this gravitationally dominant component.

Up to at least $z\sim 2.5$, galaxy populations exhibit a bimodality in their star formation rate (SFR) distribution \citep[e.g.][]{Bell2004RSevolution, Brinchmann2004, Brammer2011, Muzzin2012} and a corresponding bimodality in the distribution of rest-frame colours \citep[e.g.][]{Strateva2001, Baldry2004, Bell2004, Williams2009, Foltz2018, muzzin2013evolution, Taylor2015}. Observations of these two populations at different redshifts show that the number density and shape of the stellar mass function (SMF) for quiescent (red) galaxies has evolved dramatically, while the SMF of star-forming (blue) galaxies has remained nearly unchanged since $z\sim 3.5$ \citep[e.g.][]{Faber2007, muzzin2013evolution,2021MNRAS.tmp..735M}. This indicates that the latter population eventually stops forming stars, in a process generically called "quenching". The distinct bimodality in the colour and SFR distribution indicates that this quenching must be fairly rapid \citep[e.g.][]{Balogh2004bimodality, Wetzel2012}.

The fraction of quiescent galaxies increases strongly with stellar mass, for $M_{\rm stellar}>10{^9}~\mathrm{M}_\odot$ \citep[e.g.][]{Kauffmann2003, Kauffmann2004, Brinchmann2004, Baldry2006, Weinmann2006, Kimm2009, Muzzin2012, vanderBurg2018}. At fixed stellar mass, the quiescent fraction  declines with increasing redshift, and is higher in denser environments. This has been thoroughly demonstrated in works studying dense environments of rich groups and clusters in the local $z\sim 0$ universe \citep{Kauffmann2004, Gomez2003, Balogh2004, Hou2014}, as well as intermediate $0<z<1$ redshifts \citep{DeLucia2004, Wilman2005, Cooper2006, McGee2011DawnOfRed, Giodini2012, vanderBurg2018, Dennis2019} and at $z>1$ \citep{Muzzin2012, Balogh2014GEEC2, nantais2016stellar, 2019A&A...625A.112G,Strazzullo2019highzclustersQFE, vanderBurgGOGREENsmfs}. At low redshifts, the dependence of the quiescent fraction on stellar mass and environment is largely separable \citep{Baldry2006, Peng2010, Kovac2014, Balogh2016domSatQuenchingMechanism, vanderBurg2018}. This has been interpreted to indicate that the dominant physical mechanisms are also separable \citep[e.g.][]{Peng2010}. However, this is not necessarily the case \citep{DeLucia2012, PintosCastro2019}, and most physically-motivated models invoked to explain environment quenching have a dependence on stellar mass \citep[e.g.][]{McGee2014MNRAS,Fillingham2015,2017MNRAS.469...80Q,2019MNRAS.487.3740W}. 

The environmental dependence of quenching provides a particularly interesting and discriminating test of models \citep[e.g.][]{2010MNRAS.406.2249W,McGee2014MNRAS,2016MNRAS.461.1760H}. Comparison of data with models is challenging, however,  in part because there are many different, commonly used definitions of environment. These include local density \citep{Cooper2006, Peng2010, Sobral2011, Darvish2016, Davidzon2016, Lemaux2019ORELSE, kawinwanichakij2017EffectOfLocalEnvAndMass, Papovich2018}, group/cluster virial masses and cluster centric distance \citep{Poggianti2006, Oman2013, Muzzin2014, 2016ApJ...816L..25P, vanderBurg2018, vanderBurgGOGREENsmfs}, and status as central or satellite in halo \citep{Muzzin2012, Mok2013, Balogh2016domSatQuenchingMechanism}. The quiescent fraction at fixed stellar mass correlates with environment in most cases, but the interpretation of the physical mechanisms behind the observed trends remains elusive, in part because of the difficulty linking these observations to theoretical models. 

To interpret the environmental dependence of the quiescent fraction, numerous works have used simple accretion models where galaxies take some amount of time to fully quench, once they enter a new environment. To reproduce the observations at $z\sim 0$ requires long timescales, of at least $\sim 2-7$ Gyr  \citep[e.g.][]{DeLucia2012,Wheeler2014}. To reconcile this with the bimodality in observed SFRs, which requires a rapid transition, it is frequently assumed to be a long delay time during which the galaxy properties remain uninfluenced by their environment, before rapid quenching sets in \citep{Wetzel2013MNRAS.432..336W, Oman2016}. These timescales are shorter at higher redshift, scaling approximately with the dynamical time \citep{2010ApJ...719...88T,2013ApJ...778...93T,Balogh2016domSatQuenchingMechanism, Foltz2018}. However, all these timescales are relative to an assumed starting point associated with a change in environment. In general this is not well defined, and depends upon assumptions about the physical mechanisms at work. 

A physically relevant definition of environment in $\Lambda$CDM models is the mass of the host halo, and location within it. The environment of a galaxy undergoes a large change when it is accreted into a more massive halo and first becomes a satellite, as it orbits within that new halo. A simple objective definition of this accretion time is the moment when a galaxy crosses $R_{200}$\footnote{$R_{\Delta}$ can be defined for a halo as the radius within which the average density is $\Delta$ times either the critical density of the universe or $\Delta$ times the background density. We use the former definition in this work.} for the first time \citep{Balogh2000}. However, simulations suggest environment first becomes important when satellites are cut off from cosmological accretion, which can happen well outside $R_{200}$ \citep{Behroozi2014, Bahe2015, Pallero2019}. On the other hand, for processes like ram pressure stripping that require a dense intracluster medium, the more relevant starting point could be well inside $R_{200}$ \citep[e.g.][]{Muzzin2014}.
Moreover, the starting time for environmental effects depends on when in the merger history hierarchy the galaxy is accreted, according to one of the above definitions. One consideration is the accretion onto the main progenitor of the final halo. Alternatively, the physically relevant definition could be the first time a galaxy is accreted onto any more massive halo and hence first becomes a satellite. The latter is often referred to as "pre-processing"; observations and simulations both indicate that ``pre-processing'' may be important for a significant proportion of galaxies and that at least some cluster galaxies had their star formation quenched in $M_{\mathrm{halo}} \geq 10^{13} \mathrm{M}_{\odot}$ groups prior to being accreted into massive clusters \citep{Zabludoff1998, Kawata2008, Berrier2009, McGee+2009, DeLucia2012, Hou2014, Pallero2019, 2021MNRAS.500.4004D}. 

To make progress requires observations of galaxies spanning a wide range in stellar mass and redshift, within a range of well-characterized environments that can be directly linked to theory. This can be achieved with samples of groups and clusters with reliable halo mass estimates. 
The GOGREEN survey \citep{balogh2017gemini,GOGREEN2021data} was designed with this goal, and provides a sample of 21 galaxy systems at $1<z<1.5$ with deep photometry and extensive spectroscopy, ranging from groups to the most massive clusters. The groups are a subset of those identified from the deep X-ray and spectroscopic observations in the COSMOS and SXDF regions \citep{finoguenov2010x,leauthaud2012integrated, Giodini2012, gozaliasl2018chandraCOSMOSgroups}. COSMOS groups at $z<1$ have already been studied in some detail. For example, \cite{Giodini2012} measured stellar mass functions for quiescent and star-forming galaxies (separated using $NUV$-$R$ colours) in these systems. Among other things, they found that the fraction of quiescent galaxies increases with halo mass, and decreases with increasing redshift, though this was not examined as a function of stellar mass. In the present paper we build upon this work, taking advantage of deeper X-ray data and additional spectroscopy to extend the group sample to $1<z<1.5$. We also make use of significantly improved photometry (deeper/more bands) to separate quiescent/star-forming galaxies and measure their stellar mass functions and quiescent fractions. Combined with the GOGREEN sample, and lower redshift comparison samples, this allows for an improved picture of quenching as a function of both stellar and halo mass, over the redshift range $0<z<1.5$.

The structure of the paper is as follows. In \S\ref{sec:data} we describe the spectroscopic and photometric data sets, as well as the group catalogues, that we use for the measurements. Results are presented in \S\ref{sec: results}. In \S\ref{sec:discussion}, we discuss our measurements in the context of the literature, compare to the BAHAMAS hydrodynamical simulation, and explore a toy model to constrain pre-processing scenarios. We then conclude and summarise in \S\ref{Conclusion}. In the Appendices, we include additional details of calibrations and robustness checks, present the spectroscopically targeted GOGREEN groups and stacked velocity dispersion measurement, as well as provide supplemental plots to our analysis and discussion of halo mass trends.

Uncertainties are given at the 1-$\sigma$ (Gaussian) level, unless stated otherwise. Logarithms with base 10 ($\log_{10}$) are written simply as ``$\log$'' throughout this work. All magnitudes are given in the AB magnitude system, all (RA, DEC) coordinates are given using the J2000 system, and a \cite{chabrierIMF} initial mass function (IMF) is assumed throughout, unless specified as otherwise. As well, a flat $\Lambda$CDM cosmology with $H_0=70~{\rm km}~{\rm s}^{-1}~{\rm Mpc}^{-1}$, $\Omega_m=0.3$, and $\Omega_{\Lambda}=0.7$, is assumed. Halo masses and radii are given as either $(M_{500c}, R_{500c})$ or $(M_{200c}, R_{200c})$, where $c$ refers to the critical density of the universe at a given redshift. Conversions of mass and radius between $500c$ and $200c$ were done using concentration parameters estimated using the redshift-dependent relation defined in \cite{munozcuartas2011}. Finally, whenever the term \textquotedblleft field\textquotedblright~is used, we are referring to an average sample of the Universe, which includes all environments.

\section{Datasets and sample selections} \label{sec:data}
The core analysis of this work is based on 21 X-ray selected groups at $1<z<1.5$, in the COSMOS and SXDF survey regions. We rely on the excellent photometric redshifts, calibrated with extensive spectroscopy, and statistical background subtraction to analyse the galaxy populations in these groups. The following subsections summarise the data and sample selections, including comparison samples at lower redshift and higher halo mass.

\subsection{Photometric data}

\subsubsection{Imaging and catalogues}

For COSMOS we use the UltraVISTA \citep{mccracken2012ultravista, muzzin2013public} survey as the source of the photometry and catalogues. The first data release (DR1, v4.1) provides a $K$-selected catalogue in 38 photometric bands, covering 1.62 deg$^2$ and with a $5\sigma$ ($2.1^{\prime\prime}$ aperture) limiting magnitude of $K_s=23.9 \mathrm{AB}$. The 95\% stellar mass-completeness limit is $10^{10}~\mathrm{M}_{\odot}$ at $z=1.5$. The catalogues include photometric redshifts and rest-frame $U-V$ and $V-J$ colours, computed using EAZY \citep{brammer2010ascl.soft10052B}. The photometric redshifts are accurate to $\delta z / (1+z)=0.013$ (68\% confidence limits), with a catastrophic outlier fraction of 1.6\% \citep{muzzin2013public}. The catalogues also include stellar masses and population parameters, which were obtained using the spectral energy distribution fitting code FAST \citep{FAST} with the \cite{Bruzual2003} models.
A subset of the COSMOS field is covered by the ultra-deep stripes of UltraVISTA DR3. This catalogue includes 50 photometric bands and covers a non-contiguous 0.8465 deg$^2$, with a $5\sigma$ ($2^{\prime\prime}$ aperture) limiting magnitude of $K_s=24.9 \mathrm{AB}$\footnote{The official UltraVISTA DR3 data release document can be accessed here: \url{https://www.eso.org/sci/observing/phase3/data_releases/uvista_dr3.pdf}}. A magnitude $K_s=23.5$ is reached at the 90\% confidence limit and the 95\% mass-completeness limit is $10^{9.58}~\mathrm{M}_{\odot}$ at $z=1.5$.
We use DR3 catalogues for the subset of groups that fall entirely within one of the stripes of that survey; otherwise we use DR1.

For SXDF we use the SPLASH-SXDF 28-band catalogue \citep{mehta2018splash}. The subset of the SXDF field with all available filters covers 0.708 deg$^2$, with a 5$\sigma (2^{\prime\prime})$ limiting magnitude of $K=25.32$ \citep{mehta2018splash}. By comparing with UltraVISTA DR3, we expect this survey to be 95\% complete above a stellar mass limit of $10^{9.4}~\mathrm{M}_{\odot}$
at $z=1.5$.
Photometric redshifts, their uncertainties, and stellar mass estimates were calculated by the SPLASH team using LePhare \citep{Arnouts1999, Ilbert2006}. Photometric redshift uncertainties are reported in terms of their $\chi^2$ fit, $\Delta \chi^2 = 1.0$ for upper and lower 68\% confidence limits.
We rely on the rest-frame $U$, $V$, and $J$ colours to classify galaxies, and we compute these using the SPLASH-SXDF v1.6 photometric catalogue, with EAZY. Redshifts were fixed to the SPLASH-SXDF photometric redshifts or spectroscopic redshift, if available.

A small correction ($<0.06$) is made to the photometric redshifts in the range $1<z<1.5$, to correct a redshift-dependent bias that is observed upon comparison with a spectroscopic sample, as described in Appendix~\ref{sec-photoz_app}. Details of the various spectroscopic redshift catalogues used for this, as well as description of how they informed our photometric redshift selection of group members and groups, can be found in Appendix~\ref{sec:group-halo-masses}.

\subsubsection{Galaxy selection}\label{sec: data-cuts}

We select galaxies from the UltraVISTA DR1 catalogue of \citet{muzzin2013public}, closely reproducing the selection used by \citet{muzzin2013evolution} to calculate the stellar mass function evolution. Specifically, we select objects identified as galaxies (rather than stars), with uncontaminated photometry and $K_s<23.4$. We impose an additional cut of ${\rm S/N} >7$ in the $K_s$ photometry. We perform a similar selection for UltraVISTA DR3, except to $K_s<24.9$. For galaxies in the SPLASH-SXDF we select galaxies with the same $S/N>7$ criterion, and $K_s<23.7$, corresponding to the $5\sigma$ depth of that survey. Stars in SXDF are excluded from the sample using the ``STAR\_FLAG'' parameter, which is based on whether the photometry is best-fit to a stellar or galaxy template, with an additional restriction that the object does not belong to the stellar sequence in $BzK$ colour-colour space.

Finally, we make a survey-dependent stellar mass cut to ensure complete, unbiased galaxy samples. For $1<z<1.5$ groups in UltraVISTA DR1, the shallowest of the survey regions,
we select galaxies with $\log{M_{\rm stellar}/\mathrm{M}_\odot}>z_{\text{group}}+8.5$, corresponding to the 
mass completeness limit shown in Figure 2 of \cite{muzzin2013evolution}. For groups in the deeper UltraVISTA DR3 and SXDF we conservatively select galaxies with 
$M_{\rm stellar}>10^{9.6}~\mathrm{M}_{\odot}$,
corresponding to the $z=1.5$ completeness limit of DR3.

\subsubsection{Classification of quiescent and star-forming galaxies} \label{sec: colour-cuts} 

\begin{figure} 
	\centering 
	\includegraphics[width=\columnwidth]{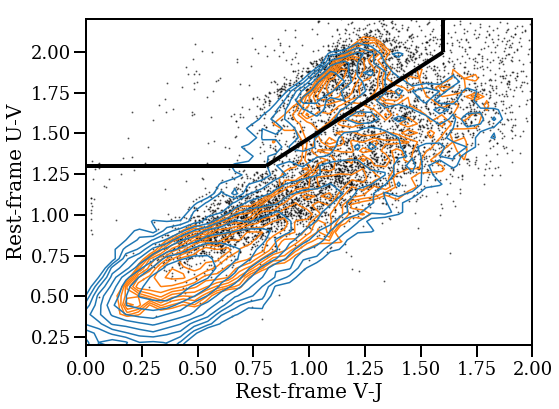}
	\caption{Rest-frame $U-V$ vs $V-J$ colour distributions for galaxies in the SXDF (black points), UltraVISTA DR1 (orange contours) and DR3 (blue contours), with the colour-colour cuts dividing quiescent and star-forming galaxies (solid black line). Small zeropoint adjustments have been made to SXDF and DR1, as described in the text. For the purposes of this figure, galaxies are limited to $1<z<1.5$ and $\log(M_{\mathrm{stellar}}/\mathrm{M}_{\odot})>10$.}
	\label{fig:restframe-colours-and-cuts}
\end{figure}

We identify quiescent\footnote{Equivalently referred to in some of the literature as "passive" or "quenched".} (\textquotedblleft red\textquotedblright) and star-forming (\textquotedblleft blue\textquotedblright) galaxies using rest-frame $UVJ$ colour-colour cuts, following \cite{muzzin2013evolution}. To ensure consistency between the three photometric catalogues we use, we apply small systematic shifts to the $U-V$ and $V-J$ colours of galaxies in UltraVISTA DR1 and SXDF, to match those of UltraVISTA DR3. Specifically, we calculate the average difference in these colours between surveys, using galaxies at $1<z<1.5$ and with stellar masses above $10^{10}~\mathrm{M}_{\odot}$. This results in a shift of $\Delta (V-J)= 0.08$ and $\Delta (U-V)= -0.05$ for UltraVISTA DR1; for SXDF the corresponding shifts are 0.10 and 0.15, respectively.
We then use the following selection, slightly modified from \cite{muzzin2013evolution} (which worked with the original UltraVISTA DR1 colours), to 
identify quiescent galaxies at $1<z<1.5$:
\begin{align}
U-V > 1.26, V-J < 1.58, \\
U-V>(V-J) \times 0.88+0.47,
\end{align}
We illustrate these cuts on the rest-frame $U-V$ vs $V-J$ distribution in Figure \ref{fig:restframe-colours-and-cuts}.

We also consider a lower redshift comparison sample of groups, at $0.5<z<0.7$, selected entirely from UltraVISTA DR1 (see \S\ref{sec:comp-intz}). Noting that the $UVJ$ colour-colour cut is weakly redshift dependent \citep{Williams2009, Whitaker2011, muzzin2013evolution} we instead adopt for these galaxies exactly the selection of
\citet{muzzin2013evolution} at $0.0<z<1.0$, namely:
\begin{align} 
U-V > 1.3, V-J < 1.5,\\
U-V > (V-J) \times 0.88+0.69 \, .
\end{align}

\subsubsection{Group selection} \label{sec:group-selection}
The COSMOS groups and SXDF groups we use for our analyses were identified in \cite{gozaliasl2018chandraCOSMOSgroups} and \cite{finoguenov2010x}, respectively. Each group in the catalogue has a quality flag ranging from 1 (best) to 5 (worst), although the precise meaning of these flags is different in the two surveys. We update quality flags for a subset of groups in the two group catalogues based on information from our GOGREEN spectroscopy, which increases the number of available spectroscopically confirmed groups. We then select only groups with quality flags $< 3$, defined in \cite{gozaliasl2018chandraCOSMOSgroups} to have both X-ray detections, photometric overdensity of galaxies, and at least one spectroscopically-confirmed member \footnote{In Table \ref{tab:groups-table}, group COSMOS-30317 is listed as having no spectroscopic members. This discrepancy with \cite{gozaliasl2018chandraCOSMOSgroups} may be either due to a different cut in redshift or a difference in the spectroscopic catalogues being used.}.

With this selection we have an initial sample of 21 groups at $1<z<1.5$: nine in UltraVISTA DR1, eight in DR3, and four in SXDF. The properties of these groups are presented in Table \ref{tab:groups-table}. All of these groups have halo masses estimated to be in the range $13.6<\text{log} (M_{\rm halo}/\mathrm{M}_{\odot})<14.0$, with an average  of $\text{log} (M_{\rm halo}/\mathrm{M}_{\odot})\approx 13.8$. For each group we calculate $R_{200c}(z)$  \citep[using][]{hearin2017forward} corresponding to this average mass (e.g. $R_{200c} = 1.044'$ in projection at $z=1.25$).

\begin{figure} 
	\centering
		\includegraphics[width=\columnwidth]{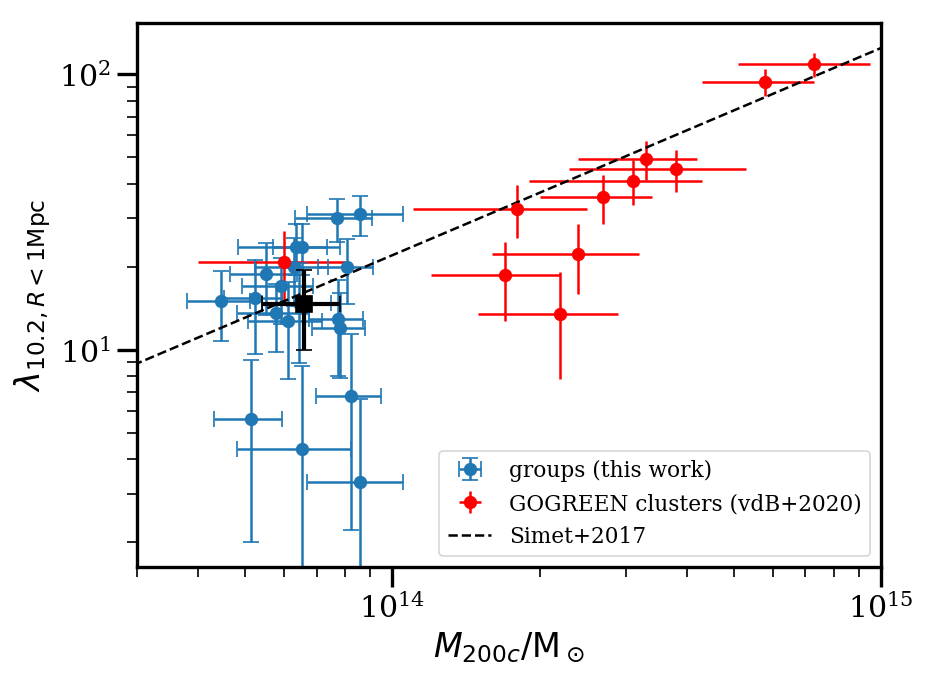}
	\caption{Richness as a function of halo mass for groups in this work (blue points; mean of sample shown with the square black point) and the GOGREEN clusters (red points). The richness, $\lambda_{\mathrm{10.2, R<1Mpc}}$, is the number of background-subtracted galaxies that have stellar masses above $10^{10.2}~\mathrm{M}_\odot$ within 1~Mpc, following \citet{vanderBurgGOGREENsmfs}. Halo masses for the groups are derived from the X-ray fluxes, while for the clusters they are based on a spectroscopic dynamical analysis from \citet{Gogreendynamics}. The richness-halo mass relation fit for clusters at $0.1<z<0.33$ from \citealt{Simet2017} is shown for comparison as a dashed line. Note that one of the groups has a formally negative richness, and lies off the bounds of this figure.}
	\label{fig:richness-vs-M200c}
\end{figure}

The masses are based on the weak lensing calibrations of the $L_{\mathrm{X}}-M_{\mathrm{halo}}$ relations in COSMOS \citep{Leauthaud2010}. The mentioned biases in the Planck 2015 paper \citep{Planck2016SZclusters} are relevant only for the hydrostatic mass estimates \citep[see also][for detailed discussion of the biases in the Planck 2015 paper]{Smith2016}. For the SZ confirmation on similar galaxy groups (near our redshift and halo mass range), there is one such measurement, at z=2 \citep{Gobat2019}. As an alternative indicator of halo mass, we calculate the richness, $\lambda_{\mathrm{10.2, R<1Mpc}}$, defined as the number of photometrically background-subtracted galaxies within a 1~Mpc radius above a stellar mass of $10^{10.2}~\mathrm{M}_\odot$. In Figure \ref{fig:richness-vs-M200c} we show the correlation between these richness values and the $M_{200c}$ masses from X-ray fluxes for our group sample. This is compared with more massive clusters from \citet[]{vanderBurgGOGREENsmfs}, discussed in \S\ref{sec-hizclus} below. For that sample, we use halo masses based on the dynamical analysis of \citet{Gogreendynamics}. Although the uncertainties on individual richness measurements are large, this comparison confirms that the group sample is systematically less rich than the cluster sample, at the level expected from their mass estimates. Only one of the groups has a richness $\lambda_{10.2, R<1{\rm Mpc}}<1$, significantly lower than expected of a truly overdense system.

The subset of $1<z<1.5$ groups with GOGREEN spectroscopy affords the opportunity to study these groups in some more detail, and to test the robustness of the statistical background subtraction. This analysis is presented in Appendix~\ref{sec:velocity-dispersions-appendix}. Where it is possible to calculate a robust velocity dispersion, we report these values in Table \ref{tab:groups-table}. The dynamical halo mass, $M_{200c,{\rm dyn}}$ (column 8 of Table \ref{tab:groups-table}), is then derived using the relation in \cite{Saro2013}. Two groups -- SXDF64 and SXDF87 -- have dynamical masses significantly higher than that based on their X-ray emission, and formally above our arbitrary threshold for low-mass haloes, of $10^{14}~\mathrm{M}_\odot$. To be conservative, therefore, we exclude these two groups from the rest of our analysis, though we have confirmed that our results are not sensitive to this choice.


\begingroup
\renewcommand{\arraystretch}{1.3}
\begin{table*}
\begin{tabular}{lrrrrrrlllc}
\toprule
Name &         RA ($\degree$) &      Dec ($\degree$) &  z &  flag &  $\log \big(\frac{M_{200c, \text{X}}}{\mathrm{M}_{\odot}}\big)$ &  K$_{s}$ limit &   $\log \big(\frac{M_{200c, \text{dyn}}}{\mathrm{M}_{\odot}}\big)$ & $\sigma_v (\,{\rm km\,s}^{-1})$ & $N_{\rm spec}$ & $\lambda_{\mathrm{10.2, R<1Mpc}}$\\ 
\midrule
\textbf{COSMOS-30221} &  150.56200 &  2.50309 &     1.197 &     0 &  13.80 &       24.9 &   12.90 $\pm$ $^{0.30}_{0.39}$ &       200 $\pm$ 50 &         9 & $23.6\pm 5.0$\\ 
COSMOS-20267 &  150.44487 &  2.75393 &     1.138 &     1 &  13.91 &       24.9 &      -- &       -- &         2 & $20.0\pm5.3$\\
COSMOS-30307 &  149.73943 &  2.34139 &     1.028 &     1 &  13.72 &       24.9 &      -- &       -- &         3 & $15.4\pm5.7$\\
\textbf{COSMOS-20028} &  149.46916 &  1.66856 &     1.316 &     0 &  13.89 &       24.9 &  13.33 $\pm$ $^{0.19}_{0.38}$ &  285 $\pm$ 75 &         10 & $30.0\pm5.4$\\
COSMOS-20057 &  150.45229 &  1.91046 &     1.179 &     1 &  13.81 &       24.9 &      -- &       -- &         1 & $23.7\pm5.0$\\
COSMOS-10155 &  150.59137 &  2.53778 &     1.138 &     1 &  13.89 &       24.9 &      -- &       -- &         3 & $12.0\pm4.1$\\
COSMOS-30317 &  150.12646 &  1.99926 &     1.019 &     1 &  13.65 &       24.9 &      -- &       -- &         0 & 15.0$\pm4.2$\\
COSMOS-20072 &  149.86012 &  1.99973 &     1.179 &     1 &  13.92 &       23.9 &      -- &       -- &         3 & $6.8\pm4.6$\\
COSMOS-20199 &  150.70682 &  2.29253 &     1.095 &     1 &  13.70 &       23.9 &      -- &       -- &         5 & $-6.5\pm2.6$\\
COSMOS-20198 &  149.59607 &  2.43788 &     1.168 &     1 &  13.81 &       23.9 &      -- &       -- &         3 & $14.8\pm5.8$\\
COSMOS-20243 &  150.26115 &  2.76857 &     1.315 &     1 &  13.80 &       23.9 &      -- &       -- &         1 & $20.0\pm5.6$\\
\textbf{COSMOS-10063} &  150.35902 &  1.93521 &     1.172 &     0 &  13.74 &       23.9 &      -- &      -- &  9 & $18.85\pm5.5$\\
COSMOS-10105 &  150.38295 &  2.10278 &     1.163 &     1 &  13.79 &       23.9 &      -- &       -- &         5 & $12.7\pm4.9$\\
\textbf{COSMOS-20125} &  150.62077 &  2.16754 &     1.404 &     0 &  13.81 &       23.9 &      -- &       -- &    8& $4.4\pm4.4$\\
COSMOS-10223 &  150.05064 &  2.47520 &     1.260 &     1 &  13.76 &       23.9 &      -- &       -- &         4 & $13.6\pm3.7$\\
COSMOS-30323 &  150.22540 &  2.55061 &     1.100 &     2 &  13.71 &       23.9 &      -- &       -- &         3 & $5.6\pm3.6$\\
\textbf{SXDF49XGG} &   34.49962 & $-5.06489$ &     1.091 &     0 &  13.77 &       25.3 &  13.25 $\pm$ $^{0.22}_{0.27}$ &  255 $\pm$ 50 &         14 & $17.0\pm4.6$\\
\textbf{SXDF64XGG$^\ast$} &   34.33188 & $-5.20675$ &     0.916 &     0 &  13.76 &       25.3 &   14.20 $\pm$ $^{0.18}_{0.21}$ &  530 $\pm$ 80 &         8& --\\
\textbf{SXDF76aXGG} &   34.74613 & $-5.30411$ &     1.459 &     0 &  13.93 &       25.3 &  14.06 $\pm$ $^{0.38}_{0.54}$ &  520 $\pm$ 180 &         6& $31.1\pm5.2$\\
\textbf{SXDF76bXGG} &   34.74743 & $-5.32348$ &     1.182 &     0 & --  &       25.3 &  12.98 $\pm$ $^{0.33}_{0.45}$ &  210 $\pm$ 65 &         7& $3.3\pm3.3$\\
\textbf{SXDF87XGG$^\ast$} &   34.53602 & $-5.06303$ &     1.406 &     0 &  13.89 &       25.3 &  14.44 $\pm$ $^{0.19}_{0.223}$ &  700$\pm$ 110 &         9& -- \\
\bottomrule
\end{tabular}
\caption{
Group names correspond to those in \citet{gozaliasl2018chandraCOSMOSgroups} and \citet{finoguenov2010x} for COSMOS and SXDF, respectively; names in boldface are those included in the GOGREEN \citep{GOGREEN2021data} spectroscopic survey. SXDF76XGG has been split into "a" and "b" to identify the foreground group; the original X-ray mass estimate for $M_{200c,X}$ has been retained only for the higher redshift system. Group redshifts are taken from the original catalogues, except where GOGREEN spectroscopy is available to provide an improved measurement (see Appendix \ref{sec:velocity-dispersions-appendix}).
Column 5 is the group quality flag. A flag value of 0 denotes a group with confirmed redshift from GOGREEN \citep{GOGREEN2021data}. Other flag numbers are based on the \citet{gozaliasl2018chandraCOSMOSgroups} catalogue: 1 for secure X-ray emission with well-defined centre and at least one spectroscopic member, and 2 for a system that has some X-ray contamination from foreground or background systems.
Column 6 gives the group halo mass estimates from the original catalogues, derived from observed X-ray luminosities.
Column 7 gives the $K_s$-band limiting magnitude for the survey from which each group is drawn, as described in the text.
Column 9 gives the (spectroscopic) velocity dispersion we determined for our GOGREEN targeted groups in Appendix \ref{sec:velocity-dispersions-appendix}, and column 8 shows the corresponding dynamical masses. Column 9 gives the number of spectroscopic group galaxies in each group, within a radius of 2$R_{200c}$. The final column gives the richness for groups in our photometric sample, with richness defined as the number of group members with $\log (M_{\rm stellar}/\mathrm{M}_\odot)>10.2$ found within a 1 Mpc circular aperture (see Figure \ref{fig:richness-vs-M200c} for a comparison of these values with the GOGREEN clusters sample from \citet{vanderBurgGOGREENsmfs}). The two groups indicated with a $\ast$ are excluded from the analysis in this paper, as their dynamical masses suggest they may exceed our threshold definition for low-mass haloes.}
\label{tab:groups-table}
\end{table*}
\endgroup

The mean and median redshift of the group sample (1.179 and 1.170, respectively) is somewhat lower than that of the field sample (1.236 and 1.39). We have verified that our conclusions are unchanged if we divide the group and field samples into two redshift bins and conclude that our findings are not sensitive to this difference.

\subsection{Comparison samples}\label{sec:comp-samp}

\subsubsection{Higher halo masses at $1<z<1.5$}\label{sec-hizclus}

We contrast our low halo-mass systems at $1<z<1.5$ with eleven higher mass clusters in the same redshift range from GOGREEN. Our measurements are similar to those in \cite{vanderBurgGOGREENsmfs}, but recalculated to include only galaxies within $R_{200c}$. Halo masses are determined dynamically \citep{Gogreendynamics}, and we show the correlation between these and the cluster richness in Figure~\ref{fig:richness-vs-M200c}.
This sample is divided into two bins of halo mass, though the highest mass bin contains only two clusters at the lower end of the target redshift range: SPT-CL J0546-5345 ($z=1.068$) and SPT-CL J2106-5844 ($z=1.126$). We note that several clusters (SpARCS-1051, SpARCS-1638, SpARCS-1034, SpARCS-0219) have low richness values more typical of our groups. 

\subsubsection{Intermediate redshift $0.5<z<0.7$}\label{sec:comp-intz}
The galaxy populations in the X-ray selected groups at $z<1$ have been extensively studied, notably by \cite{Giodini2012}. We use a similar sample but redo the analysis to ensure consistent methodology when comparing with our higher redshift sample. Specifically, we select fourteen groups in the redshift range  $0.5<z<0.7$ from the COSMOS field, and use the UltraVISTA DR1 catalogue. Since the average rest-frame $UVJ$ colours and photometric redshifts do not noticeably differ between UltraVISTA DR1 and DR3, we do not make the adjustments to colours or photometric redshifts that we applied to the higher redshift sample. 
The groups are required to be robustly identified (quality flags $< 3$) and in the halo mass range $13.6<\text{log}(M_{\rm halo}/\mathrm{M}_{\odot})<14.0$. The average halo mass of the sample is  $\text{log} (M_{\rm halo}/\mathrm{M}_{\odot})\approx 13.78$, comparable to the mass of our higher redshift group sample. The photometric redshift selection, and statistical background subtraction, is done in an analogous way to that for the $1<z<1.5$ sample.

For higher mass clusters at this redshift we use the published measurements and uncertainties in the stellar mass functions for 21 clusters selected based on their Sunyaev-Zeldovich (SZ) signal, from the Planck all sky survey, from \cite{vanderBurg2018}. These clusters span the halo mass range $14.5<\text{log}(M_{\rm halo}/\mathrm{M}_{\odot})<15.1$. These were analysed in a very similar way to the clusters from \cite{vanderBurgGOGREENsmfs} that we use at higher redshift. 

The field sample we compare with at this redshift is comprised of all UltraVISTA DR1 galaxies with photometric redshifts in the range $0.5<z<0.7$.

\subsubsection{Low redshift $0.01<z<0.2$}
At $0.01<z<0.2$ we use the SDSS-DR7 measurements from \citep{Omand2014}. Galaxy groups are selected from the \cite{YangDR12groups} friends-of-friends catalogue. Halo masses were determined through abundance matching, using the total group luminosity to rank them. We select haloes in the same mass ranges as at other redshifts, without any evolution correction; the final sample includes 13806/3282/483 haloes in the low/intermediate/high mass bins. All galaxies associated with a halo are included as members, with no additional selection based on clustercentric radius. Stellar masses are computed following the procedure described in \cite{Brinchmann2004}, with a small (10 per cent) correction to convert from a \cite{KroupaIMF} to a \cite{chabrierIMF} initial mass function using \cite{Madaureview}. Quiescent galaxies are identified as those with specific star formation rates $\mathrm{sSFR}<-0.24\log({M_{\rm stellar}/\mathrm{M}_{\odot}})-8.50$, chosen to lie below and parallel to the star-forming main sequence identified in \citet{Omand2014}.

\section{Results} \label{sec: results}

\subsection{Stellar mass functions} \label{sec:SMFs}

The photometric redshift uncertainties in both UltraVISTA and SPLASH-SXDF are still large enough that galaxies cannot be unambiguously identified as members of groups or clusters. We therefore rely on statistical background subtraction, using our representative field sample, to calculate the stellar mass functions.

The number of group members of a given type (quiescent or star-forming), and within a given stellar mass bin, is calculated as the number of galaxies $N_C$ within a circular aperture $A_C$ around the group centre, and with photometric redshifts such that $|z-z_g|<\Delta z$ relative to the group redshift, $z_g$, minus the corresponding average number of galaxies in the field within that same aperture and redshift slice.
For each galaxy sub-population (e.g. quiescent or star-forming) the average number of galaxies per group that we find is described by the following expression:
\begin{align}
\phi(M) 
= \frac{1}{N_G} \sum_g \left[ N_{C,z_g}(M) - N_{\text{survey},z_g}(M)\times \left( \frac{A_C}{A_{\text{survey}}} \right) \right],
\end{align}
where $N_G$ is the number of groups, $g$ is a given group, $N_{\mathrm{survey}}$, $A_{\rm survey}$ are the number of field galaxies and the total area of the survey from which each group is drawn, respectively. We  use $M=M_{\rm stellar}$ in the above expression for brevity. The aperture size is chosen to be $R_{200c}$, and a photometric redshift cut of $\Delta z=0.126$ was chosen for all three survey regions (see Appendix~\ref{sec:photo-z-cut-choices} for explanation of this cut choice). As well, since the area of a group aperture at a given redshift is negligible relative to the rest of the given survey region at that redshift, we refer to the overall survey area/volume as the \textquotedblleft field\textquotedblright.

The error on the number of background-subtracted galaxies in a group is given by summing in quadrature the Poisson counting error and the Poisson error term for the field contribution $(A_C/A_{\text{survey}})$, which simplifies to $\sigma \approx \sqrt{N_C(M)}$. The total error for the number of galaxies in a given mass bin is then
\begin{align}
    \sigma(M) = \frac{1}{\phi(M)} \sqrt{\sum_g N_{C,z_g}^2(M)},
\end{align}
where $N_{C,z_g}$ is the number of galaxies in the circular aperture, $C$, around a given group, $g$, at redshift $z_g$.

\begin{figure}
	\centering
		\includegraphics[width=\columnwidth]{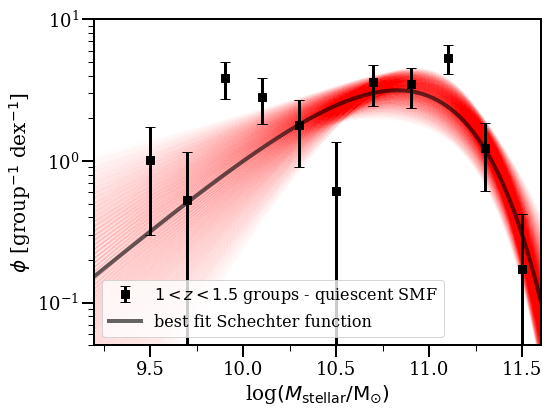}
		\includegraphics[width=\columnwidth]{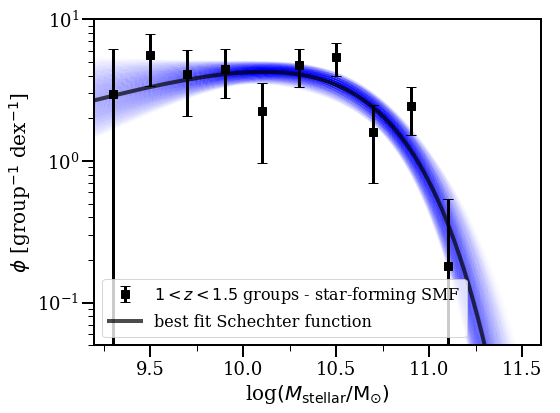}
		\includegraphics[width=\columnwidth]{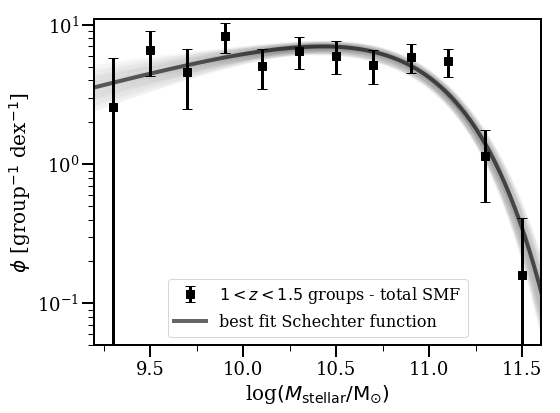}
	\caption{Background-subtracted stellar mass function of quiescent (top panel), star-forming (middle), and total (bottom) group galaxies at $1<z<1.5$. Overlaid on each plot are the Schechter function fits to the group data (solid line), normalized to match the number of group galaxies per dex (bin size $\Delta \log (M_{\mathrm{stellar}}/\mathrm{M}_\odot)=0.2$), and with shaded regions indicating the 68\% confidence interval on the fit parameters, computed as described in the text. Error bars shown represent the Poisson shot noise.} 
	\label{fig:combined-SMFs}
\end{figure}
\begin{figure}
	\centering
		\includegraphics[width=\columnwidth]{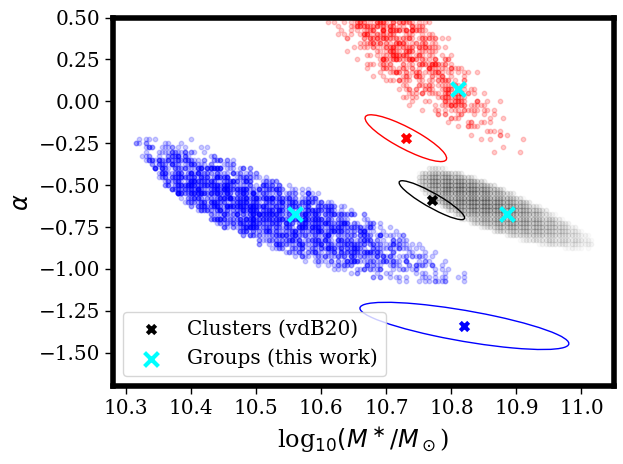}
	\caption{We show two of the Schechter function fit parameters for the $1<z<1.5$ galaxy group population, for the total sample (black), the quiescent galaxies (red) and the star-forming galaxies (blue). Points represent draws of parameters within the 68 per cent confidence limits of the fits. The cyan crosses indicate the best fit parameters. The $\alpha$ parameter for the quiescent population is effectively unbounded: high values of $\alpha>2$ provide acceptable fits within $2\sigma$. The crosses with solid ellipses represent the fit parameters and 68 per cent confidence limits for the massive cluster population at the same redshifts, from \citealt{vanderBurgGOGREENsmfs}.}
	\label{fig:mstaralpha}
\end{figure}

In Figure \ref{fig:combined-SMFs} we present the background-subtracted stellar mass functions for the full sample of $1<z<1.5$ groups, and separately for the quiescent and star-forming populations. Each bin is weighted by the number of contributing groups, such that the resulting values are the average number of galaxies per group, per dex in stellar mass. 

The stellar mass functions are fit with \cite{schechter1976analytic} functions  of the form

\begin{align}
\Phi (M) = \frac{dN}{dM}dM=\Phi^* \bigg(\frac{M}{M^*}\bigg)^{\alpha} e^{-M/M^*}dM,
\end{align}
where $M$ is the stellar mass, $M^*$ is the characteristic mass where the Schechter function transitions between a power law and an exponential cut-off, and $\alpha$ is the logarithmic slope of the faint-end power-law. We fit all three parameters, separately for each sample, by minimizing the $\chi^2$. Where needed, we arbitrarily increase the uncertainties to ensure $\chi^2/\nu=1$ for the best fit model,  where $\nu$ is the number of degrees of freedom. This is only important for the quiescent population, for which a single Schechter function does not provide a good fit ($\chi^2/\nu=2.56$); uncertainties are therefore increased by a factor $\sim 1.6$. With this adjustment we calculate the 68\% confidence limits from the $\chi^2$ distribution and determine all parameter combinations that provide a $\chi^2$ within these limits. All points are included in the fits, including those with contributions from fewer than the maximum number of groups. The best fit parameters and their uncertainties are given in Table \ref{tab-fitparams}, and the mass functions are shown in Figure~\ref{fig:combined-SMFs}.

\begin{table}
    \centering
    \begin{tabular}{llll}
        \multicolumn{4}{c}{\bf Groups} \\
        \hline
          Population & $\log({M^\ast/\mathrm{M}_{\odot}})$&$\alpha$&$\phi^\ast$\\
                     &  & &$[\mathrm{group}^{-1}~\mathrm{dex}^{-1}]$\\
                    \hline
                    \\
                    Quiescent & $10.8^{+0.2}_{-0.3}$& $0.1^{+0.7}_{-0.4}$&$8.9^{+13.6}_{-6.2}$ \\ 
                    \\
                    Star-forming & $10.6^{+0.2}_{-0.2}$&$-0.7^{+0.5}_{-0.4}$&$8.6^{+24.0}_{-6.7}$\\ 
                    \\
                    Total &$10.9^{+0.1}_{-0.1}$ &$-0.7^{+0.2}_{-0.2}$ &$14.0^{+15.3}_{-7.8}$\\ 
            \\
        \hline \\
    \end{tabular}
    \begin{tabular}{llll}
        \multicolumn{4}{c}{\bf Field} \\
        \hline
          Population & $\log({M^\ast/\mathrm{M}_{\odot}})$&$\alpha$&$\phi^\ast$\\
                     &  & &$[10^{-3}~\mathrm{Mpc}^{-3}~\mathrm{dex}^{-1}]$\\
                    \hline
                    \\
                    Quiescent & $10.63^{+0.04}_{-0.03}$& $0.06^{+0.07}_{-0.07}$& $1.42^{+0.28}_{-0.24}$ \\
                    \\
                    Star-forming & $10.82^{+0.03}_{-0.04}$&$-1.27^{+0.02}_{-0.02}$& $1.31^{+0.33}_{-0.27}$ \\
                    \\
                    Total &$10.90^{+0.04}_{-0.04}$ &$-1.12^{+0.03}_{-0.03}$ & $1.98^{+0.60}_{-0.46}$ \\
            \\
        \hline \\
    \end{tabular}
    \caption{Best fit Schechter function parameters and their 68\% confidence limits, for the low-mass halo (group) population and the combined field (UltraVISTA DR1, DR2, and SXDF). The normalization parameter for the group galaxies Schechter fits, $\phi^\ast$, reproduces the curves in Figure~\ref{fig:combined-SMFs}; it has units of number of galaxies per group per dex.
    }
    \label{tab-fitparams}
\end{table}

\begin{figure}
	\centering
		\includegraphics[width=\columnwidth]{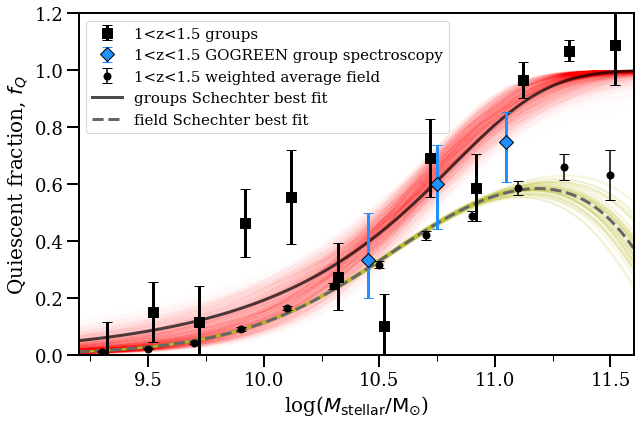}
	\caption{Measured quiescent fractions for the group (large black squares, offset slightly horizontally for clarity) and field (small black points). The red shaded region reflects fits drawn from the 68\% confidence limits on the fits to the quiescent and star-forming stellar mass functions (constrained to be within the 68\% confidence limits on the total stellar mass function fit). The quiescent fraction for the field derived from Schechter fits for the field is also shown, as the black dashed curve with yellow shading showing the 68\% confidence limits. Points with $f_Q>1$ are a result of uncertainty on the statistical background subtraction, which can lead to a formally negative abundance of star-forming galaxies. We also show the quiescent fraction for the spectroscopic group members (blue diamonds), i.e. within 2$R_{200c}$.}
	\label{fig:quiescent-fractions-vs-stellar-mass}
\end{figure}
The parameters $M^\ast$ and $\alpha$ are compared with the corresponding fits to the GOGREEN massive cluster population from \cite{vanderBurgGOGREENsmfs}, in Figure~\ref{fig:mstaralpha}. The shape of the total stellar mass function is in excellent agreement with that measured in more massive clusters. Both the quiescent and star forming populations prefer a higher $\alpha$ slope than the clusters, but the Schechter function fit parameter combinations in the two samples are still consistent at the $2\sigma$ level. Importantly, we do not observe the excess of low-mass quiescent group members that is seen at low redshifts \citep{Peng2010} and we rule out a steep  ($\alpha<-1$) low-mass slope for this population at the 99 per cent confidence level.

\subsection{Quiescent fraction and quiescent fraction excess} \label{results:fQ-and-QFE}
We use the stellar mass functions in the previous section to compute the quiescent fraction, defined as:
\begin{align}
f_Q(M) \equiv \frac{N_{\text{Q}}(M)}{N_{\text{Q}}(M)+N_{\text{SF}}(M)},
\end{align}
where $N_{\text{Q}}$ and $N_{\text{SF}}$ are the number of quiescent and star-forming galaxies, respectively, as identified in $UVJ$ colour space (see \S\ref{sec: colour-cuts}).

\begin{figure}
	\centering
		\includegraphics[width=\columnwidth]{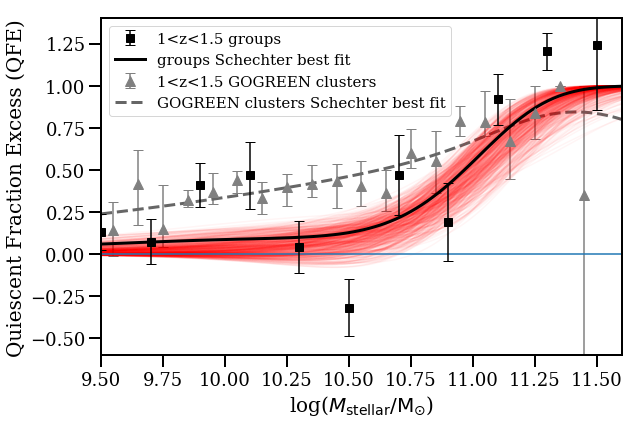}
	\caption{Quiescent fraction excess is shown as a function of stellar mass for our $1<z<1.5$ galaxy group sample (black squares). The quiescent fraction excess is significantly nonzero only for $M\gtrsim 10^{11}~\mathrm{M}_{\odot}$. The red shaded region represents the 68\% confidence limits derived from the Schechter function fits to our stellar mass functions. Quiescent fraction excess values (grey triangles) and best fit line (gray dashed line) are also plotted for the GOGREEN $1<z<1.5$ clusters in \citep{vanderBurgGOGREENsmfs}. The clusters show a strong trend in QFE with stellar mass, particularly above $\log(M^*/\mathrm{M}_\odot)\approx 10.75$. Additionally, the clusters display significant quiescent fraction excess at lower stellar masses, with a trend in the data consistent with an approximately flat relation for log($M_{\mathrm{stellar}}/\mathrm{M}_{\odot}$)<log($M^*/\mathrm{M}_{\odot})$.}
	\label{fig:QFE-vs-Mstellar-GG-groups}
\end{figure}

\begin{figure*}
	\centering
		\includegraphics[width=2\columnwidth]{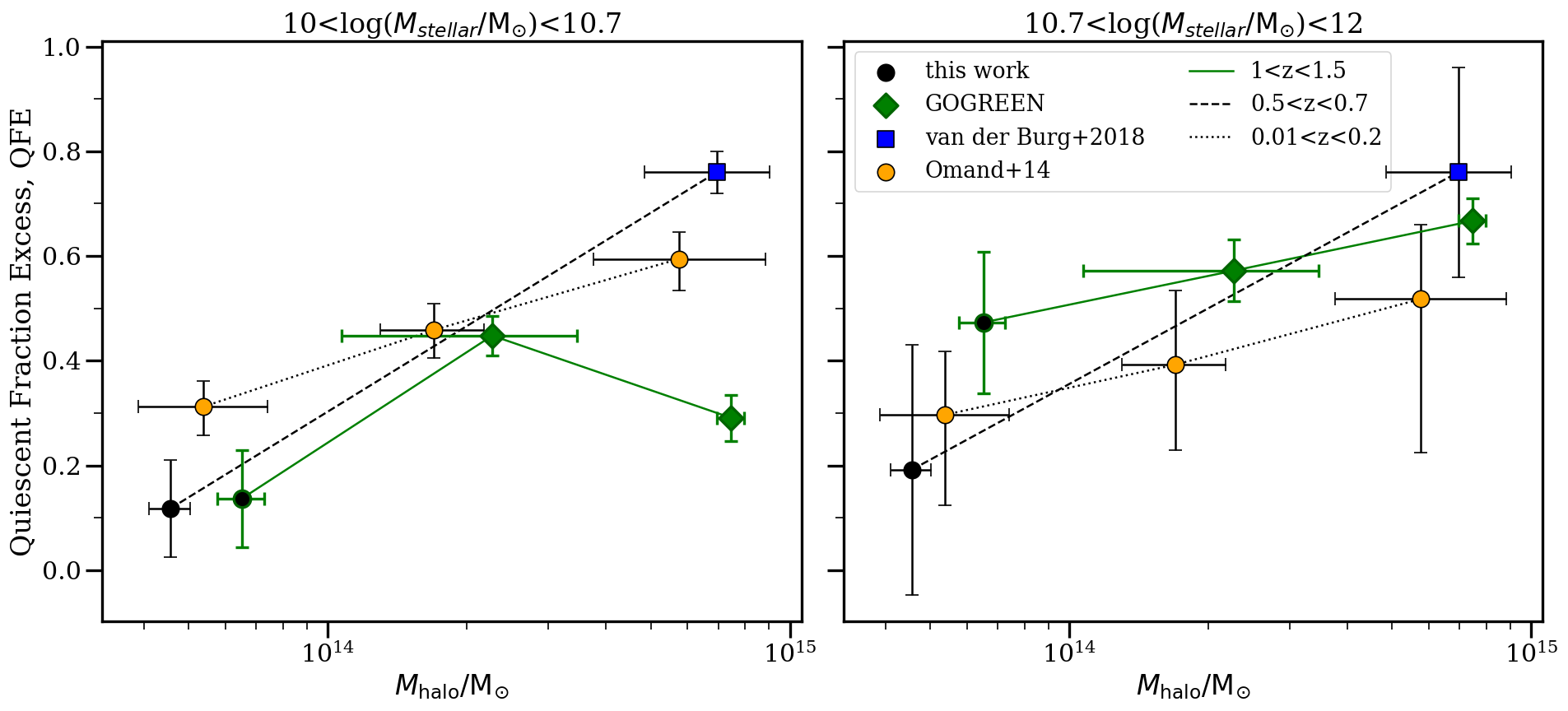}
	\caption{Quiescent fraction excess (QFE) shown as a function of halo mass ($M_{200c}$), $M_{\rm halo} / \mathrm{M}_{\odot}$, for two galaxy stellar mass bins, and for samples at three different redshift ranges, as indicated. Our new measurements of low-mass haloes at $1<z<1.5$ are shown as the green points connected by a green solid line. The other samples are described in \S\ref{sec:comp-samp}. The halo-mass dependence of the QFE has a similar slope at all redshifts and stellar mass bins, of $\sim 0.24$ (see text for details). See Figure \ref{fig:fQ-and-QE-vs-Mhalo-redshift-evolution-literature-summary-mstellar-breakdown-full} for a more detailed breakdown of both $f_Q$ and QFE by stellar mass.}
	\label{fig:fQ-and-QE-vs-Mhalo-redshift-evolution-literature-summary-mstellar-breakdown}
\end{figure*}

We show $f_Q$ as a function of stellar mass for our group sample in Figure \ref{fig:quiescent-fractions-vs-stellar-mass}. Uncertainties on the binned data are computed assuming the quiescent and star-forming stellar mass populations are independent. However, we correctly account for the covariance when deriving the quiescent fraction from the Schechter function fits, which are overlaid as the shaded region. We calculate this by taking random draws from fits within the 68\% confidence limits of the quiescent and star-forming populations, and only keep those for which the sum is in agreement with the total SMF within the same 68\% confidence. There is a strong dependence on stellar mass, with $f_Q$ increasing from near zero to unity over the full range. For high stellar masses, $M_{\rm stellar}>10^{11}~\mathrm{M}_{\odot}$, $f_Q$ is systematically larger in the group sample than the field. At lower masses the excess is both smaller and statistically not significant. To complement this comparison, we also consider the quiescent fraction for the spectroscopic members of the seven GOGREEN groups (i.e. excluding the two with high dynamical masses, that are excluded from all our analysis). This has the advantage that it does not rely on statistical background subtraction. However, the smaller sample size means we can only consider three stellar mass bins, and we also include all spectroscopic group members within $2R_{200c}$, a larger aperture than for the photometric sample. These spectroscopic members are from GOGREEN and any of the available surveys described in \S\ref{sec:spec-z-data}. As shown in the Figure, the quiescent fractions for this spectroscopic subsample are fully consistent with our full group sample.

To better characterize any difference in $f_Q$ in groups relative to the field, we calculate the quiescent fraction excess (QFE)\footnote{We choose the terminology, QFE, for consistency with recent prior works \citep{Wetzel2012, BaheHydrangea2017, vanderBurgGOGREENsmfs} and a more intuitive meaning than a variety of other synonymous terms used in the literature. Other terms synonymous with QFE used in the literature include ``transition fraction'' \citep{vandenBosh2008satQuenching}, ``conversion fraction'' \citep{Balogh2016domSatQuenchingMechanism, Fossati2017}, and ``environmental quenching efficiency'' \citep{Peng2010, Wetzel2015localSatDwarfQuenching, Nantais2017, vanderBurg2018}. \label{footnote:QFE}}.
This is defined as
\begin{align}
\text{QFE} \equiv \frac{f_{Q, \text{cluster}} - f_{Q, \text{field}}}{1-f_{\text{Q}, \text{field}}},
\end{align}
where $f_{\text{Q}, \text{field}}$ and $f_{\text{Q}, \text{cluster}}$ are the fractions of quiescent galaxies in the field and cluster, respectively. In a naive infall interpretation, this quantity represents the fraction of galaxies accreted from the field that have been transformed into quiescent galaxies to match the observed group population \citep[e.g.][]{vandenBosh2008satQuenching,Peng2010,Wetzel2012,Balogh2016domSatQuenchingMechanism,BaheHydrangea2017}. We note that QFE$<0$ is a physical solution even in the presence of environmental quenching processes, since the field is defined as the global population, including massive haloes; there will therefore be a halo mass scale below which most ``field'' galaxies reside in more massive haloes.

The QFE values for our group sample are shown as a function of stellar mass in Figure \ref{fig:QFE-vs-Mstellar-GG-groups}. The average QFE is significantly greater than zero and shows an increasing trend with stellar mass, similar to those for the $1<z<1.5$ clusters in \cite{vanderBurgGOGREENsmfs}, also shown in Figure \ref{fig:QFE-vs-Mstellar-GG-groups}. These results are also similar to those published by \cite{Balogh2016domSatQuenchingMechanism}, for groups at a slightly lower redshift $0.8<z<1$, though we find a larger QFE at the highest stellar masses. Interestingly, both groups and clusters are consistent with a mass-independent QFE below $M^*$ (QFE $\approx 0$ for groups, QFE $\approx 0.35-0.40$ for clusters) and a significant jump in QFE towards QFE $\approx 1$ above $M^*$.

\subsection{The halo mass dependence of galaxy quenching} \label{sec:QFE-halo-mass-dependence}

We now present QFE as a function of halo mass and stellar mass, in Figure \ref{fig:fQ-and-QE-vs-Mhalo-redshift-evolution-literature-summary-mstellar-breakdown}. The group and cluster samples at each redshift are described in \S\ref{sec:comp-samp}, as are the definitions of the field samples. Motivated by the results in Figure~\ref{fig:QFE-vs-Mstellar-GG-groups} we show the results in two stellar mass bins, below and above the jump in QFE. Our conclusions are unchanged if we use a smaller binning, which we demonstrate in Appendix~\ref{sec-rebin_app}.

In general we find the QFE increases with increasing stellar mass and halo mass, with at most modest redshift evolution when those parameters are fixed. Most notably, the dependence of QFE on the logarithm of halo mass appears to be similar in all stellar mass and redshift bins. To further quantify this, we fit a linear regression model to QFE as a function of $M_{\rm halo}$ for all the data, with a single slope but different intercepts for each redshift and stellar mass bin. We find a slope of $m= 0.24 \pm 0.04$ with a reduced $\chi^2$ of $\sim 1.06$ indicating an acceptable fit. 
The points that appear most discrepant with this simple scaling are for the highest halo masses in the stellar mass range $10^{10}<M_{\rm stellar}/\mathrm{M}_{\odot}<10^{10.7}$. The QFE for the GOGREEN data are actually lower than that measured at intermediate halo masses at the same redshift, though they are consistent at the $1.6\sigma$ level as determined by a two-tailed split-Gaussian hypothesis test. Though this appears to differ from the simple scaling derived above, we note that this approximate independence of QFE on halo masses above $\sim 2\times 10^{14}M_{\rm halo}$ is in fact consistent with what we observe at other redshift and stellar mass bins, and also with the simulation predictions discussed below, in \S\ref{sec:simulations}. Although there are only two clusters contributing to this bin, sample variance is unlikely to be large enough to explain the large difference between this measurement and the measurement of similarly massive clusters at lower redshift, given the observed homogeneity of cluster systems \citep[e.g.][]{Trudeau_2020}. On the other hand, the QFE observed in the massive Planck-selected clusters at $0.5<z<0.7$ is significantly higher than even the $z=0$ sample at the same mass, and implies a steeper logarithmic slope than we fit for the sample as a whole. It is possible that this reflects a bias resulting from the SZ-selection, or a difference in the field samples near those clusters, but we do not have a good explanation for the result. It would be useful to include more cluster samples at intermediate redshift in a future analysis.

\section{Discussion} \label{sec:discussion}

It is well known that the quiescent fraction of galaxies in clusters shows a general decrease with increasing redshift \citep[e.g.][]{ButcherOemler1984, Haines2013LowAndMidzOemlerEffect} and quiescent populations of galaxies are now well studied for clusters at $z>1$ \citep[e.g.][]{2013ApJ...779..138B,nantais2016stellar,LeeBrown2017,kawinwanichakij2017EffectOfLocalEnvAndMass,Foltz2018,Strazzullo2019highzclustersQFE,Trudeau_2020, vanderBurgGOGREENsmfs}. In this work we have used the wide halo mass range of the GOGREEN and COSMOS/SXDF cluster catalogues to demonstrate (Figure~\ref{fig:fQ-and-QE-vs-Mhalo-redshift-evolution-literature-summary-mstellar-breakdown}) that QFE correlates (increases) with both stellar mass and halo mass at $1<z<1.5$.

\begin{figure}
	\centering
		\includegraphics[width=\columnwidth]{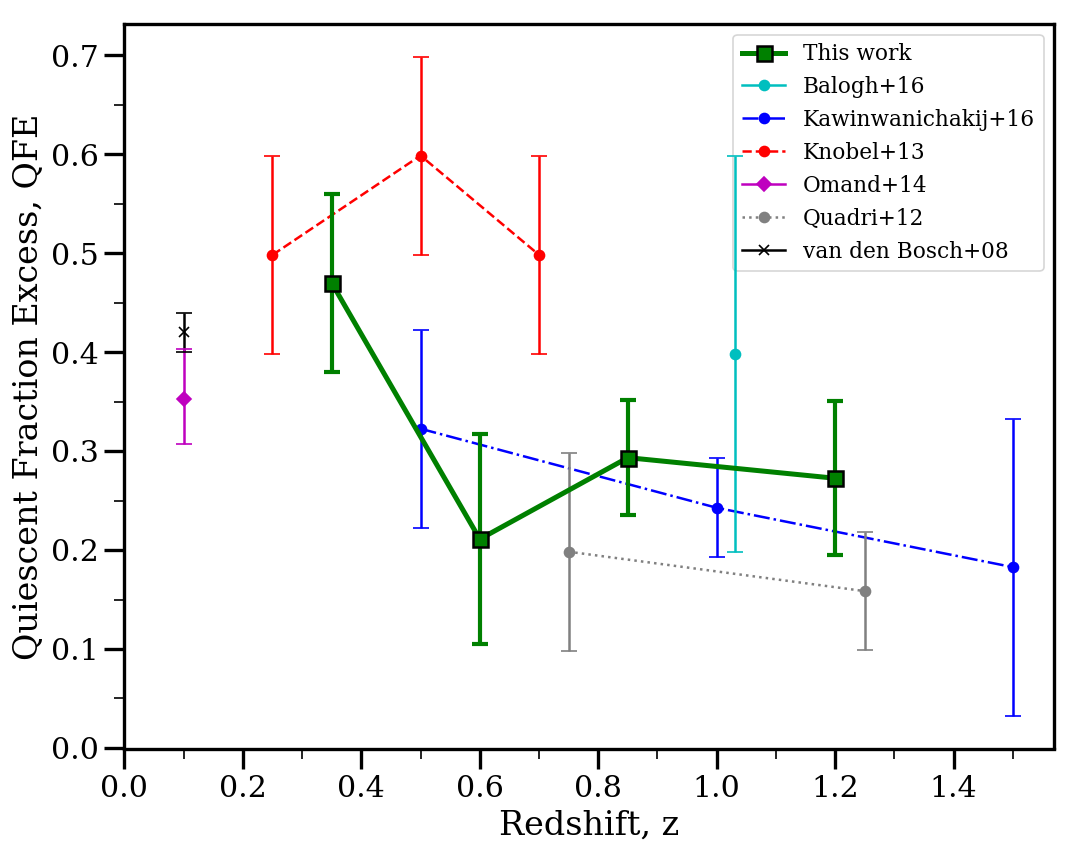}
	\caption{Compilation of quiescent fraction excess measurements as a function of redshift for groups of galaxies, adapted from \citet{nantais2016stellar}. Lines connect points from the same study. Our sample (green squares) include all galaxies with $M_{\rm stellar}>10^{10}~\mathrm{M}_\odot$, the 95\% stellar mass completeness limit at $z=1.5$. Overall the compilation appears to indicate a gradual redshift evolution of QFE. We caution, however, that the different analyses shown here are not fully consistent in their methodology or sample selections.
    }
	\label{fig:compilation-of-group_QE-with-redshift} 
\end{figure}

A compilation of QFE values (integrated over a broad stellar mass range) for low  halo mass ``groups'' at various redshifts was presented by \cite{nantais2016stellar}. We show an updated version of their figure including our background-subtracted group measurements in Figure \ref{fig:compilation-of-group_QE-with-redshift}, for all galaxies with $M_{\rm stellar}>10^{10}~\mathrm{M}_\odot$. Overall, our work is broadly consistent with the published literature: even within low-mass haloes the galaxy population exhibits enhanced quenching relative to the field. A possible redshift trend of QFE with redshift may be apparent in this compilation. However, we resist drawing any strong conclusions from a further quantitative comparison, given significant methodological differences between studies. Moreover, the stellar mass dependence of QFE complicates any physical interpretation of these integrated values. 

Our results build on the earlier work of
\cite{vanderBurgGOGREENsmfs}, who measured the stellar mass function of the GOGREEN cluster sample and found that, while the fraction of quiescent galaxies is much higher in the clusters than the field, the shape of the stellar mass function for quiescent galaxies is identical in both environments. The same is true for the star-forming population. This is a puzzling result and it indicates that, unlike at low redshift, environmental quenching is {\it not} separable from the stellar mass dependence. This is reflected in the fact that the QFE strongly increases with increasing stellar mass, from $\sim 30\%$ at $\sim 10^{10}~\mathrm{M}_{\odot}$ to $\sim 80\%$ at $>10^{11}~\mathrm{M}_{\odot}$, in contrast with studies in the local universe. A possible explanation for this, as described in \cite{vanderBurgGOGREENsmfs}, is that the quenching mechanism in these $z>1$ clusters is an accelerated version of the same process affecting field galaxies.
	
However, this interpretation would naively lead to the prediction that cluster galaxies should be substantially older than field galaxies, which contradicts the findings of \cite{WebbGOGREEN2020}. In that work, SFHs were measured for 331 quiescent galaxies in the same GOGREEN cluster and field samples, using the PROSPECTOR Bayesian inference code \citep{2017ApJ...837..170L, Prospector2019ascl.soft05025J} to fit the photometric and spectroscopic observations\footnote{The full posteriors of \cite{WebbGOGREEN2020}'s PROSPECTOR fits are available from the Canadian Advanced Network for Astronomical Research (CANFAR), at \url{www.canfar.net/storage/list/AstroDataCitationDOI/CISTI.CANFAR/20.0009/data}; DOI:10.11570/20.0009.}. 
They find that galaxies in clusters in the stellar mass range $10^{10}-10^{11.8}~\mathrm{M}_{\odot}$ are older than field galaxies, but only by $\lesssim 0.3$ Gyr.

As dark matter haloes grow, it is unclear exactly how or when environmental quenching processes become important \citep[e.g.][]{2013MNRAS.430.3017B}. In particular, if environmental processes are important in low-mass haloes, the quenching of star formation may take place long before galaxies are finally accreted onto massive clusters. In the following two subsections, we first explore how well hydrodynamic simulations reproduce our observations and then we use what we have learned about the halo mass dependence of the quiescent fraction at $1<z<1.5$ to explore the extent to which pre-processing could reconcile the \cite{vanderBurgGOGREENsmfs} and \cite{WebbGOGREEN2020} results. 

\subsection{Comparison with BAHAMAS hydrodynamic simulation predictions} \label{sec:simulations}

\begin{figure*} 
	\centering
		\includegraphics[width=2\columnwidth]{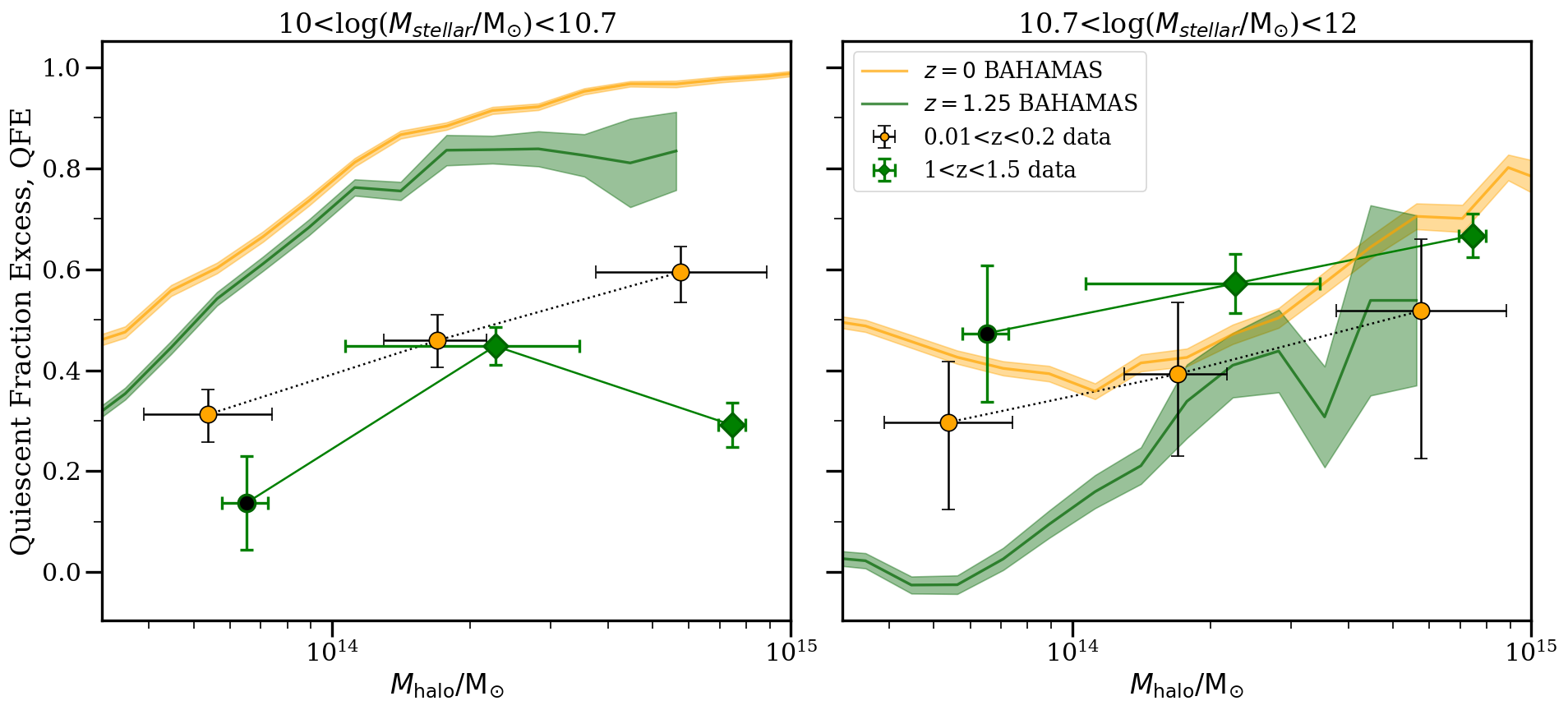}
	\caption{We show the quiescent fraction excess from the BAHAMAS hydrodynamic simulation as a function of stellar and halo mass, $M_{\rm halo} / \mathrm{M}_{\odot}$ at two redshift snapshots as indicated. The corresponding data from Figure \ref{fig:fQ-and-QE-vs-Mhalo-redshift-evolution-literature-summary-mstellar-breakdown} are reproduced here for comparison. In the simulations, both $f_Q$ and QFE decrease with increasing stellar mass, in stark contrast with the data. However, the correlation with halo mass and redshift is qualitatively similar to the trends observed in the data. See Figure \ref{fig:fQ-and-QE-vs-Mhalo-redshift-evolution-BAHAMAS-mstellar-breakdown-full} for a more detailed breakdown of both $f_Q$ and QFE by stellar mass.}
	\label{fig:fQ-and-QE-vs-Mhalo-redshift-evolution-BAHAMAS-mstellar-breakdown}
\end{figure*}

We begin considering the physical implications of our result by determining the extent to which these correlations are naturally predicted by hydrodynamic simulations. For this we use the BAHAMAS cluster simulations \citep{mccarthy2016bahamas, 2018MNRAS.476.2999M}, which were run with the standard Planck 2013 cosmology \citep{ade2014planck}, using 2$\times 1024^3$ particles in a large cosmological volume, 400 Mpc $h^{-1}$ on a side. Dark matter and (initial) baryon particles masses of $\approx4.44 \times 10^9~h^{-1}~\mathrm{M}_{\odot}$ and $8.11\times 10^8~h^{-1}~\mathrm{M}_{\odot}$ are used, respectively. These simulations implement various subgrid physics models, including metal-dependent radiative cooling in the presence of a uniform photoionising UV/X-ray background, star formation  stellar evolution and chemical enrichment, and stellar and AGN feedback \citep[see][and references therein for a detailed description of the subgrid implementations]{2010MNRAS.402.1536S}. Consistent with our work and compiled literature results, a \cite{chabrierIMF} IMF is assumed. The parameters of these prescriptions were adjusted to reproduce the observed Kennicutt-Schmidt law, the observed galaxy stellar mass function (GSMF), and the amplitude of group/cluster gas mass fraction--halo mass relation at $z \sim 0$. Thus, these simulations are distinguished from some others by the fact that they are deliberately calibrated to ensure the correct total baryon content in haloes. This is important when considering environmental effects on group scales where, for example, hydrodynamic interactions with the hot gas may be important.

We select all BAHAMAS groups with halo masses log$(M_{200c}/\mathrm{M}_{\odot})>13.2$ (no explicit upper halo mass limit). The group selection is based on a Friends-of-Friends group-finding algorithm applied to two separate snapshots of the BAHAMAS simulation, at $z=0$ and $z=1.25$, respectively. For each identified group, all galaxies within $R_{200c}$ (in 3D space) are considered to be group members. For each group, the field sample is taken to be all galaxies outside $2.5 R_{200c}$. To separate quiescent from star-forming galaxies, we use the sSFR threshold from \citet{franx2008structure}: $\mathrm{sSFR}>0.3/t_{H(z)}$, where $t_{H(z)}$ is the Hubble time at a given redshift. At $z\sim 1.25$ $t_{H(z)}\sim 5$ Gyr, so this corresponds to  $\mathrm{sSFR}>6\times 10^{-11}\mbox{ yr}^{-1}.$ Although this is different from the $UVJ$ selection made in the data, we have confirmed that the qualitative trends of $f_Q$ and QFE are stable for large variations in the choice of star-forming threshold -- a multiplicative factor of 2-3 to the sSFR cut does not qualitatively change our conclusions.

We focus our attention primarily on the QFE trends with halo mass in BAHAMAS, shown in Figure \ref{fig:fQ-and-QE-vs-Mhalo-redshift-evolution-BAHAMAS-mstellar-breakdown} for the same stellar mass bins as in Figure~\ref{fig:QFE-vs-Mstellar-GG-groups}. The quiescent fractions themselves, and an alternative stellar mass binning, are provided in Appendix~\ref{sec-rebin_app}. We first consider the high stellar mass sample, in the right panel. The BAHAMAS predictions at $z=0$ are in quite good agreement with the data, reproducing both the absolute value of the QFE and its dependence on halo mass. The simulations predict that this halo mass dependence becomes significantly steeper at $z\sim 1$, in contrast with the observations. There is reasonable agreement at high halo masses ($M_{\rm halo}>10^{14}~{\rm M}_\odot$), though the modest redshift evolution is in the opposite sense to the observations. On group scales, however, the models predict no significant QFE at $z\sim 1$, significantly below our measured QFE $= 0.48\pm0.15$.

Turning now to the lower stellar mass bin, $10<\log({M_{\rm stellar}/\mathrm{M}_{\odot}})<10.7$, the model generally predicts a steep increase in QFE with halo mass, before flattening around $M_{\rm halo}\sim 2\times 10^{14}~{\rm M}_{\odot}$. Over the whole mass range, the dependence of QFE on halo mass has a logarithmic slope of $\sim 0.3$ at both redshifts (i.e. a factor of 10 increase in halo mass results in an increase of $\sim 0.3$ in QFE), remarkably similar to our measurement in \S\ref{sec:QFE-halo-mass-dependence}. The sense and magnitude of the redshift evolution is also in good agreement with the observations. Despite these successes, the absolute value of the QFE itself is too high, for all halo masses at both $z=0$ and $z=1.25$. This reflects the difficulties faced by many simulations and models, and is likely due to an incomplete understanding of feedback (see Kukstas et al. 2021, in preparation). The result is also sensitive to choices in how quiescent galaxies are defined, and the aperture within which star formation rates and masses are measured in the simulations \citep[e.g.][]{Furlong2015,Donnari2019,2021MNRAS.500.4004D}. The fact that the simulations predict a halo mass dependence of QFE that is similar to what we observe over a wide range in redshift and halo mass is encouraging, and suggests that they may be correctly capturing the relevant physics associated with the impact of large scale structure growth on galaxy evolution, even if the feedback prescriptions themselves are not sufficiently accurate to reproduce the observed dependence of QFE on stellar mass. 

\subsection{Toy models} \label{sec:toy-models}

As described earlier, simple toy models of galaxy clusters at $1<z<1.5$, in which environmental quenching occurs after accretion onto the main progenitor, are unable to simultaneously match the observed quiescent fractions and relative ages of quiescent galaxies \citep{vanderBurgGOGREENsmfs,WebbGOGREEN2020}. In particular, \cite{vanderBurgGOGREENsmfs} note that the stellar mass-dependence of environmental quenching needs to be similar to that of the quenching process in the general field to result in the observed SMFs, and they propose that clusters experience an early accelerated form of that same phenomenon during the protocluster phase. However, \cite{WebbGOGREEN2020} find mass-weighted ages in cluster galaxies that are only slightly older than field galaxies. \cite{WebbGOGREEN2020} then use a simple infall model to demonstrate that neither a simple head-start to formation time for the cluster galaxies nor a simple quenching time delay (i.e. time since infall into a cluster) alone can explain both the enhanced quiescent fraction and very similar mass-weighted ages (MWAs) for the cluster and field populations.

\begin{figure*} 
	\centering
		\includegraphics[width=2\columnwidth]{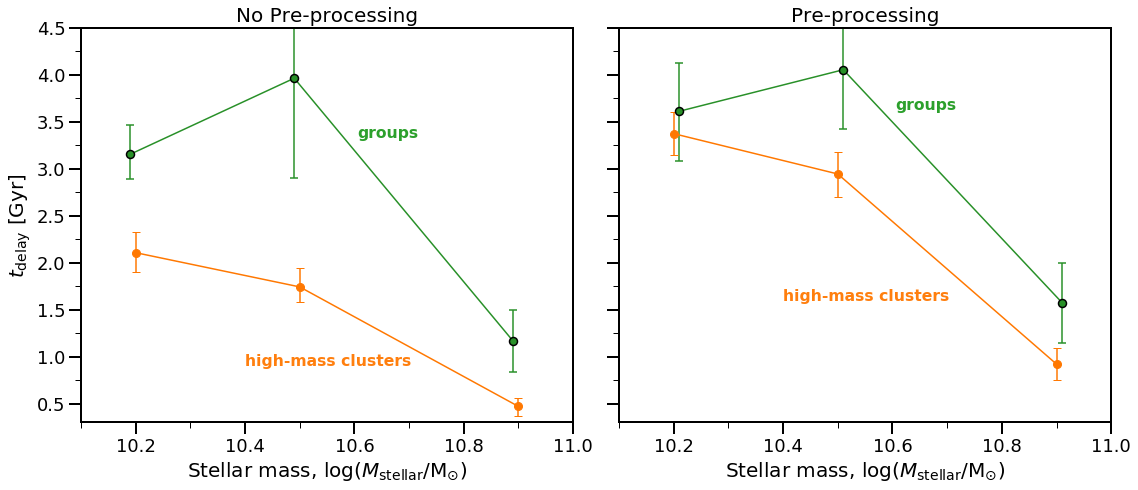}\\
	\caption{The time delay parameter values for our pre-processing and no pre-processing toy models are plotted as a function of stellar mass (points with error bars) for both groups (green) and clusters (orange). These time delays are directly constrained by the group/cluster quiescent fractions in each stellar mass bin; the uncertainties are a result of uncertainties on the observed quiescent fractions (i.e. propagated through the model).}
	\label{fig:toy-models-mstellar-dependence-tdelay}
\end{figure*}

Our results, which show how galaxy populations correlate with environment in haloes with masses well below that of massive clusters, suggest that ``pre-processing'' is likely to be important. The BAHAMAS simulations include a more complete treatment of halo growth and hydrodynamic processes, including pre-processing in a physically motivated way. The fact that those simulations predict a similar trend of QFE with halo mass to that observed is encouraging, and suggests that the failures of the toy models discussed above may lie in their simplified definitions of accretion time. We aim in this section to fit and contrast two toy models, one with pre-processing and one without, using the quiescent fraction excess in a range of halo masses at $1<z<1.5$. We can then use these models to predict group/cluster galaxies' average stellar mass-weighted ages and compare to values derived from GOGREEN observations.

\subsubsection{Toy model descriptions}\label{sec:toy-model-descriptions}

The infall-based quenching model we use here is the same as that in \cite{WebbGOGREEN2020},  but with changes to the accretion history. More explicitly, we use the \cite{Schreiber+2015} SFR evolution for star-forming galaxies. When galaxies are quenched their SF is immediately truncated. We track the number of star-forming and quiescent galaxies from $z=10$ to $z=1.2$ and then compare galaxies which have stellar masses between $10^{10}$--$10^{12}~\mathrm{M}_{\odot}$ at $z=1.2$. Galaxies that are self-quenched are modelled using the self-quenching efficiency proposed by \cite{Peng2010}, i.e.: the quenching probability is $\propto$ SFR/$M_{\mathrm{stellar}}$, using the SFR for a galaxy of stellar mass $M_{\mathrm{stellar}}$. We also assume that the shape of star-forming SMFs in our model do not differ between field and clusters, as found by \cite{vanderBurgGOGREENsmfs}. Finally, it is assumed that all galaxies can undergo satellite quenching, with star-forming galaxies in clusters completely quenching after a $t_{\mathrm{delay}}$ amount of time has elapsed from the time of first ``infall'', which we will define below.

\begin{figure*} 
	\centering
		\includegraphics[width=2\columnwidth]{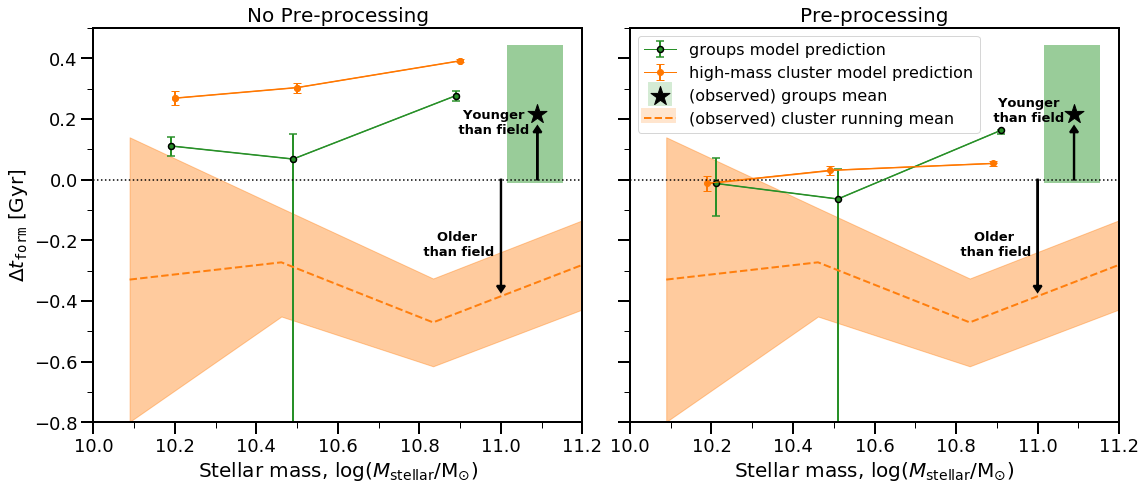}\\
	\caption{Using our fit $t_{\mathrm{delay}}$ values (Figure \ref{fig:toy-models-mstellar-dependence-tdelay}), we predict the difference in average formation time between the groups/high-mass clusters and the field. For a given galaxy in a given sample, $t_{\mathrm{form}}$ is defined as $t_{\mathrm{obs}}-$MWA, where MWA is the mass-weighted age from the stellar population synthesis modelling of quiescent galaxies in the GOGREEN spectroscopic sample \citep{WebbGOGREEN2020}. Model prediction values are shown with simple points with errorbars and connecting lines for the groups and clusters in green and orange, respectively. Observationally-derived average values are also shown, with the high-mass clusters running average shown as the dashed orange line (shaded regions for the boot-strapped standard deviation on the mean) and the groups are shown as a single point (black star) with green shaded region. The horizontal width of the green groups shaded region is the bootstrapped standard deviation on the mean stellar mass. Quiescent galaxies in groups are younger than those in clusters, ruling out the predictions of our simple model without pre-processing. Although the time delay exhibits a strong dependence on stellar mass, the dependence of mean MWA on stellar mass in our models is weak. The MWA has a significant halo mass dependence only for the no pre-processing model. For a plot showing the $t_{\mathrm{obs}}-$MWA values for individual galaxies in all samples, see Figure \ref{fig:observed-tobs-minus-MWA-appendix}.}
	\label{fig:tform-model-vs-observations}
\end{figure*}

For this quenching model there are therefore two parameters. The self-quenching efficiency normalization is set by reproducing the observationally measured field quiescent fraction (i.e. with $t_{\mathrm{delay}}$ fixed to $t_{\mathrm{delay}}=0$).
The $t_{\mathrm{delay}}$ parameter is then iteratively fit to reproduce the fraction of quiescent galaxies in a given cluster (see specifics described further in \S\ref{sec-modelresults}). As this represents quenching in excess of the field population, the delay time corresponds most directly to the QFE. If $t_{\mathrm{delay}}=0$, all cluster galaxies would be quenched; increasing the parameter reduces the QFE as fewer galaxies have been in the cluster long enough to quench. Finally, we neglect mergers, which are included in \cite{vanderBurgGOGREENsmfs}, for simplicity.

To explore whether simple pre-processing alleviates tension between quiescent fraction and ages, we consider two extreme definitions of galaxy ``infall time''. The first defines infall as the first time a central galaxy becomes a satellite, using the models of \cite{McGee+2009} (which were applied to N-body dark matter simulations) as published in \cite{Balogh2016domSatQuenchingMechanism}. In their findings, this amounts to a roughly constant accretion rate, with clusters starting their accretion $\sim 0.5$ Gyr before groups.
For the other extreme we assume galaxies are only accreted once they cross the virial radius of the most massive progenitor halo, using the halo mass accretion rate in \cite{Bouche+2010accretion}. We assume that the number of infalling galaxies in a given timestep is proportional to the mass accreted in a given timestep. We will refer to these two models as the ``pre-processing model'' and ``no pre-processing model'', respectively.

For simplicity, we only compare our ``groups'' sample (lowest halo-mass bin) with the high-mass clusters (highest halo mass bin) at $1<z<1.5$ presented in our results section (see \S\ref{sec:QFE-halo-mass-dependence}).

\subsubsection{Toy model results}\label{sec-modelresults}

We start by considering the stellar mass dependence of the $t_{\text{delay}}$ parameter, in 
Figure \ref{fig:toy-models-mstellar-dependence-tdelay}.
This parameter is effectively calculated from the observed quiescent fraction excess (see \S\ref{sec:toy-model-descriptions}), which has a strong stellar mass trend; thus, $t_{\text{delay}}$ similarly shows a strong trend with stellar mass, such that $t_{\text{delay}}$ decreases with increasing stellar mass in both models. As well, we observe the expected difference between the pre-processing and no pre-processing infall models. Pre-processing models require a longer $t_{\text{delay}}$ to reproduce the same quiescent fraction, given that first accretion happens earlier. The values of $t_{\text{delay}}$ that we find are broadly consistent with similar work at these redshifts \citep[e.g.][]{Balogh2016domSatQuenchingMechanism}, and with measurements at lower redshift \citep[e.g.][]{Wetzel2013MNRAS.432..336W} assuming they evolve proportionally to the dynamical time. As we are primarily interested in the trends with stellar and halo mass, we do not comment further on the absolute value of this parameter.

Most relevant to our discussion here, we find that the halo mass dependence of $t_{\text{delay}}$ depends on the accretion model. In the no pre-processing case there is a significant dependence on halo mass. Shorter delay times are needed in higher mass clusters, to reproduce our observations that the quiescent fractions are higher in those systems. In the pre-processing model, galaxies accreted into a cluster effectively get a head-start, and this largely accounts for the difference in quiescent fraction. Thus, we find that in a pre-processing model, the observed dependence of QFE on halo mass (at fixed stellar mass) can be explained with a $t_{\text{delay}}$ that has at most a weak dependence on halo mass. In this case the variation in QFE with halo mass derives primarily from the fact that the accretion time distribution is a function of halo mass (see e.g. \cite{DeLucia2012} for further discussion of this).

We now try to break this degeneracy by considering the mass-weighted ages of the quiescent galaxies in the two models. To show this, we define a formation time, $t_{\mathrm{form}}=t_{\text{obs}}-\text{MWA}$, which is the difference between the age of the universe (at a given galaxy's observed redshift) and the determined MWA value. This value reflects how long after the Big Bang it took until the mass-weighted bulk of stars had formed. We show our MWA predictions in the form of the difference in $t_{\mathrm{form}}$ between groups/clusters and the field in
Figure \ref{fig:tform-model-vs-observations}. The no pre-processing model predicts $t_{\mathrm{form}}$ should increase (ie: MWA decreases) with increasing halo mass, by about $\sim 150-200$ Myr between the lowest and highest halo masses shown here. In contrast, the pre-processing model predicts no significant dependence of MWA on halo mass.

We now compare these predictions directly with observed measurements\footnote{Specifically, here we show the medians of the posteriors.} of MWA from \cite{WebbGOGREEN2020}, shown in Figure \ref{fig:tform-model-vs-observations}. The running mean $t_{\mathrm{form}}$ for the highest halo mass sample (``high-mass clusters'') is shown; for the group sample, which includes only 15 quiescent galaxies, we show the mean and standard deviation (green shaded region) of the whole stellar mass range ($10<\log(M_{\mathrm{stellar}}/\mathrm{M}_\odot)<12$). A more detailed version of this Figure, showing age measurements for individual galaxies in all samples, is given in Appendix~\ref{sec-MWA_app}.

The no pre-processing model predicts that quiescent galaxies in groups should be $\sim 200$ Myr {\it older} than galaxies in clusters. Although our sample of group galaxies is small, it is inconsistent with this prediction: if anything, the quiescent group galaxies are {\it younger} than their counterparts in more massive systems. The data are more consistent with the pre-processing model. In this case, dependence on halo mass is weak, but in the observed direction for the highest stellar mass bin we consider, which also corresponds most closely to the mean stellar mass of our data.

In summary, including pre-processing does a reasonable job of explaining the halo mass dependence of quiescent galaxy ages, with a $t_{\rm delay}$ parameter that is nearly independent of halo mass. This is broadly consistent with a picture where the environmental quenching is caused by the same physical mechanism in groups and clusters. The data suggest, however, that quiescent galaxies in clusters may be even older than can be explained in the pre-processing model. One possible explanation for this would be if galaxies in rich proto-cluster environments undergo earlier quenching than primordial environments for group galaxies, as discussed in \cite{vanderBurgGOGREENsmfs} and \cite{WebbGOGREEN2020}. It seems increasingly likely that a significant portion of the GOGREEN cluster galaxy population was subject to an accelerated quenching mechanism at $z\sim 3-4$. This is additionally compatible with recent high redshift work showing that quiescent galaxies exist at redshifts as high as e.g. $z\sim 3-5$ \citep{2020ApJ...889...93V, 2020ApJ...903...47F}. 

\section{Conclusions} \label{Conclusion}

We use photometric redshifts and statistical background subtraction to measure stellar mass functions of galaxies in low mass haloes at $1<z<1.5$ (``groups''). These groups are selected from COSMOS and SXDF surveys, based on X-ray and sparse spectroscopy. We compute the quiescent fraction ($f_Q$) and quiescent fraction excess (QFE) for these systems, as a function of stellar mass. The result is then compared with higher mass clusters at $1<z<1.5$ from the GOGREEN survey \citep{GOGREEN2021data}, and a compilation of lower redshift samples at $0<z<0.2$ and $0.5<z<0.7$ that span a similar range of halo mass as our $1<z<1.5$ samples.

Observationally, we find:
\begin{itemize}
    \item Excess quenching in $1<z<1.5$ groups relative to the field, with an overall QFE of $\sim 20\%$ for galaxies with $\log(M_{\mathrm{stellar}}/\mathrm{M}_\odot)>10$.
    \item Unlike at low redshift, environmental quenching is \textit{not} separable from the stellar mass dependence. This can be seen as an increase of the QFE in groups, from $\sim 10\%$  below $M^*$ to $\sim 100\%$ above $M^*$. A similar trend is present in more massive clusters, where the QFE increases from $\sim 40\%$ to $\sim 85\%$.
    \item When controlling for stellar mass, both $f_Q$ and QFE correlate (increase) with halo mass. Observations at all redshifts and stellar masses are consistent with a single logarithmic slope of $\mathrm{d}(\mathrm{QFE})/\mathrm{d}\log (M_{\mathrm{halo}}) \sim 0.24 \pm 0.04$.
\end{itemize}

In our discussion, we compare to the BAHAMAS hydrodynamical simulation and also with toy models in which galaxies quench star formation upon infall, after some time delay. For the latter we contrast a pre-processing model, where galaxies begin this quenching time delay upon infall into any larger halo, and a no pre-processing model in which the time delay only begins when the galaxy is accreted into the main progenitor. 

From this analysis, we find:
\begin{itemize}
   \item The BAHAMAS hydrodynamic simulation reproduces the trend of quiescent fraction excess with halo mass 
   seen in the data. Specifically they show a steep increase in QFE with halo mass, which then flattens to a near constant value for halo masses $M_{\rm halo}\gtrsim 2\times 10^{14}~\mathrm{M}_\odot$. When fit with a straight line, the average trend is $\mathrm{d}(\mathrm{QFE})/\mathrm{d}\log (M_{\mathrm{halo}}) \sim 0.30$, which compares well with the observed $0.24\pm 0.04$. This suggests the simulation may be capturing the relevant physics behind the role of large scale structure growth on galaxy evolution. 
   \item Both the quiescent fraction and the quiescent fraction excess predicted by BAHAMAS decreases with increasing stellar mass, opposite to what is observed. This probably indicates an incomplete model of subgrid feedback and/or star formation at galaxy scales in the BAHAMAS simulation.
   \item From the toy models, we find the time delay until quenching begins must depend on stellar mass, reflecting the strong dependence of group/cluster quiescent fractions on stellar mass. In the absence of pre-processing, this delay time also has a strong dependence on halo mass, decreasing with increasing mass.
   \item We find pre-processing reduces the discrepancy with the observed halo mass dependence of quiescent galaxy mass-weighted ages. Specifically, assuming quenching occurs when a galaxy first becomes a satellite increases the average age of quiescent cluster galaxies, relative to a model without pre-processing. However, the data suggest that quiescent galaxies in clusters at $1<z<1.5$ may still be older than can be explained in this simple pre-processing model.
\end{itemize}

These observations further demonstrate that galaxy evolution depends on more than just stellar mass, in a nontrivial way that is still not fully captured by models. The environment, at least through the host halo mass, plays an important role at all redshifts $z<1.5$. This effect, however, is not separable from the dependence on stellar mass. Moreover, it is important even in low-mass haloes at $z\sim 1$, and thus likely not solely due to extreme physics like ram pressure stripping of cold gas reservoirs. The most natural physical mechanism that is expected to operate on all scales probed in this work is the shutoff of cosmological accretion onto satellites, and the subsequent overconsumption of gas reservoirs \citep[e.g.][]{McGee2014MNRAS}. This physics should be included with reasonable fidelity in hydrodynamic simulations, and it is encouraging that the BAHAMAS simulations are able to reproduce the observed halo-mass dependence, even while there remain problems on small-scales. Forthcoming, homogeneous surveys with large telescopes -- particularly those with highly multiplexed spectroscopy -- will make these statements much more precise and useful for constraining models. Observations of high redshift protoclusters, with JWST and other facilities, will determine whether or not there are additional effects that accelerate star formation quenching in these environments, as hinted at indirectly by our data.

\section*{Acknowledgements}

We thank the native Hawaiians for the use of Mauna Kea, as observations from Gemini, CFHT, and Subaru were all used as part of our survey.

Data products were used from observations made with ESO Telescopes at the La Silla Paranal Observatory under ESO programme ID 179.A-2005 and on data products produced by TERAPIX and the Cambridge Astronomy Survey Unit on behalf of the UltraVISTA consortium. As well, this study makes use of observations taken by the 3D-HST Treasury Program (GO 12177 and 12328) with the NASA/ESA HST, which is operated by the Association of Universities for Research in Astronomy, Inc., under NASA contract NAS5-26555.
MB gratefully acknowledges support from the NSERC Discovery Grant program.
BV acknowledges financial contribution  from the grant PRIN MIUR 2017 n.20173ML3WW\_001 (PI Cimatti) and from the INAF main-stream funding programme (PI Vulcani). GW acknowledges support from the National Science Foundation through grant AST-1517863, HST program number GO-15294, and grant number 80NSSC17K0019 issued through the NASA Astrophysics Data Analysis Program (ADAP). Support for program number GO-15294 was provided by NASA through a grant from the Space Telescope Science Institute, which is operated by the Association of Universities for Research in Astronomy, Incorporated, under NASA contract NAS5-26555. GR thanks the International Space Science Institute (ISSI) for providing financial support and a meeting facility that inspired insightful discussions for team “COSWEB: The Cosmic Web and Galaxy Evolution”. GR acknowledges support from the National Science Foundation grants AST-1517815, AST-1716690, and AST-1814159,  NASA HST grant AR-14310, and NASA ADAP grant 80NSSC19K0592. GR also acknowledges the support of an ESO visiting science fellowship. This work was supported in part by NSF grants AST-1815475 and AST-1518257. 

We thank M. Salvato for allowing us to use the master spectroscopic catalog used within the COSMOS collaboration.

This work made use of data products derived from Prospector \citep{2017ApJ...837..170L, Prospector2019ascl.soft05025J}, python-fsps and FSPS \citep{2010ascl.soft10043C, 2009ApJ...699..486C}. We also used the following Python \citep{van1995python} software packages: Astropy \citep{robitaille2013astropy, price2018astropy}, matplotlib \citep{Hunter:2007}, scipy \citep{2020SciPy-NMeth}, ipython \citep{PER-GRA:2007}, numpy \citep{harris2020array}, pandas \citep{reback2020pandas, mckinney-proc-scipy-2010}, and emcee \citep{2013PASP..125..306F}.

\section{Data availability}

Group catalogues are publicly available for both the COSMOS and SXDF group samples at \url{https://academic.oup.com/mnras/article/483/3/3545/5211093?login=true#supplementary-data} and in \cite{finoguenov2010x}, respectively.

The photometric datasets used for the $1<z<1.5$ groups analysis in this work are from sources in the public domain; UltraVISTA DR1 and DR3 at \url{http://ultravista.org/}, and SPLASH SXDF at \url{https://homepages.spa.umn.edu/~mehta074/splash/}.

Spectroscopic release information and datasets are also available in the public domain, 3D-HST at \url{https://archive.stsci.edu/prepds/3d-hst/}, UDSz at \url{https://www.nottingham.ac.uk/astronomy/UDS/UDSz/}, XMM-LSS survey at \url{https://heasarc.gsfc.nasa.gov/W3Browse/all/xmmlssclas.html}, VANDELS at \url{https://www.eso.org/sci/publications/announcements/sciann17248.html}. The compilation of published redshifts in the COSMOS field was obtained from a catalogue curated by M. Salvato; further information about a future public release of this catalogue can be found at \url{https://cosmos.astro.caltech.edu/page/specz}.

Access to the GOGREEN and GCLASS data release of spectroscopy, photometry, and derived data products, including jupyter python3 notebooks for reading and using the data, is available at the CADC (\url{https://www.cadc-ccda.hia-iha. nrc-cnrc.gc.ca/en/community/gogreen}), and NSF’s NOIR-Lab (\url{https://datalab.noao.edu/gogreendr1/}). Future releases, science results and other updates will be announced via the GOGREEN website at \url{http://gogreensurvey.ca/}. The full posteriors of \cite{WebbGOGREEN2020}'s PROSPECTOR fits are available from the Canadian Advanced Network for Astronomical Research (CANFAR), at \url{www.canfar.net/storage/list/AstroDataCitationDOI/CISTI.CANFAR/20.0009/data}; DOI:10.11570/20.0009.




\bibliographystyle{mnras}

\begin{thebibliography}{}
	\makeatletter
	\relax
	\def\mn@urlcharsother{\let\do\@makeother \do\$\do\&\do\#\do\^\do\_\do\%\do\~}
	\def\mn@doi{\begingroup\mn@urlcharsother \@ifnextchar [ {\mn@doi@}
		{\mn@doi@[]}}
	\def\mn@doi@[#1]#2{\def\@tempa{#1}\ifx\@tempa\@empty \href
		{http://dx.doi.org/#2} {doi:#2}\else \href {http://dx.doi.org/#2} {#1}\fi
		\endgroup}
	\def\mn@eprint#1#2{\mn@eprint@#1:#2::\@nil}
	\def\mn@eprint@arXiv#1{\href {http://arxiv.org/abs/#1} {{\tt arXiv:#1}}}
	\def\mn@eprint@dblp#1{\href {http://dblp.uni-trier.de/rec/bibtex/#1.xml}
		{dblp:#1}}
	\def\mn@eprint@#1:#2:#3:#4\@nil{\def\@tempa {#1}\def\@tempb {#2}\def\@tempc
		{#3}\ifx \@tempc \@empty \let \@tempc \@tempb \let \@tempb \@tempa \fi \ifx
		\@tempb \@empty \def\@tempb {arXiv}\fi \@ifundefined
		{mn@eprint@\@tempb}{\@tempb:\@tempc}{\expandafter \expandafter \csname
			mn@eprint@\@tempb\endcsname \expandafter{\@tempc}}}
	
	\bibitem[\protect\citeauthoryear{Ade et~al.,}{Ade et~al.}{2014}]{ade2014planck}
	Ade P.~A.,  et~al., 2014, Astronomy \& Astrophysics, 571, A16
	
	\bibitem[\protect\citeauthoryear{{Arnouts}, {Cristiani}, {Moscardini},
		{Matarrese}, {Lucchin}, {Fontana}  \& {Giallongo}}{{Arnouts}
		et~al.}{1999}]{Arnouts1999}
	{Arnouts} S.,  {Cristiani} S.,  {Moscardini} L.,  {Matarrese} S.,  {Lucchin}
	F.,  {Fontana} A.,   {Giallongo} E.,  1999, \mn@doi [\mnras]
	{10.1046/j.1365-8711.1999.02978.x}, \href
	{https://ui.adsabs.harvard.edu/abs/1999MNRAS.310..540A} {310, 540}
	
	\bibitem[\protect\citeauthoryear{{Bah{\'e}} \& {McCarthy}}{{Bah{\'e}} \&
		{McCarthy}}{2015}]{Bahe2015}
	{Bah{\'e}} Y.~M.,  {McCarthy} I.~G.,  2015, \mn@doi [\mnras]
	{10.1093/mnras/stu2293}, \href
	{https://ui.adsabs.harvard.edu/abs/2015MNRAS.447..969B} {447, 969}
	
	\bibitem[\protect\citeauthoryear{{Bah{\'e}}, {McCarthy}, {Balogh}  \&
		{Font}}{{Bah{\'e}} et~al.}{2013}]{2013MNRAS.430.3017B}
	{Bah{\'e}} Y.~M.,  {McCarthy} I.~G.,  {Balogh} M.~L.,   {Font} A.~S.,  2013,
	\mn@doi [\mnras] {10.1093/mnras/stt109}, \href
	{https://ui.adsabs.harvard.edu/abs/2013MNRAS.430.3017B} {430, 3017}
	
	\bibitem[\protect\citeauthoryear{{Bah{\'e}} et~al.,}{{Bah{\'e}}
		et~al.}{2017}]{BaheHydrangea2017}
	{Bah{\'e}} Y.~M.,  et~al., 2017, \mn@doi [\mnras] {10.1093/mnras/stx1403},
	\href {https://ui.adsabs.harvard.edu/abs/2017MNRAS.470.4186B} {470, 4186}
	
	\bibitem[\protect\citeauthoryear{{Baldry}, {Glazebrook}, {Brinkmann},
		{Ivezi{\'c}}, {Lupton}, {Nichol}  \& {Szalay}}{{Baldry}
		et~al.}{2004}]{Baldry2004}
	{Baldry} I.~K.,  {Glazebrook} K.,  {Brinkmann} J.,  {Ivezi{\'c}} {\v{Z}}.,
	{Lupton} R.~H.,  {Nichol} R.~C.,   {Szalay} A.~S.,  2004, \mn@doi [\apj]
	{10.1086/380092}, \href
	{https://ui.adsabs.harvard.edu/abs/2004ApJ...600..681B} {600, 681}
	
	\bibitem[\protect\citeauthoryear{{Baldry}, {Balogh}, {Bower}, {Glazebrook},
		{Nichol}, {Bamford}  \& {Budavari}}{{Baldry} et~al.}{2006}]{Baldry2006}
	{Baldry} I.~K.,  {Balogh} M.~L.,  {Bower} R.~G.,  {Glazebrook} K.,  {Nichol}
	R.~C.,  {Bamford} S.~P.,   {Budavari} T.,  2006, \mn@doi [\mnras]
	{10.1111/j.1365-2966.2006.11081.x}, \href
	{https://ui.adsabs.harvard.edu/abs/2006MNRAS.373..469B} {373, 469}
	
	\bibitem[\protect\citeauthoryear{{Balogh}, {Navarro}  \& {Morris}}{{Balogh}
		et~al.}{2000}]{Balogh2000}
	{Balogh} M.~L.,  {Navarro} J.~F.,   {Morris} S.~L.,  2000, \mn@doi [\apj]
	{10.1086/309323}, \href
	{https://ui.adsabs.harvard.edu/abs/2000ApJ...540..113B} {540, 113}
	
	\bibitem[\protect\citeauthoryear{{Balogh} et~al.,}{{Balogh}
		et~al.}{2004a}]{Balogh2004}
	{Balogh} M.,  et~al., 2004a, \mn@doi [\mnras]
	{10.1111/j.1365-2966.2004.07453.x}, \href
	{https://ui.adsabs.harvard.edu/abs/2004MNRAS.348.1355B} {348, 1355}
	
	\bibitem[\protect\citeauthoryear{{Balogh}, {Baldry}, {Nichol}, {Miller},
		{Bower}  \& {Glazebrook}}{{Balogh} et~al.}{2004b}]{Balogh2004bimodality}
	{Balogh} M.~L.,  {Baldry} I.~K.,  {Nichol} R.,  {Miller} C.,  {Bower} R.,
	{Glazebrook} K.,  2004b, \mn@doi [\apjl] {10.1086/426079}, \href
	{https://ui.adsabs.harvard.edu/abs/2004ApJ...615L.101B} {615, L101}
	
	\bibitem[\protect\citeauthoryear{{Balogh} et~al.,}{{Balogh}
		et~al.}{2014}]{Balogh2014GEEC2}
	{Balogh} M.~L.,  et~al., 2014, \mn@doi [\mnras] {10.1093/mnras/stu1332}, \href
	{https://ui.adsabs.harvard.edu/abs/2014MNRAS.443.2679B} {443, 2679}
	
	\bibitem[\protect\citeauthoryear{{Balogh} et~al.,}{{Balogh}
		et~al.}{2016}]{Balogh2016domSatQuenchingMechanism}
	{Balogh} M.~L.,  et~al., 2016, \mn@doi [\mnras] {10.1093/mnras/stv2949}, \href
	{https://ui.adsabs.harvard.edu/abs/2016MNRAS.456.4364B} {456, 4364}
	
	\bibitem[\protect\citeauthoryear{{Balogh} et~al.,}{{Balogh}
		et~al.}{2017}]{balogh2017gemini}
	{Balogh} M.~L.,  et~al., 2017, \mn@doi [\mnras] {10.1093/mnras/stx1370}, \href
	{https://ui.adsabs.harvard.edu/abs/2017MNRAS.470.4168B} {470, 4168}
	
	\bibitem[\protect\citeauthoryear{{Balogh} et~al.,}{{Balogh}
		et~al.}{2021}]{GOGREEN2021data}
	{Balogh} M.~L.,  et~al., 2021, \mn@doi [\mnras] {10.1093/mnras/staa3008}, \href
	{https://ui.adsabs.harvard.edu/abs/2021MNRAS.500..358B} {500, 358}
	
	\bibitem[\protect\citeauthoryear{{Behroozi}, {Wechsler}, {Lu}, {Hahn}, {Busha},
		{Klypin}  \& {Primack}}{{Behroozi} et~al.}{2014}]{Behroozi2014}
	{Behroozi} P.~S.,  {Wechsler} R.~H.,  {Lu} Y.,  {Hahn} O.,  {Busha} M.~T.,
	{Klypin} A.,   {Primack} J.~R.,  2014, \mn@doi [\apj]
	{10.1088/0004-637X/787/2/156}, \href
	{https://ui.adsabs.harvard.edu/abs/2014ApJ...787..156B} {787, 156}
	
	\bibitem[\protect\citeauthoryear{{Bell} et~al.,}{{Bell}
		et~al.}{2004a}]{Bell2004}
	{Bell} E.~F.,  et~al., 2004a, \mn@doi [\apjl] {10.1086/381388}, \href
	{https://ui.adsabs.harvard.edu/abs/2004ApJ...600L..11B} {600, L11}
	
	\bibitem[\protect\citeauthoryear{{Bell} et~al.,}{{Bell}
		et~al.}{2004b}]{Bell2004RSevolution}
	{Bell} E.~F.,  et~al., 2004b, \mn@doi [\apj] {10.1086/420778}, \href
	{https://ui.adsabs.harvard.edu/abs/2004ApJ...608..752B} {608, 752}
	
	\bibitem[\protect\citeauthoryear{{Berrier}, {Stewart}, {Bullock}, {Purcell},
		{Barton}  \& {Wechsler}}{{Berrier} et~al.}{2009}]{Berrier2009}
	{Berrier} J.~C.,  {Stewart} K.~R.,  {Bullock} J.~S.,  {Purcell} C.~W.,
	{Barton} E.~J.,   {Wechsler} R.~H.,  2009, \mn@doi [\apj]
	{10.1088/0004-637X/690/2/1292}, \href
	{https://ui.adsabs.harvard.edu/abs/2009ApJ...690.1292B} {690, 1292}
	
	\bibitem[\protect\citeauthoryear{{Biviano} et~al.,}{{Biviano}
		et~al.}{2021}]{Gogreendynamics}
	{Biviano} A.,  et~al., 2021, arXiv e-prints, \href
	{https://ui.adsabs.harvard.edu/abs/2021arXiv210401183B} {p. arXiv:2104.01183}
	
	\bibitem[\protect\citeauthoryear{{Bouch{\'e}} et~al.,}{{Bouch{\'e}}
		et~al.}{2010}]{Bouche+2010accretion}
	{Bouch{\'e}} N.,  et~al., 2010, \mn@doi [\apj] {10.1088/0004-637X/718/2/1001},
	\href {https://ui.adsabs.harvard.edu/abs/2010ApJ...718.1001B} {718, 1001}
	
	\bibitem[\protect\citeauthoryear{{Bower}, {Schaye}, {Frenk}, {Theuns},
		{Schaller}, {Crain}  \& {McAlpine}}{{Bower} et~al.}{2017}]{Bower2017}
	{Bower} R.~G.,  {Schaye} J.,  {Frenk} C.~S.,  {Theuns} T.,  {Schaller} M.,
	{Crain} R.~A.,   {McAlpine} S.,  2017, \mn@doi [\mnras]
	{10.1093/mnras/stw2735}, \href
	{https://ui.adsabs.harvard.edu/abs/2017MNRAS.465...32B} {465, 32}
	
	\bibitem[\protect\citeauthoryear{{Bradshaw} et~al.,}{{Bradshaw}
		et~al.}{2013}]{bradshaw2013high}
	{Bradshaw} E.~J.,  et~al., 2013, \mn@doi [\mnras] {10.1093/mnras/stt715}, \href
	{https://ui.adsabs.harvard.edu/abs/2013MNRAS.433..194B} {433, 194}
	
	\bibitem[\protect\citeauthoryear{{Brammer}, {van Dokkum}  \& {Coppi}}{{Brammer}
		et~al.}{2010}]{brammer2010ascl.soft10052B}
	{Brammer} G.~B.,  {van Dokkum} P.~G.,   {Coppi} P.,  2010, {EAZY: A Fast,
		Public Photometric Redshift Code} (\mn@eprint {ascl} {1010.052})
	
	\bibitem[\protect\citeauthoryear{{Brammer} et~al.,}{{Brammer}
		et~al.}{2011}]{Brammer2011}
	{Brammer} G.~B.,  et~al., 2011, \mn@doi [\apj] {10.1088/0004-637X/739/1/24},
	\href {https://ui.adsabs.harvard.edu/abs/2011ApJ...739...24B} {739, 24}
	
	\bibitem[\protect\citeauthoryear{Brammer et~al.,}{Brammer
		et~al.}{2012}]{brammer20123d}
	Brammer G.~B.,  et~al., 2012, The Astrophysical Journal Supplement Series, 200,
	13
	
	\bibitem[\protect\citeauthoryear{{Brinchmann}, {Charlot}, {White}, {Tremonti},
		{Kauffmann}, {Heckman}  \& {Brinkmann}}{{Brinchmann}
		et~al.}{2004}]{Brinchmann2004}
	{Brinchmann} J.,  {Charlot} S.,  {White} S.~D.~M.,  {Tremonti} C.,  {Kauffmann}
	G.,  {Heckman} T.,   {Brinkmann} J.,  2004, \mn@doi [\mnras]
	{10.1111/j.1365-2966.2004.07881.x}, \href
	{https://ui.adsabs.harvard.edu/abs/2004MNRAS.351.1151B} {351, 1151}
	
	\bibitem[\protect\citeauthoryear{{Brodwin} et~al.,}{{Brodwin}
		et~al.}{2013}]{2013ApJ...779..138B}
	{Brodwin} M.,  et~al., 2013, \mn@doi [\apj] {10.1088/0004-637X/779/2/138},
	\href {https://ui.adsabs.harvard.edu/abs/2013ApJ...779..138B} {779, 138}
	
	\bibitem[\protect\citeauthoryear{{Bruzual} \& {Charlot}}{{Bruzual} \&
		{Charlot}}{2003}]{Bruzual2003}
	{Bruzual} G.,  {Charlot} S.,  2003, \mn@doi [\mnras]
	{10.1046/j.1365-8711.2003.06897.x}, \href
	{https://ui.adsabs.harvard.edu/abs/2003MNRAS.344.1000B} {344, 1000}
	
	\bibitem[\protect\citeauthoryear{{Butcher} \& {Oemler}}{{Butcher} \&
		{Oemler}}{1984}]{ButcherOemler1984}
	{Butcher} H.,  {Oemler} A. J.,  1984, \mn@doi [\apj] {10.1086/162519}, \href
	{https://ui.adsabs.harvard.edu/abs/1984ApJ...285..426B} {285, 426}
	
	\bibitem[\protect\citeauthoryear{{Chabrier}}{{Chabrier}}{2003}]{chabrierIMF}
	{Chabrier} G.,  2003, \mn@doi [\pasp] {10.1086/376392}, \href
	{https://ui.adsabs.harvard.edu/abs/2003PASP..115..763C} {115, 763}
	
	\bibitem[\protect\citeauthoryear{{Chiappetti} et~al.,}{{Chiappetti}
		et~al.}{2013}]{2013MNRAS.429.1652C}
	{Chiappetti} L.,  et~al., 2013, \mn@doi [\mnras] {10.1093/mnras/sts453}, \href
	{https://ui.adsabs.harvard.edu/abs/2013MNRAS.429.1652C} {429, 1652}
	
	\bibitem[\protect\citeauthoryear{{Conroy} \& {Gunn}}{{Conroy} \&
		{Gunn}}{2010}]{2010ascl.soft10043C}
	{Conroy} C.,  {Gunn} J.~E.,  2010, {FSPS: Flexible Stellar Population
		Synthesis} (\mn@eprint {ascl} {1010.043})
	
	\bibitem[\protect\citeauthoryear{{Conroy}, {Gunn}  \& {White}}{{Conroy}
		et~al.}{2009}]{2009ApJ...699..486C}
	{Conroy} C.,  {Gunn} J.~E.,   {White} M.,  2009, \mn@doi [\apj]
	{10.1088/0004-637X/699/1/486}, \href
	{https://ui.adsabs.harvard.edu/abs/2009ApJ...699..486C} {699, 486}
	
	\bibitem[\protect\citeauthoryear{{Cooper} et~al.,}{{Cooper}
		et~al.}{2006}]{Cooper2006}
	{Cooper} M.~C.,  et~al., 2006, \mn@doi [\mnras]
	{10.1111/j.1365-2966.2006.10485.x}, \href
	{https://ui.adsabs.harvard.edu/abs/2006MNRAS.370..198C} {370, 198}
	
	\bibitem[\protect\citeauthoryear{{Darvish}, {Mobasher}, {Sobral}, {Rettura},
		{Scoville}, {Faisst}  \& {Capak}}{{Darvish} et~al.}{2016}]{Darvish2016}
	{Darvish} B.,  {Mobasher} B.,  {Sobral} D.,  {Rettura} A.,  {Scoville} N.,
	{Faisst} A.,   {Capak} P.,  2016, \mn@doi [\apj]
	{10.3847/0004-637X/825/2/113}, \href
	{https://ui.adsabs.harvard.edu/abs/2016ApJ...825..113D} {825, 113}
	
	\bibitem[\protect\citeauthoryear{{Dav{\'e}}, {Finlator}  \&
		{Oppenheimer}}{{Dav{\'e}} et~al.}{2012}]{Dave2012}
	{Dav{\'e}} R.,  {Finlator} K.,   {Oppenheimer} B.~D.,  2012, \mn@doi [\mnras]
	{10.1111/j.1365-2966.2011.20148.x}, \href
	{https://ui.adsabs.harvard.edu/abs/2012MNRAS.421...98D} {421, 98}
	
	\bibitem[\protect\citeauthoryear{{Davidzon} et~al.,}{{Davidzon}
		et~al.}{2016}]{Davidzon2016}
	{Davidzon} I.,  et~al., 2016, \mn@doi [\aap] {10.1051/0004-6361/201527129},
	\href {https://ui.adsabs.harvard.edu/abs/2016A&A...586A..23D} {586, A23}
	
	\bibitem[\protect\citeauthoryear{{De Lucia} et~al.,}{{De Lucia}
		et~al.}{2004}]{DeLucia2004}
	{De Lucia} G.,  et~al., 2004, \mn@doi [\apjl] {10.1086/423373}, \href
	{https://ui.adsabs.harvard.edu/abs/2004ApJ...610L..77D} {610, L77}
	
	\bibitem[\protect\citeauthoryear{{De Lucia}, {Weinmann}, {Poggianti},
		{Arag{\'o}n-Salamanca}  \& {Zaritsky}}{{De Lucia} et~al.}{2012}]{DeLucia2012}
	{De Lucia} G.,  {Weinmann} S.,  {Poggianti} B.~M.,  {Arag{\'o}n-Salamanca} A.,
	{Zaritsky} D.,  2012, \mn@doi [\mnras] {10.1111/j.1365-2966.2012.20983.x},
	\href {https://ui.adsabs.harvard.edu/abs/2012MNRAS.423.1277D} {423, 1277}
	
	\bibitem[\protect\citeauthoryear{{Donnari} et~al.,}{{Donnari}
		et~al.}{2019}]{Donnari2019}
	{Donnari} M.,  et~al., 2019, \mn@doi [\mnras] {10.1093/mnras/stz712}, \href
	{https://ui.adsabs.harvard.edu/abs/2019MNRAS.485.4817D} {485, 4817}
	
	\bibitem[\protect\citeauthoryear{{Donnari} et~al.,}{{Donnari}
		et~al.}{2021}]{2021MNRAS.500.4004D}
	{Donnari} M.,  et~al., 2021, \mn@doi [\mnras] {10.1093/mnras/staa3006}, \href
	{https://ui.adsabs.harvard.edu/abs/2021MNRAS.500.4004D} {500, 4004}
	
	\bibitem[\protect\citeauthoryear{{Faber} et~al.,}{{Faber}
		et~al.}{2007}]{Faber2007}
	{Faber} S.~M.,  et~al., 2007, \mn@doi [\apj] {10.1086/519294}, \href
	{https://ui.adsabs.harvard.edu/abs/2007ApJ...665..265F} {665, 265}
	
	\bibitem[\protect\citeauthoryear{{Fillingham}, {Cooper}, {Wheeler},
		{Garrison-Kimmel}, {Boylan-Kolchin}  \& {Bullock}}{{Fillingham}
		et~al.}{2015}]{Fillingham2015}
	{Fillingham} S.~P.,  {Cooper} M.~C.,  {Wheeler} C.,  {Garrison-Kimmel} S.,
	{Boylan-Kolchin} M.,   {Bullock} J.~S.,  2015, \mn@doi [\mnras]
	{10.1093/mnras/stv2058}, \href
	{https://ui.adsabs.harvard.edu/abs/2015MNRAS.454.2039F} {454, 2039}
	
	\bibitem[\protect\citeauthoryear{{Finlator} \& {Dav{\'e}}}{{Finlator} \&
		{Dav{\'e}}}{2008}]{Finlator2008}
	{Finlator} K.,  {Dav{\'e}} R.,  2008, \mn@doi [\mnras]
	{10.1111/j.1365-2966.2008.12991.x}, \href
	{https://ui.adsabs.harvard.edu/abs/2008MNRAS.385.2181F} {385, 2181}
	
	\bibitem[\protect\citeauthoryear{{Finoguenov} et~al.,}{{Finoguenov}
		et~al.}{2010}]{finoguenov2010x}
	{Finoguenov} A.,  et~al., 2010, \mn@doi [\mnras]
	{10.1111/j.1365-2966.2010.16256.x}, \href
	{https://ui.adsabs.harvard.edu/abs/2010MNRAS.403.2063F} {403, 2063}
	
	\bibitem[\protect\citeauthoryear{{Foltz} et~al.,}{{Foltz}
		et~al.}{2018}]{Foltz2018}
	{Foltz} R.,  et~al., 2018, \mn@doi [\apj] {10.3847/1538-4357/aad80d}, \href
	{https://ui.adsabs.harvard.edu/abs/2018ApJ...866..136F} {866, 136}
	
	\bibitem[\protect\citeauthoryear{{Foreman-Mackey}, {Hogg}, {Lang}  \&
		{Goodman}}{{Foreman-Mackey} et~al.}{2013}]{2013PASP..125..306F}
	{Foreman-Mackey} D.,  {Hogg} D.~W.,  {Lang} D.,   {Goodman} J.,  2013, \mn@doi
	[\pasp] {10.1086/670067}, \href
	{https://ui.adsabs.harvard.edu/abs/2013PASP..125..306F} {125, 306}
	
	\bibitem[\protect\citeauthoryear{{Forrest} et~al.,}{{Forrest}
		et~al.}{2020}]{2020ApJ...903...47F}
	{Forrest} B.,  et~al., 2020, \mn@doi [\apj] {10.3847/1538-4357/abb819}, \href
	{https://ui.adsabs.harvard.edu/abs/2020ApJ...903...47F} {903, 47}
	
	\bibitem[\protect\citeauthoryear{{Fossati} et~al.,}{{Fossati}
		et~al.}{2017}]{Fossati2017}
	{Fossati} M.,  et~al., 2017, \mn@doi [\apj] {10.3847/1538-4357/835/2/153},
	\href {https://ui.adsabs.harvard.edu/abs/2017ApJ...835..153F} {835, 153}
	
	\bibitem[\protect\citeauthoryear{Franx, van Dokkum, Schreiber, Wuyts, Labb{\'e}
		\& Toft}{Franx et~al.}{2008}]{franx2008structure}
	Franx M.,  van Dokkum P.~G.,  Schreiber N. M.~F.,  Wuyts S.,  Labb{\'e} I.,
	Toft S.,  2008, The Astrophysical Journal, 688, 770
	
	\bibitem[\protect\citeauthoryear{{Furlong} et~al.,}{{Furlong}
		et~al.}{2015}]{Furlong2015}
	{Furlong} M.,  et~al., 2015, \mn@doi [\mnras] {10.1093/mnras/stv852}, \href
	{https://ui.adsabs.harvard.edu/abs/2015MNRAS.450.4486F} {450, 4486}
	
	\bibitem[\protect\citeauthoryear{{Giodini} et~al.,}{{Giodini}
		et~al.}{2012}]{Giodini2012}
	{Giodini} S.,  et~al., 2012, \mn@doi [\aap] {10.1051/0004-6361/201117696},
	\href {https://ui.adsabs.harvard.edu/abs/2012A&A...538A.104G} {538, A104}
	
	\bibitem[\protect\citeauthoryear{{Gobat} et~al.,}{{Gobat}
		et~al.}{2019}]{Gobat2019}
	{Gobat} R.,  et~al., 2019, \mn@doi [\aap] {10.1051/0004-6361/201935862}, \href
	{https://ui.adsabs.harvard.edu/abs/2019A&A...629A.104G} {629, A104}
	
	\bibitem[\protect\citeauthoryear{{G{\'o}mez} et~al.,}{{G{\'o}mez}
		et~al.}{2003}]{Gomez2003}
	{G{\'o}mez} P.~L.,  et~al., 2003, \mn@doi [\apj] {10.1086/345593}, \href
	{https://ui.adsabs.harvard.edu/abs/2003ApJ...584..210G} {584, 210}
	
	\bibitem[\protect\citeauthoryear{{Gozaliasl} et~al.,}{{Gozaliasl}
		et~al.}{2019}]{gozaliasl2018chandraCOSMOSgroups}
	{Gozaliasl} G.,  et~al., 2019, \mn@doi [\mnras] {10.1093/mnras/sty3203}, \href
	{https://ui.adsabs.harvard.edu/abs/2019MNRAS.483.3545G} {483, 3545}
	
	\bibitem[\protect\citeauthoryear{{Guglielmo} et~al.,}{{Guglielmo}
		et~al.}{2019}]{2019A&A...625A.112G}
	{Guglielmo} V.,  et~al., 2019, \mn@doi [\aap] {10.1051/0004-6361/201834970},
	\href {https://ui.adsabs.harvard.edu/abs/2019A&A...625A.112G} {625, A112}
	
	\bibitem[\protect\citeauthoryear{{Haines} et~al.,}{{Haines}
		et~al.}{2013}]{Haines2013LowAndMidzOemlerEffect}
	{Haines} C.~P.,  et~al., 2013, \mn@doi [\apj] {10.1088/0004-637X/775/2/126},
	\href {https://ui.adsabs.harvard.edu/abs/2013ApJ...775..126H} {775, 126}
	
	\bibitem[\protect\citeauthoryear{Harris et~al.,}{Harris
		et~al.}{2020}]{harris2020array}
	Harris C.~R.,  et~al., 2020, \mn@doi [Nature] {10.1038/s41586-020-2649-2}, 585,
	357
	
	\bibitem[\protect\citeauthoryear{Hearin et~al.,}{Hearin
		et~al.}{2017}]{hearin2017forward}
	Hearin A.~P.,  et~al., 2017, The Astronomical Journal, 154, 190
	
	\bibitem[\protect\citeauthoryear{{Hirschmann}, {De Lucia}  \&
		{Fontanot}}{{Hirschmann} et~al.}{2016}]{2016MNRAS.461.1760H}
	{Hirschmann} M.,  {De Lucia} G.,   {Fontanot} F.,  2016, \mn@doi [\mnras]
	{10.1093/mnras/stw1318}, \href
	{https://ui.adsabs.harvard.edu/abs/2016MNRAS.461.1760H} {461, 1760}
	
	\bibitem[\protect\citeauthoryear{{Hou}, {Parker}  \& {Harris}}{{Hou}
		et~al.}{2014}]{Hou2014}
	{Hou} A.,  {Parker} L.~C.,   {Harris} W.~E.,  2014, \mn@doi [\mnras]
	{10.1093/mnras/stu829}, \href
	{https://ui.adsabs.harvard.edu/abs/2014MNRAS.442..406H} {442, 406}
	
	\bibitem[\protect\citeauthoryear{Hunter}{Hunter}{2007}]{Hunter:2007}
	Hunter J.~D.,  2007, \mn@doi [Computing in Science \& Engineering]
	{10.1109/MCSE.2007.55}, 9, 90
	
	\bibitem[\protect\citeauthoryear{{Ilbert} et~al.,}{{Ilbert}
		et~al.}{2006}]{Ilbert2006}
	{Ilbert} O.,  et~al., 2006, \mn@doi [\aap] {10.1051/0004-6361:20065138}, \href
	{https://ui.adsabs.harvard.edu/abs/2006A&A...457..841I} {457, 841}
	
	\bibitem[\protect\citeauthoryear{{Johnson}, {Leja}, {Conroy}  \&
		{Speagle}}{{Johnson} et~al.}{2019}]{Prospector2019ascl.soft05025J}
	{Johnson} B.~D.,  {Leja} J.~L.,  {Conroy} C.,   {Speagle} J.~S.,  2019,
	{Prospector: Stellar population inference from spectra and SEDs} (\mn@eprint
	{ascl} {1905.025})
	
	\bibitem[\protect\citeauthoryear{{Just} et~al.,}{{Just}
		et~al.}{2019}]{Dennis2019}
	{Just} D.~W.,  et~al., 2019, \mn@doi [\apj] {10.3847/1538-4357/ab44a0}, \href
	{https://ui.adsabs.harvard.edu/abs/2019ApJ...885....6J} {885, 6}
	
	\bibitem[\protect\citeauthoryear{{Kauffmann} et~al.,}{{Kauffmann}
		et~al.}{2003}]{Kauffmann2003}
	{Kauffmann} G.,  et~al., 2003, \mn@doi [\mnras]
	{10.1111/j.1365-2966.2003.07154.x}, \href
	{https://ui.adsabs.harvard.edu/abs/2003MNRAS.346.1055K} {346, 1055}
	
	\bibitem[\protect\citeauthoryear{{Kauffmann}, {White}, {Heckman}, {M{\'e}nard},
		{Brinchmann}, {Charlot}, {Tremonti}  \& {Brinkmann}}{{Kauffmann}
		et~al.}{2004}]{Kauffmann2004}
	{Kauffmann} G.,  {White} S. D.~M.,  {Heckman} T.~M.,  {M{\'e}nard} B.,
	{Brinchmann} J.,  {Charlot} S.,  {Tremonti} C.,   {Brinkmann} J.,  2004,
	\mn@doi [\mnras] {10.1111/j.1365-2966.2004.08117.x}, \href
	{https://ui.adsabs.harvard.edu/abs/2004MNRAS.353..713K} {353, 713}
	
	\bibitem[\protect\citeauthoryear{{Kawata} \& {Mulchaey}}{{Kawata} \&
		{Mulchaey}}{2008}]{Kawata2008}
	{Kawata} D.,  {Mulchaey} J.~S.,  2008, \mn@doi [\apjl] {10.1086/526544}, \href
	{https://ui.adsabs.harvard.edu/abs/2008ApJ...672L.103K} {672, L103}
	
	\bibitem[\protect\citeauthoryear{{Kawinwanichakij} et~al.,}{{Kawinwanichakij}
		et~al.}{2017}]{kawinwanichakij2017EffectOfLocalEnvAndMass}
	{Kawinwanichakij} L.,  et~al., 2017, \mn@doi [\apj] {10.3847/1538-4357/aa8b75},
	\href {https://ui.adsabs.harvard.edu/abs/2017ApJ...847..134K} {847, 134}
	
	\bibitem[\protect\citeauthoryear{{Kimm} et~al.,}{{Kimm}
		et~al.}{2009}]{Kimm2009}
	{Kimm} T.,  et~al., 2009, \mn@doi [\mnras] {10.1111/j.1365-2966.2009.14414.x},
	\href {https://ui.adsabs.harvard.edu/abs/2009MNRAS.394.1131K} {394, 1131}
	
	\bibitem[\protect\citeauthoryear{{Kova{\v{c}}} et~al.,}{{Kova{\v{c}}}
		et~al.}{2010}]{Kovac2010zCOSMOS}
	{Kova{\v{c}}} K.,  et~al., 2010, \mn@doi [\apj] {10.1088/0004-637X/708/1/505},
	\href {https://ui.adsabs.harvard.edu/abs/2010ApJ...708..505K} {708, 505}
	
	\bibitem[\protect\citeauthoryear{{Kova{\v{c}}} et~al.,}{{Kova{\v{c}}}
		et~al.}{2014}]{Kovac2014}
	{Kova{\v{c}}} K.,  et~al., 2014, \mn@doi [\mnras] {10.1093/mnras/stt2241},
	\href {https://ui.adsabs.harvard.edu/abs/2014MNRAS.438..717K} {438, 717}
	
	\bibitem[\protect\citeauthoryear{{Kriek}, {van Dokkum}, {Labb{\'e}}, {Franx},
		{Illingworth}, {Marchesini}  \& {Quadri}}{{Kriek} et~al.}{2009}]{FAST}
	{Kriek} M.,  {van Dokkum} P.~G.,  {Labb{\'e}} I.,  {Franx} M.,  {Illingworth}
	G.~D.,  {Marchesini} D.,   {Quadri} R.~F.,  2009, \mn@doi [\apj]
	{10.1088/0004-637X/700/1/221}, \href
	{https://ui.adsabs.harvard.edu/abs/2009ApJ...700..221K} {700, 221}
	
	\bibitem[\protect\citeauthoryear{{Kroupa}}{{Kroupa}}{2001}]{KroupaIMF}
	{Kroupa} P.,  2001, \mn@doi [\mnras] {10.1046/j.1365-8711.2001.04022.x}, \href
	{https://ui.adsabs.harvard.edu/abs/2001MNRAS.322..231K} {322, 231}
	
	\bibitem[\protect\citeauthoryear{{Leauthaud} et~al.,}{{Leauthaud}
		et~al.}{2010}]{Leauthaud2010}
	{Leauthaud} A.,  et~al., 2010, \mn@doi [\apj] {10.1088/0004-637X/709/1/97},
	\href {https://ui.adsabs.harvard.edu/abs/2010ApJ...709...97L} {709, 97}
	
	\bibitem[\protect\citeauthoryear{Leauthaud et~al.,}{Leauthaud
		et~al.}{2012}]{leauthaud2012integrated}
	Leauthaud A.,  et~al., 2012, The Astrophysical Journal, 746, 95
	
	\bibitem[\protect\citeauthoryear{{Lee-Brown} et~al.,}{{Lee-Brown}
		et~al.}{2017}]{LeeBrown2017}
	{Lee-Brown} D.~B.,  et~al., 2017, \mn@doi [\apj] {10.3847/1538-4357/aa7948},
	\href {https://ui.adsabs.harvard.edu/abs/2017ApJ...844...43L} {844, 43}
	
	\bibitem[\protect\citeauthoryear{{Leja}, {Johnson}, {Conroy}, {van Dokkum}  \&
		{Byler}}{{Leja} et~al.}{2017}]{2017ApJ...837..170L}
	{Leja} J.,  {Johnson} B.~D.,  {Conroy} C.,  {van Dokkum} P.~G.,   {Byler} N.,
	2017, \mn@doi [\apj] {10.3847/1538-4357/aa5ffe}, \href
	{http://adsabs.harvard.edu/abs/2017ApJ...837..170L} {837, 170}
	
	\bibitem[\protect\citeauthoryear{{Lemaux} et~al.,}{{Lemaux}
		et~al.}{2019}]{Lemaux2019ORELSE}
	{Lemaux} B.~C.,  et~al., 2019, \mn@doi [\mnras] {10.1093/mnras/stz2661}, \href
	{https://ui.adsabs.harvard.edu/abs/2019MNRAS.490.1231L} {490, 1231}
	
	\bibitem[\protect\citeauthoryear{{Madau} \& {Dickinson}}{{Madau} \&
		{Dickinson}}{2014}]{Madaureview}
	{Madau} P.,  {Dickinson} M.,  2014, \mn@doi [\araa]
	{10.1146/annurev-astro-081811-125615}, \href
	{https://ui.adsabs.harvard.edu/abs/2014ARA&A..52..415M} {52, 415}
	
	\bibitem[\protect\citeauthoryear{{McCarthy}, {Schaye}, {Bird}  \& {Le
			Brun}}{{McCarthy} et~al.}{2017}]{mccarthy2016bahamas}
	{McCarthy} I.~G.,  {Schaye} J.,  {Bird} S.,   {Le Brun} A. M.~C.,  2017,
	\mn@doi [\mnras] {10.1093/mnras/stw2792}, \href
	{https://ui.adsabs.harvard.edu/abs/2017MNRAS.465.2936M} {465, 2936}
	
	\bibitem[\protect\citeauthoryear{{McCarthy}, {Bird}, {Schaye},
		{Harnois-Deraps}, {Font}  \& {van Waerbeke}}{{McCarthy}
		et~al.}{2018}]{2018MNRAS.476.2999M}
	{McCarthy} I.~G.,  {Bird} S.,  {Schaye} J.,  {Harnois-Deraps} J.,  {Font}
	A.~S.,   {van Waerbeke} L.,  2018, \mn@doi [\mnras] {10.1093/mnras/sty377},
	\href {https://ui.adsabs.harvard.edu/abs/2018MNRAS.476.2999M} {476, 2999}
	
	\bibitem[\protect\citeauthoryear{McCracken et~al.,}{McCracken
		et~al.}{2012}]{mccracken2012ultravista}
	McCracken H.,  et~al., 2012, Astronomy \& Astrophysics, 544, A156
	
	\bibitem[\protect\citeauthoryear{{McGee}, {Balogh}, {Bower}, {Font}  \&
		{McCarthy}}{{McGee} et~al.}{2009}]{McGee+2009}
	{McGee} S.~L.,  {Balogh} M.~L.,  {Bower} R.~G.,  {Font} A.~S.,   {McCarthy}
	I.~G.,  2009, \mn@doi [\mnras] {10.1111/j.1365-2966.2009.15507.x}, \href
	{https://ui.adsabs.harvard.edu/abs/2009MNRAS.400..937M} {400, 937}
	
	\bibitem[\protect\citeauthoryear{{McGee}, {Balogh}, {Wilman}, {Bower},
		{Mulchaey}, {Parker}  \& {Oemler}}{{McGee} et~al.}{2011}]{McGee2011DawnOfRed}
	{McGee} S.~L.,  {Balogh} M.~L.,  {Wilman} D.~J.,  {Bower} R.~G.,  {Mulchaey}
	J.~S.,  {Parker} L.~C.,   {Oemler} A.,  2011, \mn@doi [\mnras]
	{10.1111/j.1365-2966.2010.18189.x}, \href
	{https://ui.adsabs.harvard.edu/abs/2011MNRAS.413..996M} {413, 996}
	
	\bibitem[\protect\citeauthoryear{{McGee}, {Bower}  \& {Balogh}}{{McGee}
		et~al.}{2014}]{McGee2014MNRAS}
	{McGee} S.~L.,  {Bower} R.~G.,   {Balogh} M.~L.,  2014, \mn@doi [\mnras]
	{10.1093/mnrasl/slu066}, \href
	{https://ui.adsabs.harvard.edu/abs/2014MNRAS.442L.105M} {442, L105}
	
	\bibitem[\protect\citeauthoryear{{McLeod}, {McLure}, {Dunlop}, {Cullen},
		{Carnall}  \& {Duncan}}{{McLeod} et~al.}{2021}]{2021MNRAS.tmp..735M}
	{McLeod} D.~J.,  {McLure} R.~J.,  {Dunlop} J.~S.,  {Cullen} F.,  {Carnall}
	A.~C.,   {Duncan} K.,  2021, \mn@doi [\mnras] {10.1093/mnras/stab731}, \href
	{https://ui.adsabs.harvard.edu/abs/2021MNRAS.tmp..735M} {}
	
	\bibitem[\protect\citeauthoryear{{McLure} et~al.,}{{McLure}
		et~al.}{2013}]{mclure2012sizes}
	{McLure} R.~J.,  et~al., 2013, \mn@doi [\mnras] {10.1093/mnras/sts092}, \href
	{https://ui.adsabs.harvard.edu/abs/2013MNRAS.428.1088M} {428, 1088}
	
	\bibitem[\protect\citeauthoryear{Mehta et~al.,}{Mehta
		et~al.}{2018}]{mehta2018splash}
	Mehta V.,  et~al., 2018, The Astrophysical Journal Supplement Series, 235, 36
	
	\bibitem[\protect\citeauthoryear{{Melnyk} et~al.,}{{Melnyk}
		et~al.}{2013}]{2013A&A...557A..81M}
	{Melnyk} O.,  et~al., 2013, \mn@doi [\aap] {10.1051/0004-6361/201220624}, \href
	{https://ui.adsabs.harvard.edu/abs/2013A%26A...557A..81M} {557, A81}
		
		\bibitem[\protect\citeauthoryear{{Mok} et~al.,}{{Mok} et~al.}{2013}]{Mok2013}
		{Mok} A.,  et~al., 2013, \mn@doi [\mnras] {10.1093/mnras/stt251}, \href
		{https://ui.adsabs.harvard.edu/abs/2013MNRAS.431.1090M} {431, 1090}
		
		\bibitem[\protect\citeauthoryear{Moster, Somerville, Newman  \& Rix}{Moster
			et~al.}{2011}]{moster2011cosmic}
		Moster B.~P.,  Somerville R.~S.,  Newman J.~A.,   Rix H.-W.,  2011, The
		Astrophysical Journal, 731, 113
		
		\bibitem[\protect\citeauthoryear{{Mu{\~n}oz-Cuartas}, {Macci{\`o}},
			{Gottl{\"o}ber}  \& {Dutton}}{{Mu{\~n}oz-Cuartas}
			et~al.}{2011}]{munozcuartas2011}
		{Mu{\~n}oz-Cuartas} J.~C.,  {Macci{\`o}} A.~V.,  {Gottl{\"o}ber} S.,   {Dutton}
		A.~A.,  2011, \mn@doi [\mnras] {10.1111/j.1365-2966.2010.17704.x}, \href
		{https://ui.adsabs.harvard.edu/abs/2011MNRAS.411..584M} {411, 584}
		
		\bibitem[\protect\citeauthoryear{{Muzzin} et~al.,}{{Muzzin}
			et~al.}{2012}]{Muzzin2012}
		{Muzzin} A.,  et~al., 2012, \mn@doi [\apj] {10.1088/0004-637X/746/2/188}, \href
		{https://ui.adsabs.harvard.edu/abs/2012ApJ...746..188M} {746, 188}
		
		\bibitem[\protect\citeauthoryear{Muzzin et~al.,}{Muzzin
			et~al.}{2013a}]{muzzin2013public}
		Muzzin A.,  et~al., 2013a, The Astrophysical Journal Supplement Series, 206, 8
		
		\bibitem[\protect\citeauthoryear{Muzzin et~al.,}{Muzzin
			et~al.}{2013b}]{muzzin2013evolution}
		Muzzin A.,  et~al., 2013b, The Astrophysical Journal, 777, 18
		
		\bibitem[\protect\citeauthoryear{{Muzzin} et~al.,}{{Muzzin}
			et~al.}{2014}]{Muzzin2014}
		{Muzzin} A.,  et~al., 2014, \mn@doi [\apj] {10.1088/0004-637X/796/1/65}, \href
		{https://ui.adsabs.harvard.edu/abs/2014ApJ...796...65M} {796, 65}
		
		\bibitem[\protect\citeauthoryear{Nantais et~al.,}{Nantais
			et~al.}{2016}]{nantais2016stellar}
		Nantais J.~B.,  et~al., 2016, Astronomy \& Astrophysics, 592, A161
		
		\bibitem[\protect\citeauthoryear{{Nantais} et~al.,}{{Nantais}
			et~al.}{2017}]{Nantais2017}
		{Nantais} J.~B.,  et~al., 2017, \mn@doi [\mnras] {10.1093/mnrasl/slw224}, \href
		{https://ui.adsabs.harvard.edu/abs/2017MNRAS.465L.104N} {465, L104}
		
		\bibitem[\protect\citeauthoryear{Navarro, Frenk  \& White}{Navarro
			et~al.}{1997}]{navarro1997universal}
		Navarro J.~F.,  Frenk C.~S.,   White S.~D.,  1997, The Astrophysical Journal,
		490, 493
		
		\bibitem[\protect\citeauthoryear{{Oman} \& {Hudson}}{{Oman} \&
			{Hudson}}{2016}]{Oman2016}
		{Oman} K.~A.,  {Hudson} M.~J.,  2016, \mn@doi [\mnras] {10.1093/mnras/stw2195},
		\href {https://ui.adsabs.harvard.edu/abs/2016MNRAS.463.3083O} {463, 3083}
		
		\bibitem[\protect\citeauthoryear{{Oman}, {Hudson}  \& {Behroozi}}{{Oman}
			et~al.}{2013}]{Oman2013}
		{Oman} K.~A.,  {Hudson} M.~J.,   {Behroozi} P.~S.,  2013, \mn@doi [\mnras]
		{10.1093/mnras/stt328}, \href
		{https://ui.adsabs.harvard.edu/abs/2013MNRAS.431.2307O} {431, 2307}
		
		\bibitem[\protect\citeauthoryear{{Omand}, {Balogh}  \& {Poggianti}}{{Omand}
			et~al.}{2014}]{Omand2014}
		{Omand} C. M.~B.,  {Balogh} M.~L.,   {Poggianti} B.~M.,  2014, \mn@doi [\mnras]
		{10.1093/mnras/stu331}, \href
		{https://ui.adsabs.harvard.edu/abs/2014MNRAS.440..843O} {440, 843}
		
		\bibitem[\protect\citeauthoryear{{Paccagnella} et~al.,}{{Paccagnella}
			et~al.}{2016}]{2016ApJ...816L..25P}
		{Paccagnella} A.,  et~al., 2016, \mn@doi [\apjl] {10.3847/2041-8205/816/2/L25},
		\href {https://ui.adsabs.harvard.edu/abs/2016ApJ...816L..25P} {816, L25}
		
		\bibitem[\protect\citeauthoryear{{Pallero}, {G{\'o}mez}, {Padilla},
			{Torres-Flores}, {Demarco}, {Cerulo}  \& {Olave-Rojas}}{{Pallero}
			et~al.}{2019}]{Pallero2019}
		{Pallero} D.,  {G{\'o}mez} F.~A.,  {Padilla} N.~D.,  {Torres-Flores} S.,
		{Demarco} R.,  {Cerulo} P.,   {Olave-Rojas} D.,  2019, \mn@doi [\mnras]
		{10.1093/mnras/stz1745}, \href
		{https://ui.adsabs.harvard.edu/abs/2019MNRAS.488..847P} {488, 847}
		
		\bibitem[\protect\citeauthoryear{{Papovich} et~al.,}{{Papovich}
			et~al.}{2018}]{Papovich2018}
		{Papovich} C.,  et~al., 2018, \mn@doi [\apj] {10.3847/1538-4357/aaa766}, \href
		{https://ui.adsabs.harvard.edu/abs/2018ApJ...854...30P} {854, 30}
		
		\bibitem[\protect\citeauthoryear{{Peng} et~al.,}{{Peng}
			et~al.}{2010}]{Peng2010}
		{Peng} Y.-j.,  et~al., 2010, \mn@doi [\apj] {10.1088/0004-637X/721/1/193},
		\href {https://ui.adsabs.harvard.edu/abs/2010ApJ...721..193P} {721, 193}
		
		\bibitem[\protect\citeauthoryear{{Pentericci}, {McLure}, {Franzetti}, {Garilli}
			\& {the VANDELS team}}{{Pentericci} et~al.}{2018}]{2018arXiv181105298P}
		{Pentericci} L.,  {McLure} R.~J.,  {Franzetti} P.,  {Garilli} B.,   {the
			VANDELS team} 2018, arXiv e-prints, \href
		{https://ui.adsabs.harvard.edu/abs/2018arXiv181105298P} {p. arXiv:1811.05298}
		
		\bibitem[\protect\citeauthoryear{P\'erez \& Granger}{P\'erez \&
			Granger}{2007}]{PER-GRA:2007}
		P\'erez F.,  Granger B.~E.,  2007, \mn@doi [Computing in Science and
		Engineering] {10.1109/MCSE.2007.53}, 9, 21
		
		\bibitem[\protect\citeauthoryear{{Pintos-Castro}, {Yee}, {Muzzin}, {Old}  \&
			{Wilson}}{{Pintos-Castro} et~al.}{2019}]{PintosCastro2019}
		{Pintos-Castro} I.,  {Yee} H.~K.~C.,  {Muzzin} A.,  {Old} L.,   {Wilson} G.,
		2019, \mn@doi [\apj] {10.3847/1538-4357/ab14ee}, \href
		{https://ui.adsabs.harvard.edu/abs/2019ApJ...876...40P} {876, 40}
		
		\bibitem[\protect\citeauthoryear{{Planck Collaboration} et~al.,}{{Planck
				Collaboration} et~al.}{2016}]{Planck2016SZclusters}
		{Planck Collaboration} et~al., 2016, \mn@doi [\aap]
		{10.1051/0004-6361/201525833}, \href
		{https://ui.adsabs.harvard.edu/abs/2016A&A...594A..24P} {594, A24}
		
		\bibitem[\protect\citeauthoryear{{Poggianti} et~al.,}{{Poggianti}
			et~al.}{2006}]{Poggianti2006}
		{Poggianti} B.~M.,  et~al., 2006, \mn@doi [\apj] {10.1086/500666}, \href
		{https://ui.adsabs.harvard.edu/abs/2006ApJ...642..188P} {642, 188}
		
		\bibitem[\protect\citeauthoryear{Price-Whelan et~al.,}{Price-Whelan
			et~al.}{2018}]{price2018astropy}
		Price-Whelan A.,  et~al., 2018, The Astronomical Journal, 156, 123
		
		\bibitem[\protect\citeauthoryear{{Qu} et~al.,}{{Qu} et~al.}{2017}]{Qu2017}
		{Qu} Y.,  et~al., 2017, \mn@doi [\mnras] {10.1093/mnras/stw2437}, \href
		{https://ui.adsabs.harvard.edu/abs/2017MNRAS.464.1659Q} {464, 1659}
		
		\bibitem[\protect\citeauthoryear{{Quilis}, {Planelles}  \&
			{Ricciardelli}}{{Quilis} et~al.}{2017}]{2017MNRAS.469...80Q}
		{Quilis} V.,  {Planelles} S.,   {Ricciardelli} E.,  2017, \mn@doi [\mnras]
		{10.1093/mnras/stx770}, \href
		{https://ui.adsabs.harvard.edu/abs/2017MNRAS.469...80Q} {469, 80}
		
		\bibitem[\protect\citeauthoryear{Robitaille et~al.,}{Robitaille
			et~al.}{2013}]{robitaille2013astropy}
		Robitaille T.~P.,  et~al., 2013, Astronomy \& Astrophysics, 558, A33
		
		\bibitem[\protect\citeauthoryear{{Saro}, {Mohr}, {Bazin}  \& {Dolag}}{{Saro}
			et~al.}{2013}]{Saro2013}
		{Saro} A.,  {Mohr} J.~J.,  {Bazin} G.,   {Dolag} K.,  2013, \mn@doi [\apj]
		{10.1088/0004-637X/772/1/47}, \href
		{https://ui.adsabs.harvard.edu/abs/2013ApJ...772...47S} {772, 47}
		
		\bibitem[\protect\citeauthoryear{{Schaye} et~al.,}{{Schaye}
			et~al.}{2010a}]{Schaye2010}
		{Schaye} J.,  et~al., 2010a, \mn@doi [\mnras]
		{10.1111/j.1365-2966.2009.16029.x}, \href
		{https://ui.adsabs.harvard.edu/abs/2010MNRAS.402.1536S} {402, 1536}
		
		\bibitem[\protect\citeauthoryear{{Schaye} et~al.,}{{Schaye}
			et~al.}{2010b}]{2010MNRAS.402.1536S}
		{Schaye} J.,  et~al., 2010b, \mn@doi [\mnras]
		{10.1111/j.1365-2966.2009.16029.x}, \href
		{https://ui.adsabs.harvard.edu/abs/2010MNRAS.402.1536S} {402, 1536}
		
		\bibitem[\protect\citeauthoryear{Schechter}{Schechter}{1976}]{schechter1976analytic}
		Schechter P.,  1976, The Astrophysical Journal, 203, 297
		
		\bibitem[\protect\citeauthoryear{{Schreiber} et~al.,}{{Schreiber}
			et~al.}{2015}]{Schreiber+2015}
		{Schreiber} C.,  et~al., 2015, \mn@doi [\aap] {10.1051/0004-6361/201425017},
		\href {https://ui.adsabs.harvard.edu/abs/2015A&A...575A..74S} {575, A74}
		
		\bibitem[\protect\citeauthoryear{{Simet}, {McClintock}, {Mandelbaum}, {Rozo},
			{Rykoff}, {Sheldon}  \& {Wechsler}}{{Simet} et~al.}{2017}]{Simet2017}
		{Simet} M.,  {McClintock} T.,  {Mandelbaum} R.,  {Rozo} E.,  {Rykoff} E.,
		{Sheldon} E.,   {Wechsler} R.~H.,  2017, \mn@doi [\mnras]
		{10.1093/mnras/stw3250}, \href
		{https://ui.adsabs.harvard.edu/abs/2017MNRAS.466.3103S} {466, 3103}
		
		\bibitem[\protect\citeauthoryear{{Skelton} et~al.,}{{Skelton}
			et~al.}{2014}]{2014ApJS..214...24S}
		{Skelton} R.~E.,  et~al., 2014, \mn@doi [\apjs] {10.1088/0067-0049/214/2/24},
		\href {https://ui.adsabs.harvard.edu/abs/2014ApJS..214...24S} {214, 24}
		
		\bibitem[\protect\citeauthoryear{{Smith} et~al.,}{{Smith}
			et~al.}{2016}]{Smith2016}
		{Smith} G.~P.,  et~al., 2016, \mn@doi [\mnras] {10.1093/mnrasl/slv175}, \href
		{https://ui.adsabs.harvard.edu/abs/2016MNRAS.456L..74S} {456, L74}
		
		\bibitem[\protect\citeauthoryear{{Sobral}, {Best}, {Smail}, {Geach},
			{Cirasuolo}, {Garn}  \& {Dalton}}{{Sobral} et~al.}{2011}]{Sobral2011}
		{Sobral} D.,  {Best} P.~N.,  {Smail} I.,  {Geach} J.~E.,  {Cirasuolo} M.,
		{Garn} T.,   {Dalton} G.~B.,  2011, \mn@doi [\mnras]
		{10.1111/j.1365-2966.2010.17707.x}, \href
		{https://ui.adsabs.harvard.edu/abs/2011MNRAS.411..675S} {411, 675}
		
		\bibitem[\protect\citeauthoryear{{Strateva} et~al.,}{{Strateva}
			et~al.}{2001}]{Strateva2001}
		{Strateva} I.,  et~al., 2001, \mn@doi [\aj] {10.1086/323301}, \href
		{https://ui.adsabs.harvard.edu/abs/2001AJ....122.1861S} {122, 1861}
		
		\bibitem[\protect\citeauthoryear{{Strazzullo} et~al.,}{{Strazzullo}
			et~al.}{2019}]{Strazzullo2019highzclustersQFE}
		{Strazzullo} V.,  et~al., 2019, \mn@doi [\aap] {10.1051/0004-6361/201833944},
		\href {https://ui.adsabs.harvard.edu/abs/2019A&A...622A.117S} {622, A117}
		
		\bibitem[\protect\citeauthoryear{{Taylor} et~al.,}{{Taylor}
			et~al.}{2015}]{Taylor2015}
		{Taylor} E.~N.,  et~al., 2015, \mn@doi [\mnras] {10.1093/mnras/stu1900}, \href
		{https://ui.adsabs.harvard.edu/abs/2015MNRAS.446.2144T} {446, 2144}
		
		\bibitem[\protect\citeauthoryear{{Tinker} \& {Wetzel}}{{Tinker} \&
			{Wetzel}}{2010}]{2010ApJ...719...88T}
		{Tinker} J.~L.,  {Wetzel} A.~R.,  2010, \mn@doi [\apj]
		{10.1088/0004-637X/719/1/88}, \href
		{https://ui.adsabs.harvard.edu/abs/2010ApJ...719...88T} {719, 88}
		
		\bibitem[\protect\citeauthoryear{{Tinker}, {Leauthaud}, {Bundy}, {George},
			{Behroozi}, {Massey}, {Rhodes}  \& {Wechsler}}{{Tinker}
			et~al.}{2013}]{2013ApJ...778...93T}
		{Tinker} J.~L.,  {Leauthaud} A.,  {Bundy} K.,  {George} M.~R.,  {Behroozi} P.,
		{Massey} R.,  {Rhodes} J.,   {Wechsler} R.~H.,  2013, \mn@doi [\apj]
		{10.1088/0004-637X/778/2/93}, \href
		{https://ui.adsabs.harvard.edu/abs/2013ApJ...778...93T} {778, 93}
		
		\bibitem[\protect\citeauthoryear{Trudeau et~al.,}{Trudeau
			et~al.}{2020}]{Trudeau_2020}
		Trudeau A.,  et~al., 2020, \mn@doi [Astronomy & Astrophysics]
		{10.1051/0004-6361/202038982}, 642, A124
		
		\bibitem[\protect\citeauthoryear{{Valentino} et~al.,}{{Valentino}
			et~al.}{2020}]{2020ApJ...889...93V}
		{Valentino} F.,  et~al., 2020, \mn@doi [\apj] {10.3847/1538-4357/ab64dc}, \href
		{https://ui.adsabs.harvard.edu/abs/2020ApJ...889...93V} {889, 93}
		
		\bibitem[\protect\citeauthoryear{Van~Rossum \& Drake~Jr}{Van~Rossum \&
			Drake~Jr}{1995}]{van1995python}
		Van~Rossum G.,  Drake~Jr F.~L.,  1995, Python tutorial.
		Centrum voor Wiskunde en Informatica Amsterdam, The Netherlands
		
		\bibitem[\protect\citeauthoryear{Virtanen et~al.,}{Virtanen
			et~al.}{2020}]{2020SciPy-NMeth}
		Virtanen P.,  et~al., 2020, \mn@doi [Nature Methods]
		{10.1038/s41592-019-0686-2}, \href {https://rdcu.be/b08Wh} {17, 261}
		
		\bibitem[\protect\citeauthoryear{{Webb} et~al.,}{{Webb}
			et~al.}{2020}]{WebbGOGREEN2020}
		{Webb} K.,  et~al., 2020, arXiv e-prints, \href
		{https://ui.adsabs.harvard.edu/abs/2020arXiv200903953W} {p. arXiv:2009.03953}
		
		\bibitem[\protect\citeauthoryear{{Weinmann}, {van den Bosch}, {Yang}  \&
			{Mo}}{{Weinmann} et~al.}{2006}]{Weinmann2006}
		{Weinmann} S.~M.,  {van den Bosch} F.~C.,  {Yang} X.,   {Mo} H.~J.,  2006,
		\mn@doi [\mnras] {10.1111/j.1365-2966.2005.09865.x}, \href
		{https://ui.adsabs.harvard.edu/abs/2006MNRAS.366....2W} {366, 2}
		
		\bibitem[\protect\citeauthoryear{{Weinmann}, {Kauffmann}, {von der Linden}  \&
			{De Lucia}}{{Weinmann} et~al.}{2010}]{2010MNRAS.406.2249W}
		{Weinmann} S.~M.,  {Kauffmann} G.,  {von der Linden} A.,   {De Lucia} G.,
		2010, \mn@doi [\mnras] {10.1111/j.1365-2966.2010.16855.x}, \href
		{https://ui.adsabs.harvard.edu/abs/2010MNRAS.406.2249W} {406, 2249}
		
		\bibitem[\protect\citeauthoryear{{W}es {M}c{K}inney}{{W}es
			{M}c{K}inney}{2010}]{mckinney-proc-scipy-2010}
		{W}es {M}c{K}inney 2010, in {S}t\'efan van~der {W}alt {J}arrod {M}illman eds,
		{P}roceedings of the 9th {P}ython in {S}cience {C}onference. pp 56 -- 61,
		\mn@doi{10.25080/Majora-92bf1922-00a}
		
		\bibitem[\protect\citeauthoryear{{Wetzel}, {Tinker}  \& {Conroy}}{{Wetzel}
			et~al.}{2012}]{Wetzel2012}
		{Wetzel} A.~R.,  {Tinker} J.~L.,   {Conroy} C.,  2012, \mn@doi [\mnras]
		{10.1111/j.1365-2966.2012.21188.x}, \href
		{https://ui.adsabs.harvard.edu/abs/2012MNRAS.424..232W} {424, 232}
		
		\bibitem[\protect\citeauthoryear{{Wetzel}, {Tinker}, {Conroy}  \& {van den
				Bosch}}{{Wetzel} et~al.}{2013}]{Wetzel2013MNRAS.432..336W}
		{Wetzel} A.~R.,  {Tinker} J.~L.,  {Conroy} C.,   {van den Bosch} F.~C.,  2013,
		\mn@doi [\mnras] {10.1093/mnras/stt469}, \href
		{https://ui.adsabs.harvard.edu/abs/2013MNRAS.432..336W} {432, 336}
		
		\bibitem[\protect\citeauthoryear{{Wetzel}, {Tollerud}  \& {Weisz}}{{Wetzel}
			et~al.}{2015}]{Wetzel2015localSatDwarfQuenching}
		{Wetzel} A.~R.,  {Tollerud} E.~J.,   {Weisz} D.~R.,  2015, \mn@doi [\apjl]
		{10.1088/2041-8205/808/1/L27}, \href
		{https://ui.adsabs.harvard.edu/abs/2015ApJ...808L..27W} {808, L27}
		
		\bibitem[\protect\citeauthoryear{{Wheeler}, {Phillips}, {Cooper},
			{Boylan-Kolchin}  \& {Bullock}}{{Wheeler} et~al.}{2014}]{Wheeler2014}
		{Wheeler} C.,  {Phillips} J.~I.,  {Cooper} M.~C.,  {Boylan-Kolchin} M.,
		{Bullock} J.~S.,  2014, \mn@doi [\mnras] {10.1093/mnras/stu965}, \href
		{https://ui.adsabs.harvard.edu/abs/2014MNRAS.442.1396W} {442, 1396}
		
		\bibitem[\protect\citeauthoryear{{Whitaker} et~al.,}{{Whitaker}
			et~al.}{2011}]{Whitaker2011}
		{Whitaker} K.~E.,  et~al., 2011, \mn@doi [\apj] {10.1088/0004-637X/735/2/86},
		\href {https://ui.adsabs.harvard.edu/abs/2011ApJ...735...86W} {735, 86}
		
		\bibitem[\protect\citeauthoryear{{White} \& {Frenk}}{{White} \&
			{Frenk}}{1991}]{WhiteFrenk1991}
		{White} S. D.~M.,  {Frenk} C.~S.,  1991, \mn@doi [\apj] {10.1086/170483}, \href
		{https://ui.adsabs.harvard.edu/abs/1991ApJ...379...52W} {379, 52}
		
		\bibitem[\protect\citeauthoryear{{White} \& {Rees}}{{White} \&
			{Rees}}{1978}]{WhiteRees1978}
		{White} S.~D.~M.,  {Rees} M.~J.,  1978, \mn@doi [\mnras]
		{10.1093/mnras/183.3.341}, \href
		{https://ui.adsabs.harvard.edu/abs/1978MNRAS.183..341W} {183, 341}
		
		\bibitem[\protect\citeauthoryear{{Williams}, {Quadri}, {Franx}, {van Dokkum}
			\& {Labb{\'e}}}{{Williams} et~al.}{2009}]{Williams2009}
		{Williams} R.~J.,  {Quadri} R.~F.,  {Franx} M.,  {van Dokkum} P.,   {Labb{\'e}}
		I.,  2009, \mn@doi [\apj] {10.1088/0004-637X/691/2/1879}, \href
		{https://ui.adsabs.harvard.edu/abs/2009ApJ...691.1879W} {691, 1879}
		
		\bibitem[\protect\citeauthoryear{{Wilman} et~al.,}{{Wilman}
			et~al.}{2005}]{Wilman2005}
		{Wilman} D.~J.,  et~al., 2005, \mn@doi [\mnras]
		{10.1111/j.1365-2966.2005.08745.x}, \href
		{https://ui.adsabs.harvard.edu/abs/2005MNRAS.358...88W} {358, 88}
		
		\bibitem[\protect\citeauthoryear{{Wright}, {Lagos}, {Davies}, {Power},
			{Trayford}  \& {Wong}}{{Wright} et~al.}{2019}]{2019MNRAS.487.3740W}
		{Wright} R.~J.,  {Lagos} C. d.~P.,  {Davies} L. J.~M.,  {Power} C.,  {Trayford}
		J.~W.,   {Wong} O.~I.,  2019, \mn@doi [\mnras] {10.1093/mnras/stz1410}, \href
		{https://ui.adsabs.harvard.edu/abs/2019MNRAS.487.3740W} {487, 3740}
		
		\bibitem[\protect\citeauthoryear{{Yang}, {Mo}, {van den Bosch}, {Zhang}  \&
			{Han}}{{Yang} et~al.}{2012}]{YangDR12groups}
		{Yang} X.,  {Mo} H.~J.,  {van den Bosch} F.~C.,  {Zhang} Y.,   {Han} J.,  2012,
		\mn@doi [\apj] {10.1088/0004-637X/752/1/41}, \href
		{https://ui.adsabs.harvard.edu/abs/2012ApJ...752...41Y} {752, 41}
		
		\bibitem[\protect\citeauthoryear{{Zabludoff} \& {Mulchaey}}{{Zabludoff} \&
			{Mulchaey}}{1998}]{Zabludoff1998}
		{Zabludoff} A.~I.,  {Mulchaey} J.~S.,  1998, \mn@doi [\apjl] {10.1086/311312},
		\href {https://ui.adsabs.harvard.edu/abs/1998ApJ...498L...5Z} {498, L5}
		
		\bibitem[\protect\citeauthoryear{pandas~development team}{pandas~development
			team}{2020}]{reback2020pandas}
		pandas~development team T.,  2020, pandas-dev/pandas: Pandas,
		\mn@doi{10.5281/zenodo.3509134}, \url
		{https://doi.org/10.5281/zenodo.3509134}
		
		\bibitem[\protect\citeauthoryear{{van den Bosch}, {Aquino}, {Yang}, {Mo},
			{Pasquali}, {McIntosh}, {Weinmann}  \& {Kang}}{{van den Bosch}
			et~al.}{2008}]{vandenBosh2008satQuenching}
		{van den Bosch} F.~C.,  {Aquino} D.,  {Yang} X.,  {Mo} H.~J.,  {Pasquali} A.,
		{McIntosh} D.~H.,  {Weinmann} S.~M.,   {Kang} X.,  2008, \mn@doi [\mnras]
		{10.1111/j.1365-2966.2008.13230.x}, \href
		{https://ui.adsabs.harvard.edu/abs/2008MNRAS.387...79V} {387, 79}
		
		\bibitem[\protect\citeauthoryear{{van der Burg} et~al.,}{{van der Burg}
			et~al.}{2013}]{vdB2013}
		{van der Burg} R.~F.~J.,  et~al., 2013, \mn@doi [\aap]
		{10.1051/0004-6361/201321237}, \href
		{https://ui.adsabs.harvard.edu/abs/2013A&A...557A..15V} {557, A15}
		
		\bibitem[\protect\citeauthoryear{{van der Burg}, {McGee}, {Aussel}, {Dahle},
			{Arnaud}, {Pratt}  \& {Muzzin}}{{van der Burg} et~al.}{2018}]{vanderBurg2018}
		{van der Burg} R. F.~J.,  {McGee} S.,  {Aussel} H.,  {Dahle} H.,  {Arnaud} M.,
		{Pratt} G.~W.,   {Muzzin} A.,  2018, \mn@doi [\aap]
		{10.1051/0004-6361/201833572}, \href
		{https://ui.adsabs.harvard.edu/abs/2018A&A...618A.140V} {618, A140}
		
		\bibitem[\protect\citeauthoryear{{van der Burg} et~al.,}{{van der Burg}
			et~al.}{2020}]{vanderBurgGOGREENsmfs}
		{van der Burg} R. F.~J.,  et~al., 2020, \mn@doi [\aap]
		{10.1051/0004-6361/202037754}, \href
		{https://ui.adsabs.harvard.edu/abs/2020A&A...638A.112V} {638, A112}
		
		\makeatother
	\end{thebibliography}


\appendix \label{sec:appendices}

\section{Photometric redshift calibration and selection}\label{sec-photoz_app}

\subsection{De-biasing photometric redshifts using available spectroscopic redshifts} \label{sec: debias-photo-z}

\begin{figure}
	\centering
		\includegraphics[width=\columnwidth]{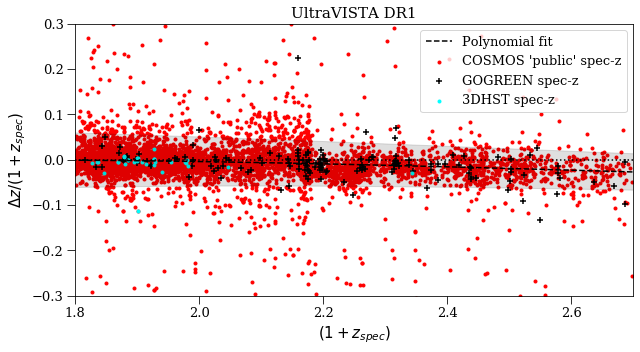}
		\includegraphics[width=\columnwidth]{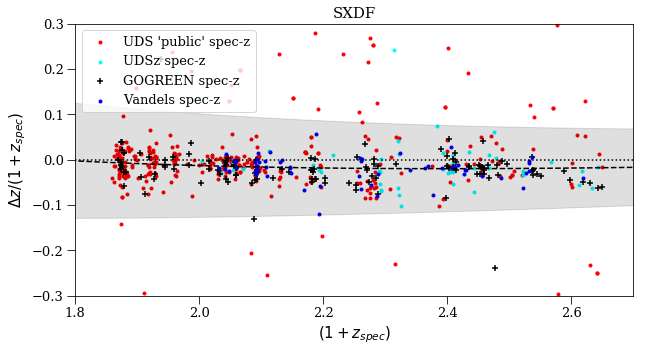}
	\caption{The difference between catalogued photometric and spectroscopic redshifts for UltraVISTA DR1 (top) and SPLASH-SXDF (bottom) is shown as a function of spectroscopic redshifts. The final quadratic fit (solid blue curve) is based on the points within the shaded region, which we determine by an iterative sigma-clipping procedure.}
	\label{fig:photo-z-debiasing}
\end{figure}

In Figure~\ref{fig:photo-z-debiasing} we show  the correlation between $\Delta z / (1+z_{\mathrm{spec}})$ and $1+z_{\mathrm{spec}}$, for all galaxies in COSMOS and SXDF with available spectroscopic redshifts, where $\Delta z=z_{\mathrm{phot}}-z_{\mathrm{spec}}$. These include deep spectroscopy focused on the $1<z<1.5$ regime by GOGREEN \citep{GOGREEN2021data}. For UltraVISTA DR3 (not shown), there is no significant bias, up to $z\sim 2$. Both UltraVISTA DR1 and SXDF show a small bias, such that the photometric redshift is lower than the spectroscopic redshift for $1<z<2$. This difference increases modestly with redshift. 

To correct for this we fit a quadratic function to this correlation. We find a fit of $z_{\mathrm{phot}}= -0.085 z_{\mathrm{spec}}^2 + 1.121 z_{\mathrm{spec}} - 0.044$ for UltraVISTA DR1 and $z_{\mathrm{phot}}= 0.087 z_{\mathrm{spec}}^2 + 0.737 z_{\mathrm{spec}} + 0.150$ for SPLASH-SXDF. We remove the bias by subtracting off the difference of this relation from a linear one-to-one relation between spectroscopic and photometric redshifts. 
Galaxies with photometric redshifts outside 2-4 $\sigma$ (depending on dataset) are considered outliers, and iteratively removed from the fit.
The lower and upper 68\% photometric redshift uncertainties are also corrected to reflect this shift in photometric redshift. We note that this does not entirely eliminate the bias from SPLASH-SXDF, as photometric redshifts at $z\sim1.3$ are still too small, on average, by $\sim 0.04$ after the correction.

For each galaxy, we also apply a small redshift-dependent stellar mass correction, to account for the corresponding change in luminosity distance. This adjustment is by a factor of $\big[ D_L (z_{\text{pert}}) / D_L (z_{\text{true}}) \big]^2$. $D_L(z)$ is the luminosity distance at a given redshift and $z_{\text{pert}}$ and $z_{\text{true}}$ are the redshifts of the galaxy originally and after being perturbed, respectively \footnote{To see how this factor comes about, consider that the original ``$M_{\mathrm{true}}$'' of the galaxy in the UltraVISTA dataset was $M_{\mathrm{true}} = 4\pi D_L^2 (z_{\mathrm{true}}) F [M/L],$ with $F$ being the observed flux of the galaxy and $M/L$ being the mass/luminosity ratio of the galaxy. The recovered stellar mass of that galaxy after having its photometric redshift perturbed will be $M_{\mathrm{rec}} = 4\pi D_L^2(z_{\text{pert}}) F [M/L]$. Taking the ratio of $M_{\mathrm{rec}} / M_{\mathrm{true}}$ then gives the desired result.}.

\subsection{Photometric redshift selection} \label{sec:photo-z-cut-choices}

\begin{figure}
	\centering
		COSMOS: UltraVISTA DR1
		\includegraphics[width=\columnwidth]{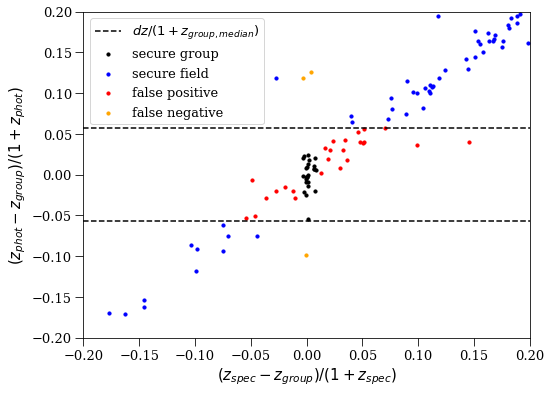}
		COSMOS: UltraVISTA DR3
		\includegraphics[width=\columnwidth]{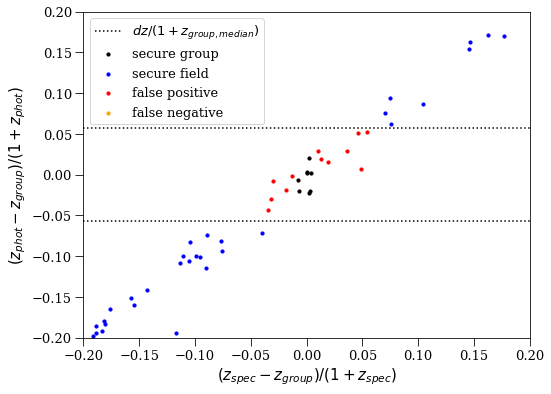}
		SPLASH-SXDF
		\includegraphics[width=\columnwidth]{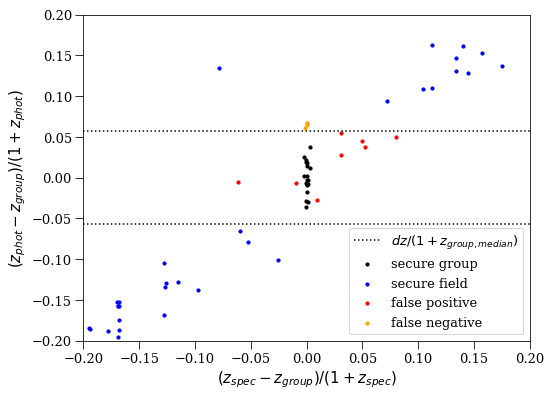}
	\caption{Verification of the photometric redshift cuts for COSMOS (UltraVISTA DR1 and DR3 UltraDeep) and SXDF using the spectroscopic redshifts. The COSMOS: UltraVISTA DR1 subplot shown here contains all 23 $1<z<1.5$ groups, regardless of quality flag, to maximize the spectroscopic and photometric redshift matches. Dashed horizontal lines show the photometric redshift cut that was chosen.}
	\label{fig:Remco-plots}
\end{figure}

As described in \S\ref{sec:SMFs}, group members are first selected to lie within a photometric redshift range of width $\Delta z=0.126$, prior to applying a statistical background subtraction. Plots of available spectroscopic redshifts matched to photometric redshifts (for UltraVISTA DR1 and SPLASH-SXDF) are shown in Figure \ref{fig:Remco-plots}, following \cite{vdB2013}. This Figure demonstrates how the size of this cut compares with the photometric redshift scatter, for galaxies with available spectroscopic redshifts.

The choice of $\Delta z=0.126$ is made as it corresponds to $\Delta z=2\times$median$(z_{u68}-z_{\text{peak}})$ for UltraVISTA DR1 and DR3. Here, $z_{u68}$ is the 84th percentile of the photometric redshift probability distribution and $z_{\text{peak}}$ is the peak of that distribution (i.e. $z_{u68}$ is the upper 1-sigma confidence interval) . Although SPLASH-SXDF has smaller photometric redshift uncertainties, thanks to greater depth in several bands, there is still a remaining bias at $z\sim1.2-1.3$ that is not fully removed from the debiasing described in Appendix~\ref{sec: debias-photo-z}. Therefore we conservatively adopt the same $\Delta z=0.126$ for all systems, to help mitigate this.

\subsection{Field stellar mass functions}

\begin{figure}
	\centering
		\includegraphics[width=\columnwidth]{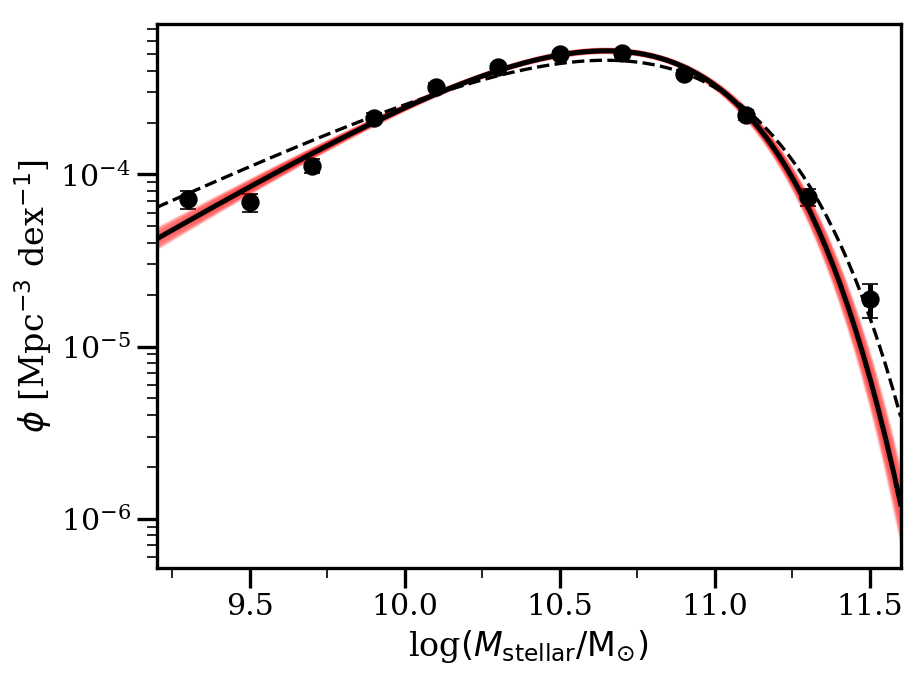}
		\includegraphics[width=\columnwidth]{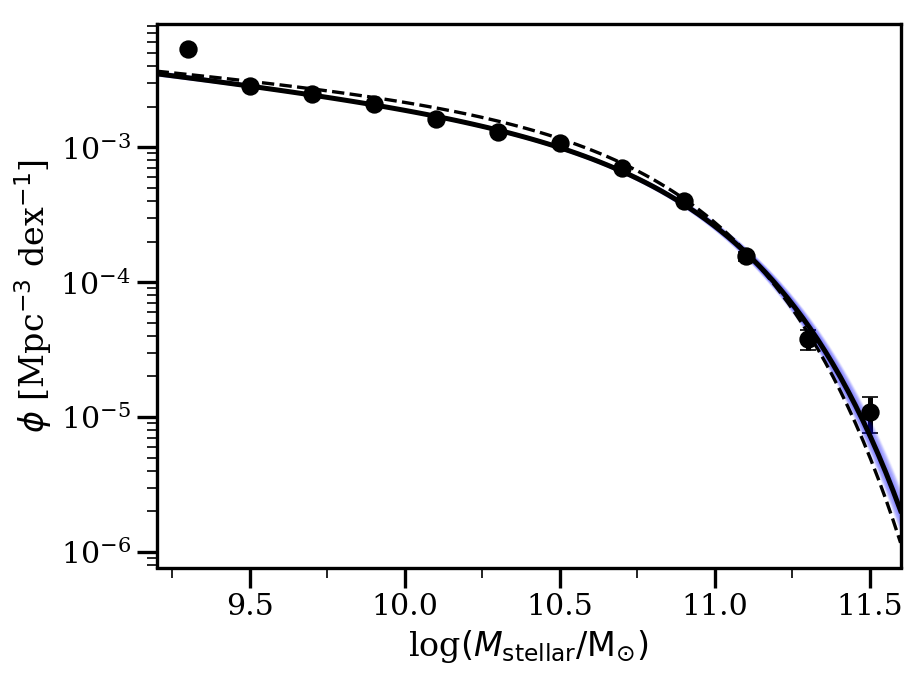}
		\includegraphics[width=\columnwidth]{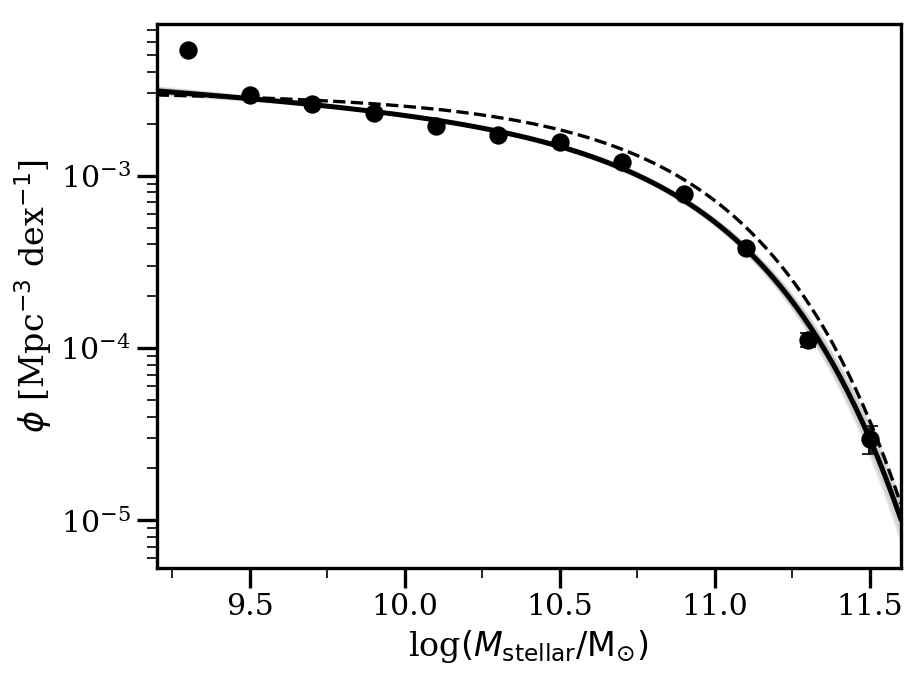}
	\caption{Stellar mass function of quiescent (top panel), star-forming (middle), and total (bottom) field galaxies at $1<z<1.5$. Error bars shown represent the Poisson shot noise. Overlaid on each plot are the Schechter function fits (solid line), normalized to match the number of field galaxies per Mpc$^3$ per dex (bin size $\Delta \log (M_{\mathrm{stellar}}/\mathrm{M}_\odot)=0.2$), and with shaded regions indicating the 68\% confidence interval on the fit parameters, computed as described in the text. The best fit from \citet{muzzin2013evolution} is shown as a dashed line.}
	\label{fig:field-SMFs}
\end{figure}

In Figure \ref{fig:field-SMFs} we show our field stellar mass functions, which were fit and described in \S\ref{sec:SMFs}. We plot Schechter fits using the best fit parameters in Table \ref{tab-fitparams} and contrast our field fits with those from \cite{muzzin2013evolution}, which exclusively measured stellar mass functions using the UltraVISTA survey region. We find similar stellar mass functions and our fits are consistent with theirs, within 2$\sigma$. The similarity in fit is expected, given that much of our survey area is UltraVISTA. We note that the Schechter fit for the quiescent population doesn't quite fit the very high mass end as well as the \cite{muzzin2013evolution} fits. This explains why the quiescent fraction curves, as plotted in Figure \ref{fig:quiescent-fractions-vs-stellar-mass}, turn over rather than flattening out for the two highest stellar mass bins. As well for the quiescent population, there is some deviation from the fit in the lowest stellar mass points. These small discrepancies in fit do not significantly affect any of our results, discussion, or conclusion.

\section{Spectroscopy and GOGREEN spectroscopic groups} \label{sec:group-halo-masses}

Nine of the groups in our $1<z<1.5$ redshift range were observed with Gemini-GMOS by the GOGREEN spectroscopic survey \citep{GOGREEN2021data}: four from COSMOS 
and five from SXDF, 
as listed in Table~\ref{tab:groups-table}.
The names given to the GOGREEN-identified groups are the same as those used in the original COSMOS \citep{gozaliasl2018chandraCOSMOSgroups} and SXDF \citep{finoguenov2010x} group catalogues. For simplicity we will refer to individual groups using a shortened name comprised of the last three digits in the formal name appended to the appropriate catalogue (e.g. COSMOS-28 instead of COSMOS-20028). Here we present additional details of the spectroscopic datasets used (Appendix~\ref{sec:spec-z-data}), definition of our field sample at $1<z<1.5$ (Appendix~\ref{sec:spec-field-sample}), as well as analysis of the dynamics (in particular, confirming the group membership and dynamical masses for a subset of our groups in Appendix~\ref{sec:velocity-dispersions-appendix}), and description of the spectroscopically-determined mass-weighted ages used for a subset of our group sample (Appendix~\ref{sec-MWA_app}).

\subsection{Spectroscopy}\label{sec:spec-z-data}

Our primary source for redshifts is the GOGREEN survey \citep{GOGREEN2021data}.
This sample was selected from galaxies with $z'<24.25$ within a $5.5\arcmin\times5.5\arcmin$ area around nine targets in these two fields. A broad colour selection was applied to reduce foreground and background galaxies. The survey provides redshifts for an average $\sim 45\%$  of the parent cluster population within 500~kpc, unbiased with respect to galaxy type for stellar masses $M_{\rm stellar}> 10^{10.2}~\mathrm{M}_{\odot}$ \citep{GOGREEN2021data}. We use 173 robust redshifts in the COSMOS field, and 198 in the SXDF field, from GOGREEN. For more details we refer to \cite{GOGREEN2021data}.

In addition, for COSMOS we use the master spectroscopic redshift catalogue of publicly available redshifts in use within the COSMOS collaboration, curated by M. Salvato (priv. comm.). We also use available 3D-HST redshifts \citep{2014ApJS..214...24S, brammer20123d}, but only for the purpose of checking and debiasing the photometric redshifts in Appendix~\ref{sec: debias-photo-z}.

For the SXDF field we supplement the GOGREEN redshifts with data from UDSz \citep{bradshaw2013high, mclure2012sizes}, the XMM-Large Scale Structure (XMM-LSS) survey \citep{2013A&A...557A..81M, 2013MNRAS.429.1652C} and VANDELS \citep{2018arXiv181105298P}. 3D-HST data are again used only for debiasing the photometric redshifts.

\subsection{Field sample at 1<\emph{z}<1.5}\label{sec:spec-field-sample}

For comparison with our sample of overdense galaxy systems, we define a reference ``field'' galaxy sample that is representative of the average galaxy population. We simply define our field sample as all galaxies in the UltraVISTA and SPLASH-SXDF catalogues with photometric redshifts in the range of interest. This includes galaxies that make up our group sample, but as they only make up $\sim 1$ per cent of the total, this has a negligible effect on our analysis. Calculated field values are an area-weighted average. Such a sample includes overdense regions, and thus provides a lower contrast to our group sample than would a comparison with low-density regions \citep{Peng2010, kawinwanichakij2017EffectOfLocalEnvAndMass, Papovich2018} or samples of ``central'' galaxies. It will also be influenced by cosmic variance  \citep{Kovac2010zCOSMOS, moster2011cosmic}, though this is small given the relatively large area of the combined surveys.


\subsection{Group membership, dynamics and masses}\label{sec:velocity-dispersions-appendix}\label{sec:halo-mass-checks}

\begin{figure} 
	\centering %
		\includegraphics[width=\columnwidth]{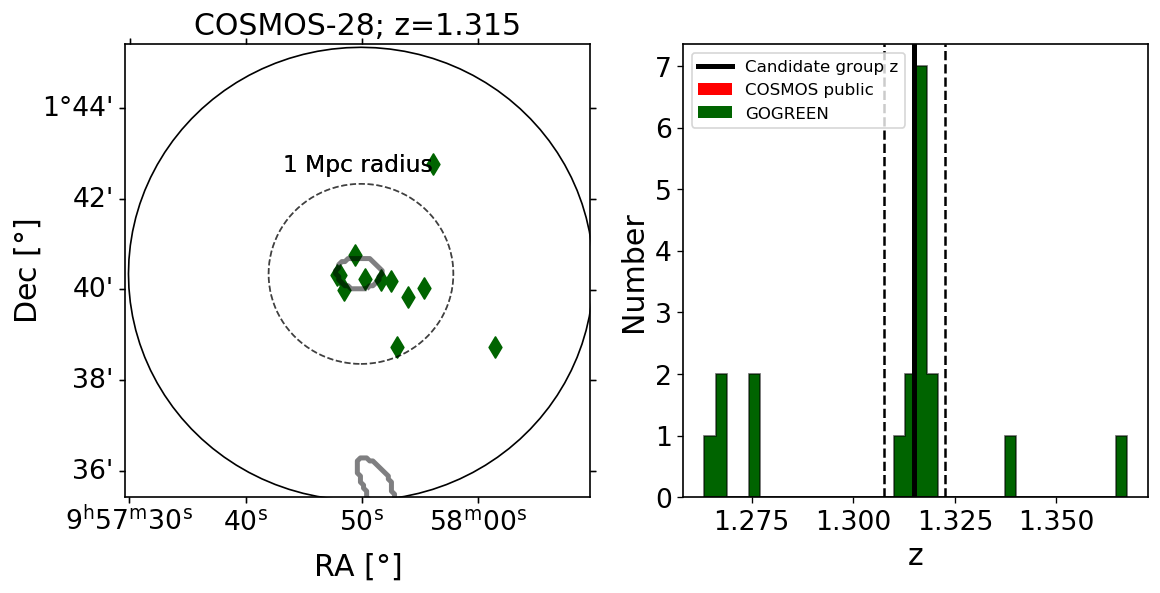}\\
		\includegraphics[width=\columnwidth]{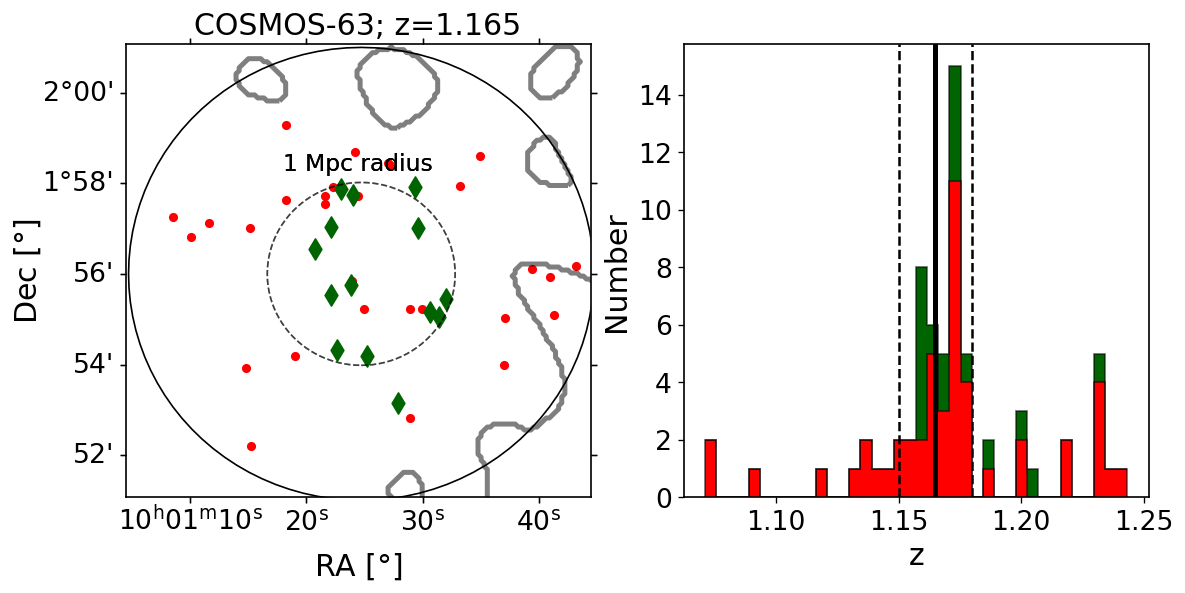}\\
		\includegraphics[width=\columnwidth]{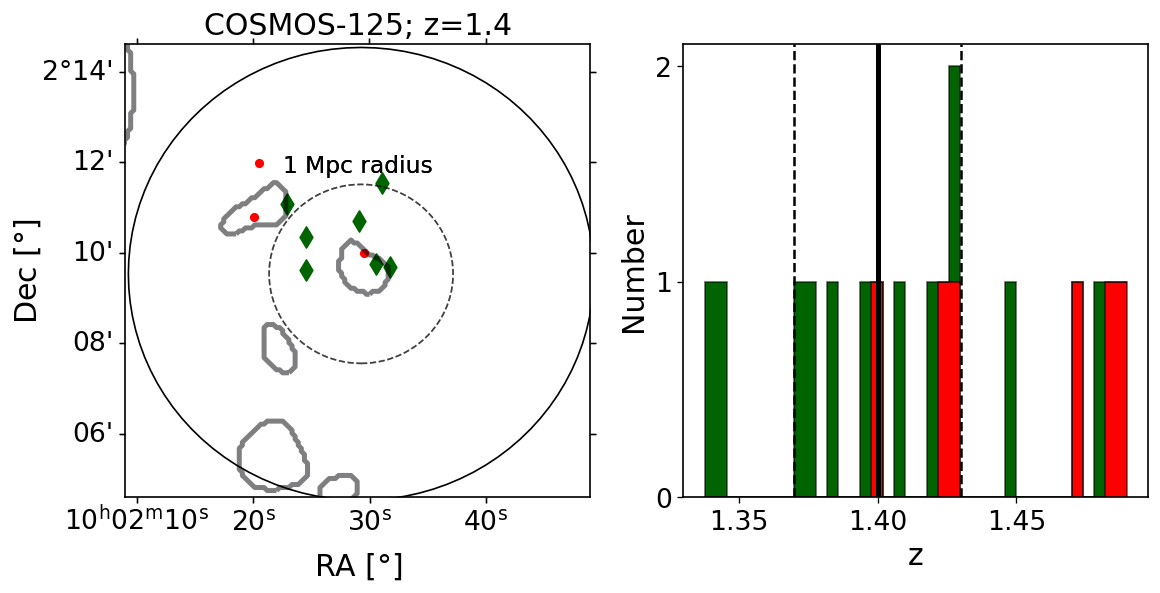}\\
		\includegraphics[width=\columnwidth]{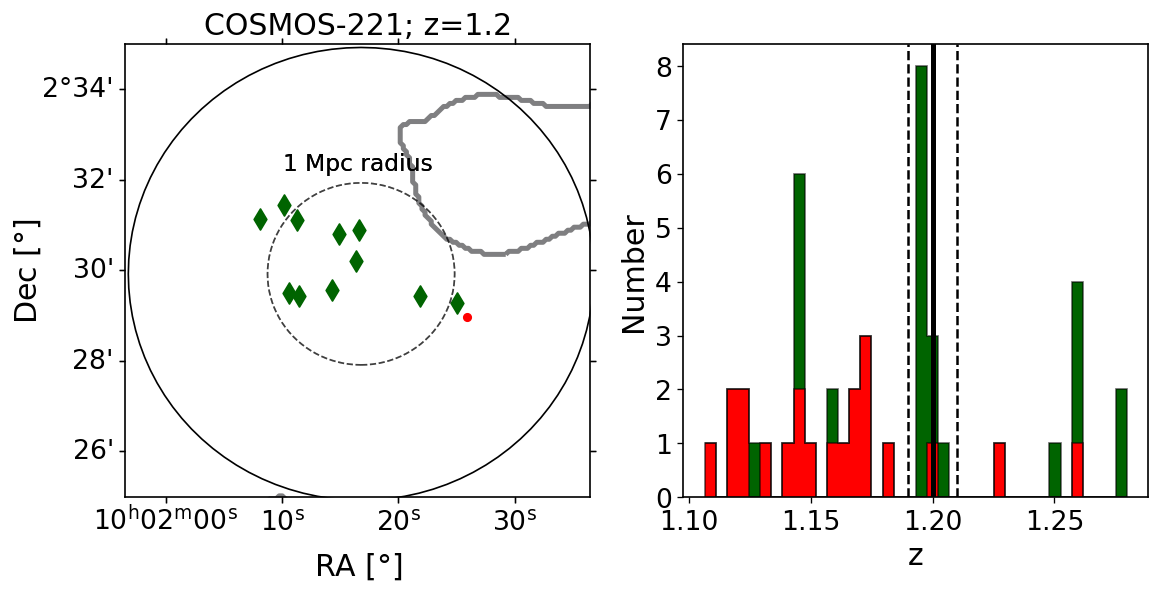}\\
	\caption{Groups in the COSMOS field spectroscopically targeted by GOGREEN. Left subplots: spatial distribution of galaxies with a spectroscopic redshift centred on the identified group centre and within the iterative velocity dispersion cut. GOGREEN targets are indicated with green diamonds. The solid grey contours indicate smoothed X-ray emission. Right subplots: distribution of spectroscopic redshifts within the circular field of view (solid line) shown on the left. The dashed vertical lines indicates the redshift selection of galaxies that are displayed on the corresponding left subplot.}
	\label{fig:gogreen-spec-and-xray-groups-cosmos}
\end{figure}

In Figure~\ref{fig:gogreen-spec-and-xray-groups-cosmos} we show the spatial and redshift distribution of galaxies with spectroscopic redshifts, in each of the 5\arcmin\ GMOS fields of view, with respect to the X-ray contours. 
In addition to GOGREEN, we include spectroscopy from available public sources as described in \S\ref{sec:spec-z-data}.
We calculate a new weighted centre for each group within 1 Mpc ($\sim 2 R_{200c}$) of the catalogued position, and then rerun an iterative 2.5$\sigma$-clipping routine to calculate the final redshift and velocity dispersion.
This procedure did not converge for two groups. In the case of COSMOS-125, the member candidates are spatially concentrated but redshift distribution does not show a clear peak. For COSMOS-63 it is the opposite, with a strong redshift overdensity but no spatial concentration of sources. 
Finally, for SXDF-76 we find two distinct groups along the line of sight. We keep both in the catalogue, and we have labelled them SXDF-76a and SXDF-76b.

\begin{figure} 
	\centering
		\includegraphics[width=\columnwidth]{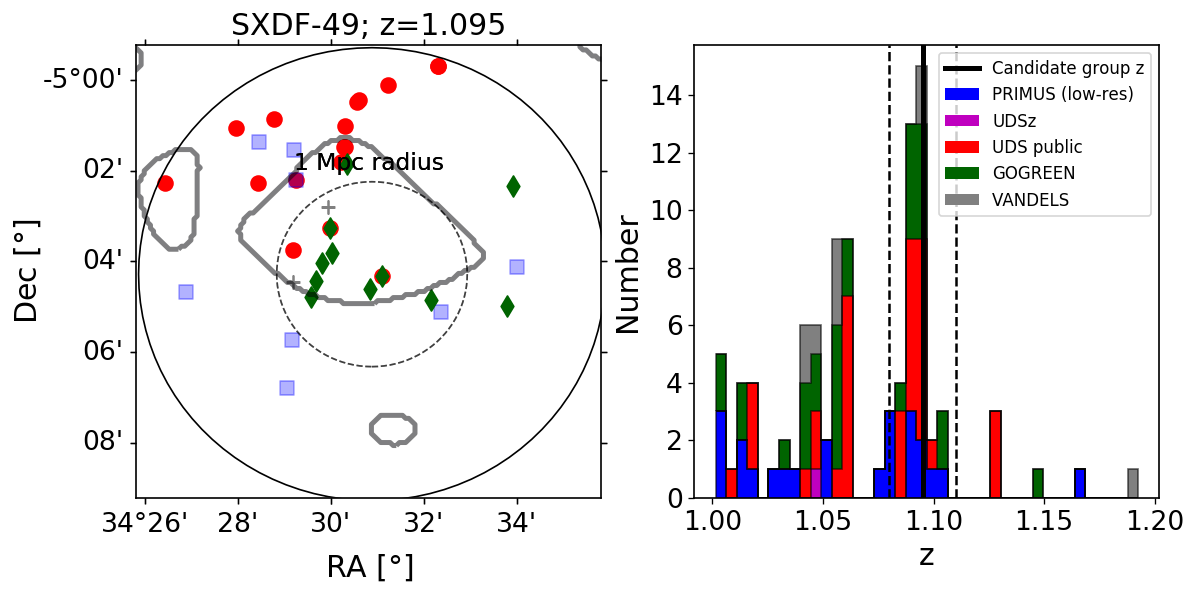}\\
		\includegraphics[width=\columnwidth]{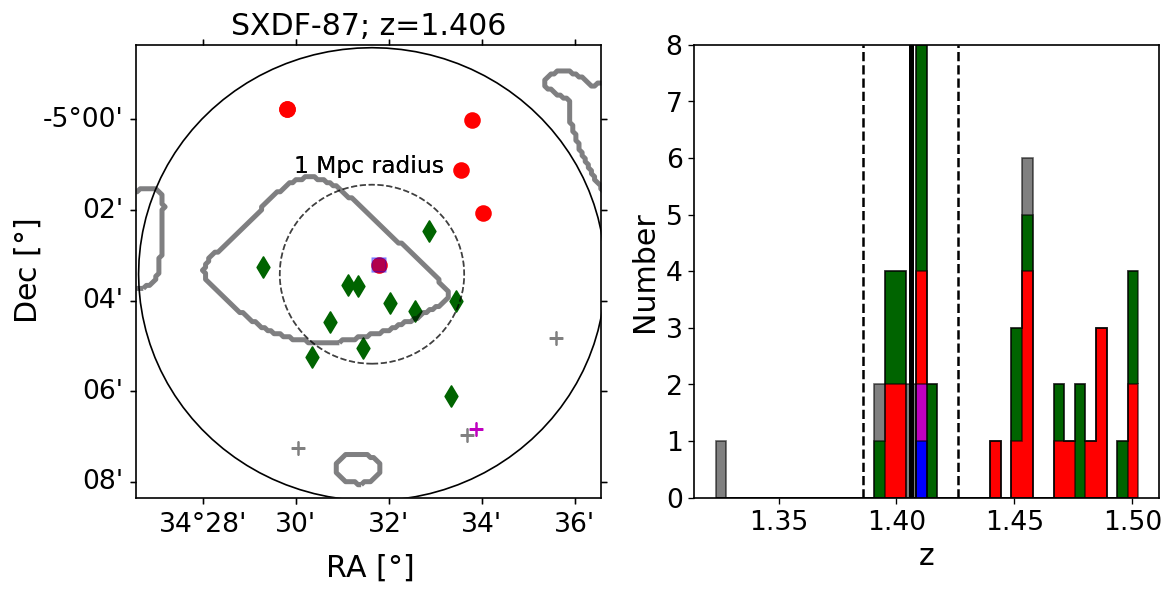}\\
		\includegraphics[width=\columnwidth]{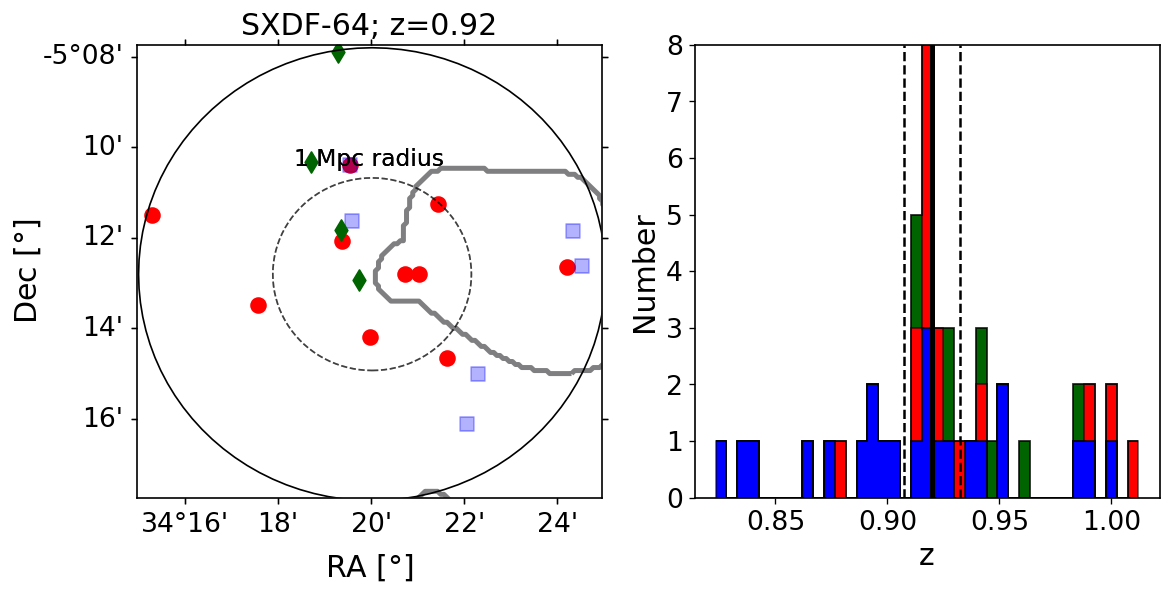}\\
		\includegraphics[width=\columnwidth]{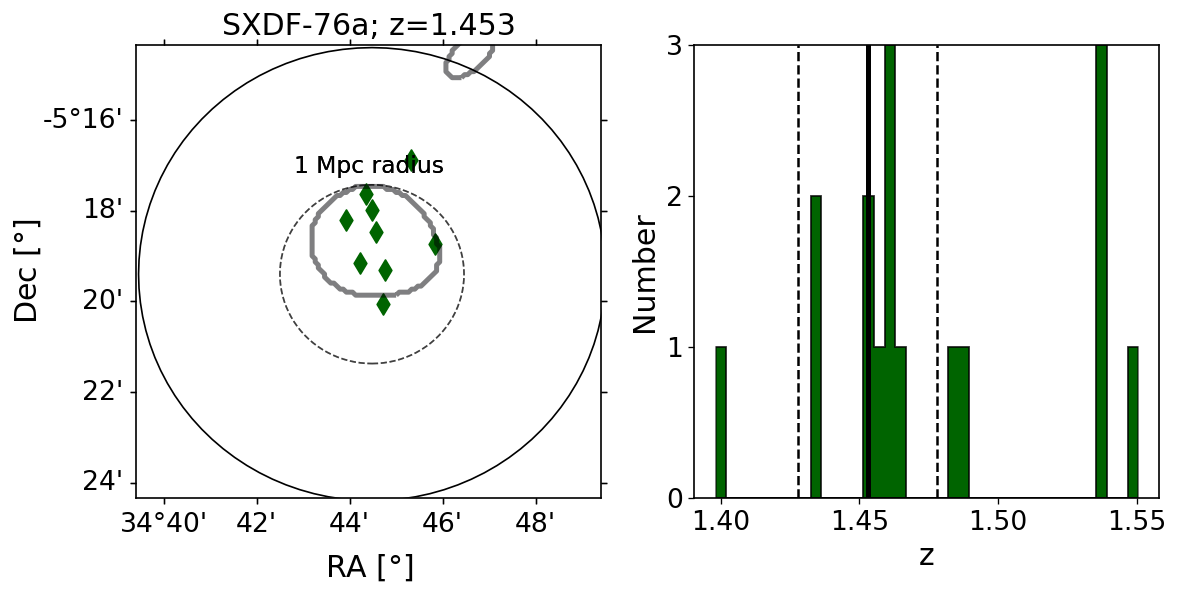}\\
		\includegraphics[width=\columnwidth]{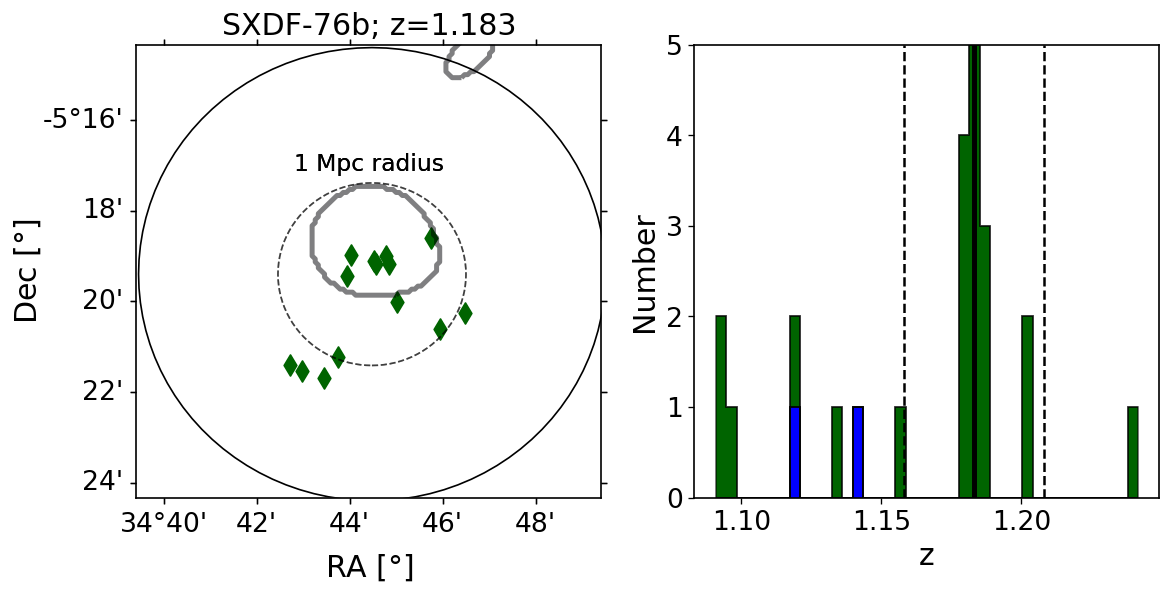}\\
	\caption{SXDF GOGREEN spectroscopically targeted groups. Analogous to Figure \ref{fig:gogreen-spec-and-xray-groups-cosmos}.}
	\label{fig:gogreen-spec-and-xray-groups-sxdf}
\end{figure}

\begin{figure} 
	\centering
		\includegraphics[width=\columnwidth]{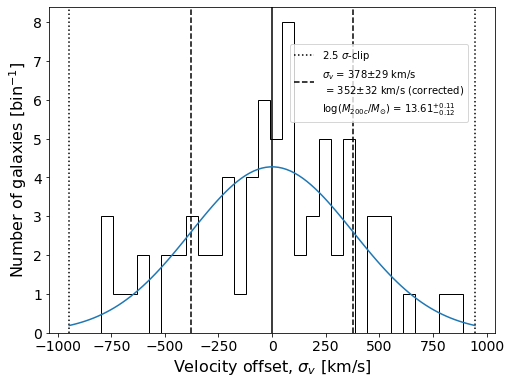}
	\caption{Distribution of rest-frame galaxy velocities in the ensemble for all 83 spectroscopic group members in 9 GOGREEN groups. Vertical lines indicate the mean of $0 \,{\rm km\,s}^{-1}$ (solid black), velocity dispersion (vertical dashed), and final 2.5$\sigma$ cut (black dotted). Thirty evenly sized bins were used, corresponding to a bin width of $\sim 56 \,{\rm km\,s}^{-1}$. The curved blue line indicates a Gaussian with mean 0 and standard deviation set to the velocity dispersion, normalized so the area corresponds to the total number of spectroscopic members.}
	\label{fig:stacked-group-velocity-dispersions}
\end{figure}

We estimate dynamical halo masses from these velocity dispersion values using the relation presented in \cite{Saro2013}. All galaxies within the 2.5$\sigma$ velocity cut and within 2$R_{200c}$ count as spectroscopic members for the purposes of Table \ref{tab:groups-table}, where we present all of these values.
We also compute an average velocity dispersion by stacking all group galaxies in an ensemble. We find $\sigma_v = 352\pm 32 \,{\rm km\,s}^{-1}$, which has been corrected by subtracting in quadrature the estimated individual redshift uncertainty of $280 \,{\rm km\,s}^{-1}$ observed-frame, from \cite{GOGREEN2021data}. This corresponds to a halo mass of log$(M_{200c}/M_{\text{solar}})=13.61 \pm^{+0.11}_{-0.12}$, again using the \cite{Saro2013} relation. The distribution of the velocities in this ensemble is shown in \ref{fig:stacked-group-velocity-dispersions}.

\subsection{Mass-weighted ages}\label{sec-MWA_app}

\begin{figure} 
	\centering
		\includegraphics[width=\columnwidth]{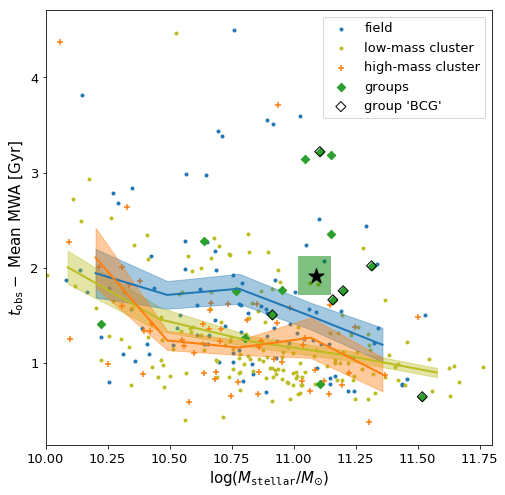}\\
	\caption{This figure is a more detailed complement to 
	Figure~\ref{fig:tform-model-vs-observations} in the main body of the paper. The points show measurements of $t_{\mathrm{obs}}-$MWA as a function of stellar mass, from stellar population synthesis modelling of quiescent galaxies in the GOGREEN spectroscopic sample \citep{WebbGOGREEN2020}. Individual $t_{\mathrm{obs}}-$MWA and stellar mass values are plotted for the field (blue dots), group (green diamonds), low-mass cluster (yellow dots), and high-mass cluster samples (small orange crosses). Groups, low-mass clusters, and high-mass clusters are the three halo mass bins explored at $1<z<1.5$ in \S\ref{sec:QFE-halo-mass-dependence}. The most massive galaxy in each group ('BCGs') are indicated with a black diamond (two groups had only one quiescent galaxy). As well, a running mean with bootstrapped standard deviation on the mean is shown for the field and cluster samples. For the groups, there are only 15 quiescent spectroscopic members so we simply plot the mean $t_{\mathrm{obs}}-$MWA and mean stellar mass of the full sample with a black star and a shaded region reflecting the bootstrapped errors. Quiescent galaxies in groups are not older than those in clusters, ruling out the predictions of our simple model without pre-processing.}
	\label{fig:observed-tobs-minus-MWA-appendix}
\end{figure}

Formation times of quiescent galaxies in the GOGREEN spectroscopic sample, from \citet{WebbGOGREEN2020}, are shown as a function of stellar mass in Figure \ref{fig:observed-tobs-minus-MWA-appendix}. This figure complements Figure \ref{fig:tform-model-vs-observations}, which only shows differences with the field and omits the individual measurements for the sake of clarity. Formation times are computed from the time of observation and the mass-weighted age, as $t_{\mathrm{form}} = t_{\mathrm{obs}}-$MWA. The running mean is additionally shown for the intermediate halo-mass bin used in the QFE vs halo mass analysis. The field, low-mass clusters, and high-mass clusters all show a declining trend of $t_{\mathrm{form}}$ with stellar mass. Both cluster sub-samples display ages about 200-300 Myr older than the field, as noted by \cite{WebbGOGREEN2020} for the entire cluster sample. The group sample, on the other hand, appears {\it younger} than the field by $\sim 150-200$ Myr, as noted in \S\ref{sec-modelresults}.

The small group sample has a high average galaxy stellar mass. To check that our results are not dominated by the most massive group galaxies, which could be central galaxies with a different formation history, we highlight these as green diamonds with black borders on Figure~\ref{fig:observed-tobs-minus-MWA-appendix}. There is no evidence that the ages of these galaxies are significantly different from other group members.

\section{Sensitivity of results to stellar mass binning}\label{sec-rebin_app}

\begin{figure*}
	\centering
		\includegraphics[width=2\columnwidth]{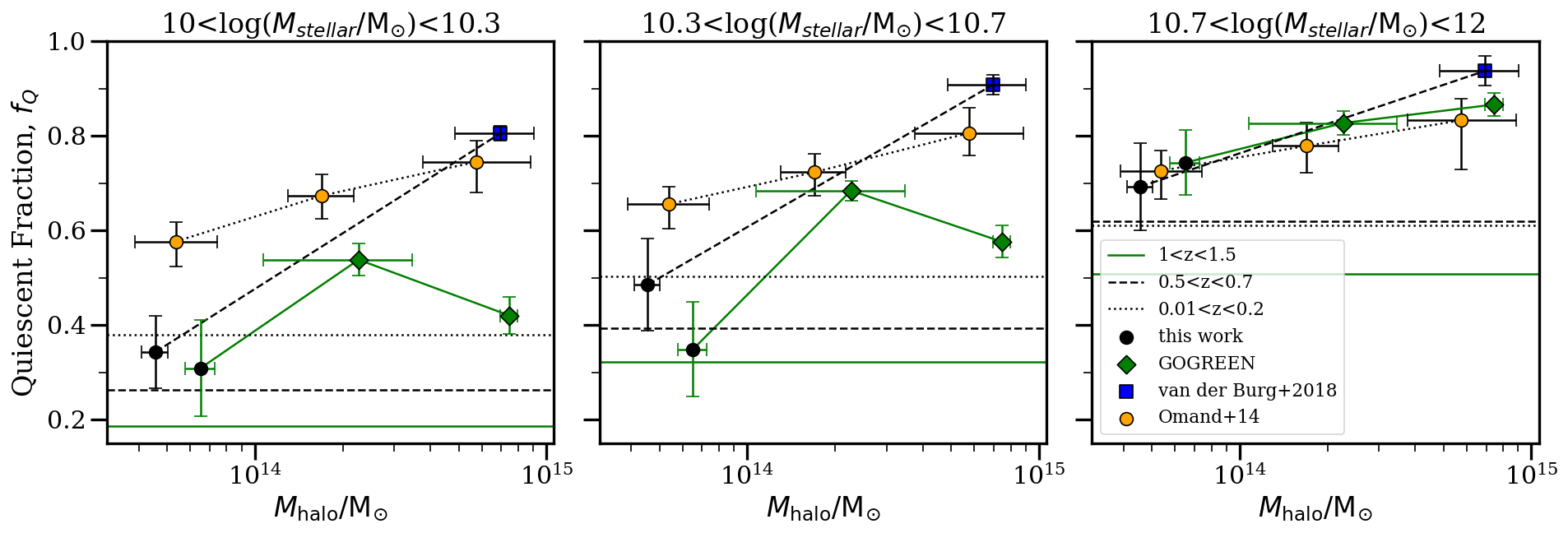}
		\includegraphics[width=2\columnwidth]{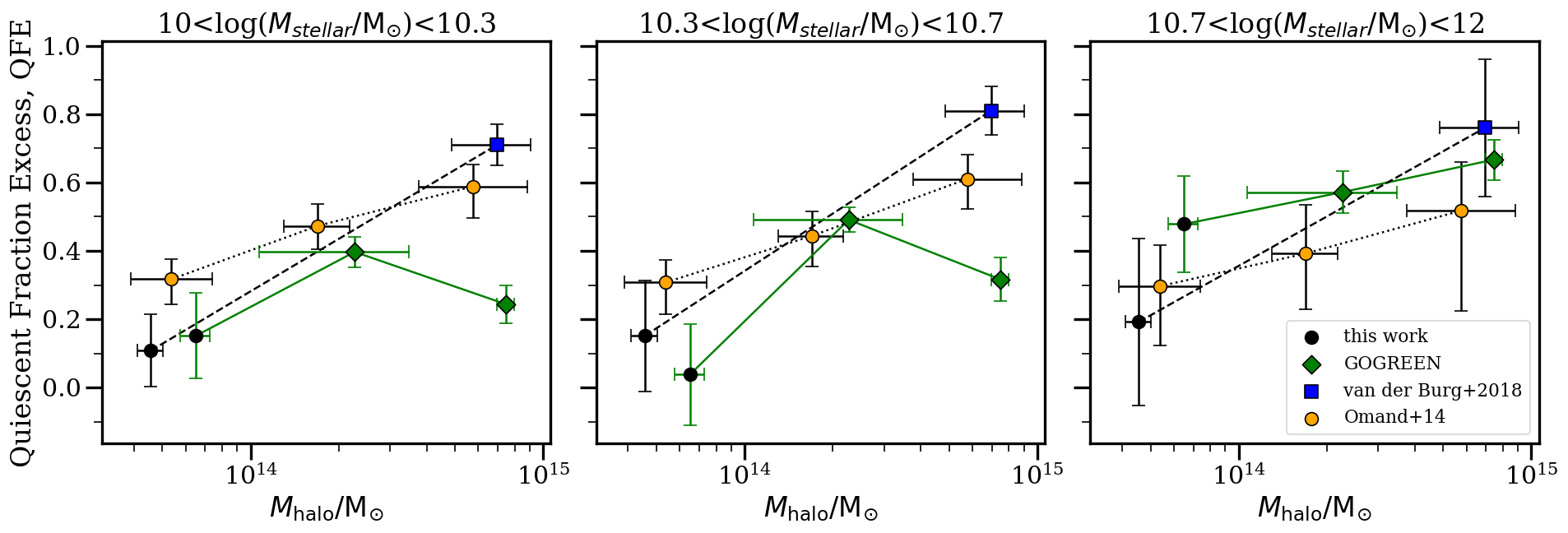}
	\caption{Quiescent fraction (top row) and quiescent fraction excess (bottom row) are shown as a function of halo mass ($M_{200c}$), $M_{\rm halo} / \mathrm{M}_{\odot}$, for three galaxy stellar mass bins (one stellar mass bin per column), and for samples at three different redshift ranges, as indicated. Our new measurements of low-mass haloes at $1<z<1.5$ are shown as the green points connected by a green solid line. The other samples are described in \S\ref{sec:comp-samp}. Horizontal lines represent the field at a given redshift range (errors are not significantly larger than the line widths on this plot). The bottom row is analogous to Figure \ref{fig:fQ-and-QE-vs-Mhalo-redshift-evolution-literature-summary-mstellar-breakdown}, with a further subdivision of that Figure's lowest stellar mass bin.
	}
	\label{fig:fQ-and-QE-vs-Mhalo-redshift-evolution-literature-summary-mstellar-breakdown-full}
\end{figure*}

In \S\ref{sec:QFE-halo-mass-dependence} we consider the halo mass dependence of the QFE in two broad stellar mass bins (Figure \ref{fig:fQ-and-QE-vs-Mhalo-redshift-evolution-literature-summary-mstellar-breakdown}), motivated by the qualitative change in QFE shown in Figure~\ref{fig:QFE-vs-Mstellar-GG-groups}. Here we subdivide the lower mass bin into two, to demonstrate explicitly that the results are similar in both bins. We show both the QFE, and the $f_Q$ values from which it is derived, in this binning in Figure~\ref{fig:fQ-and-QE-vs-Mhalo-redshift-evolution-literature-summary-mstellar-breakdown-full}. As claimed, the halo mass dependence observed in the two lowest stellar mass bins is very similar. In particular, the unusually high QFE observed for the Planck clusters at intermediate redshifts persists in both stellar mass bins. As discussed in \S\ref{sec:QFE-halo-mass-dependence}, we find the data in all three stellar mass bins are consistent with a QFE that depends on $\log(M_{\mathrm{halo}}/\mathrm{M}_\odot)$ with a slope of $m \approx 0.24 \pm 0.04$. 

\begin{figure*} 
	\centering
		\includegraphics[width=2\columnwidth]{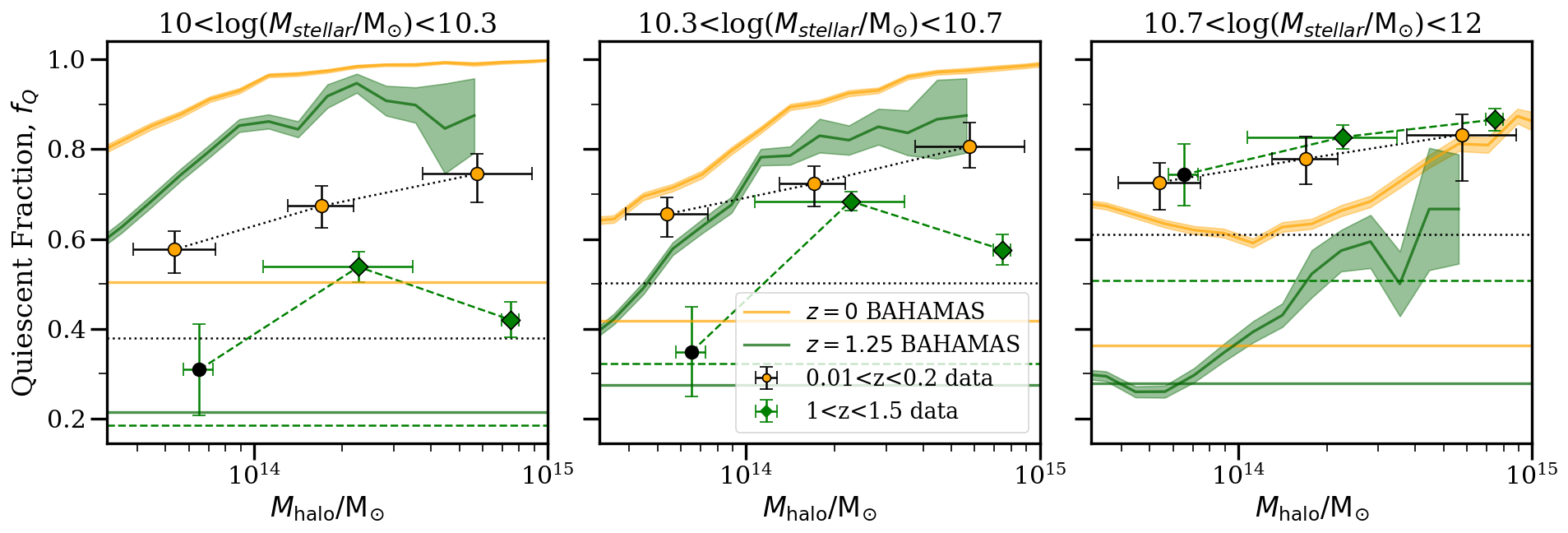}
		\includegraphics[width=2\columnwidth]{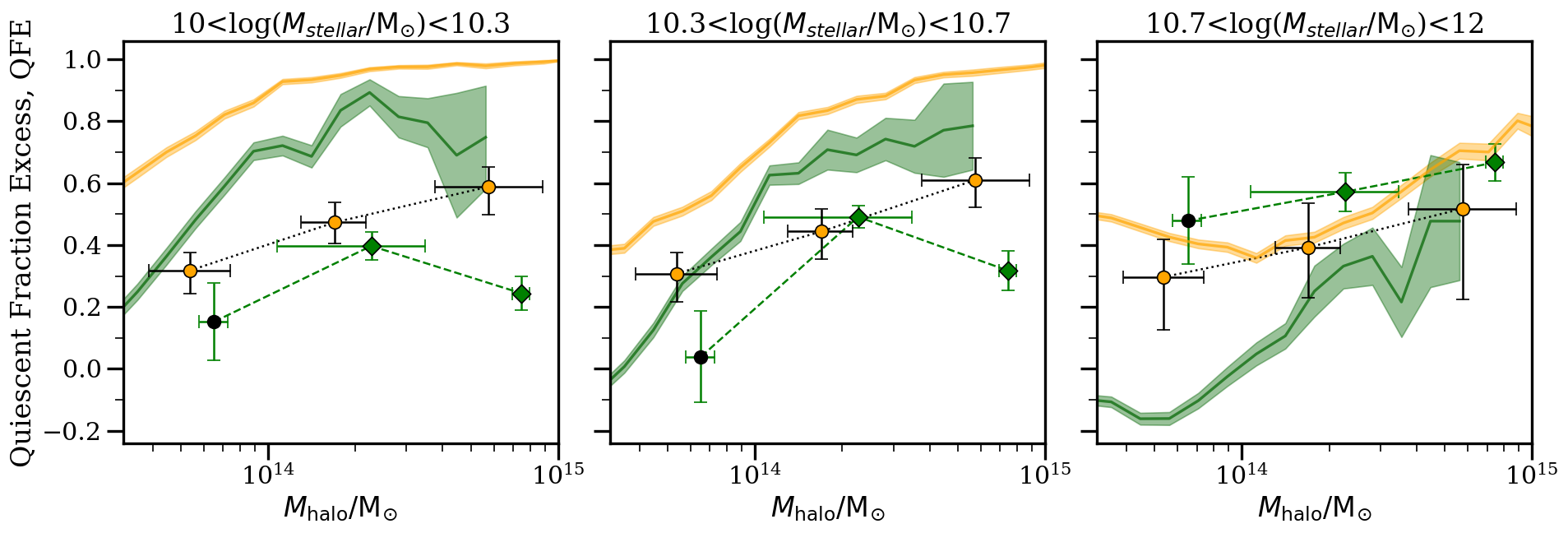}
	\caption{We show results from the BAHAMAS hydrodynamic simulation, for the quiescent fraction (top) and quiescent fraction excess (bottom) as a function of stellar and halo mass, $M_{\rm halo} / \mathrm{M}_{\odot}$ at two redshifts as indicated, with the same stellar mass binning as in Figure~\ref{fig:fQ-and-QE-vs-Mhalo-redshift-evolution-literature-summary-mstellar-breakdown-full}. The corresponding data from that Figure are shown, omitting the intermediate redshift sample for clarity.	In the simulations, both $f_Q$ and QFE decrease with increasing stellar mass, in contrast with the data. However, the correlation with halo mass and redshift is qualitatively similar to the trends observed in the data. The bottom row is analogous to Figure \ref{fig:fQ-and-QE-vs-Mhalo-redshift-evolution-BAHAMAS-mstellar-breakdown} in the main body of the paper, with the lowest stellar mass bin subdivided into two.
	}
	\label{fig:fQ-and-QE-vs-Mhalo-redshift-evolution-BAHAMAS-mstellar-breakdown-full}
\end{figure*}

In Figure~\ref{fig:fQ-and-QE-vs-Mhalo-redshift-evolution-BAHAMAS-mstellar-breakdown-full} we show the BAHAMAS simulation results (originally shown in Figure~\ref{fig:fQ-and-QE-vs-Mhalo-redshift-evolution-BAHAMAS-mstellar-breakdown}) for the same quantities and binning. It is readily apparent here that, at fixed halo mass, the predicted $f_Q$ and QFE in the simulations {\it decreases} with increasing stellar mass, in contrast with the observations. This behaviour is well-known and discussed further in Kukstas et al. (in preparation). Despite this, the correlation with halo mass is similar in both lower stellar mass bins -- it increases up to a halo mass of $\sim 2\times 10^{14}\mathrm{M}_\odot$ and becomes much shallower as halo mass increases further. This behaviour is not inconsistent with what we observe, with the possible exception of the intermediate redshift Planck-selected clusters. 

\bsp	
\label{lastpage}
\end{document}